\documentclass[12pt,preprint]{aastex}

\begin{document}
\title{X-ray Radiation Mechanisms and Beaming Effect of Hot Spots and Knots in Active Galactic Nuclear Jets}
\author{Jin Zhang\altaffilmark{1,2,5}, J. M. Bai\altaffilmark{1,4}, Liang
Chen\altaffilmark{1,5}, and Enwei Liang\altaffilmark{3}}
\altaffiltext{1}{National Astronomical Observatories/Yunnan Observatory,
Chinese Academy of Sciences, P. O. Box 110, Kunming, Yunnan, 650011, China;
zhang.jin@hotmail.com}\altaffiltext{2}{College of Physics and Electronic
Engineering, Guangxi Teachers Education University, Nanning, Guangxi, 530001,
China} \altaffiltext{3}{Department of Physics, Guangxi University, Nanning
530004, China} \altaffiltext{4}{Key Laboratory for the Structure and Evolution
of Celestial Bodies, Chinese Academy of Sciences} \altaffiltext{5}{The Graduate
School of Chinese Academy of Sciences\\}

\begin{abstract}
The observed radio-optical-X-ray spectral energy distributions (SEDs) of 22 hot
spots and 45 knots in the jets of 35 active galactic nuclei are complied from
literature and modeled with single-zone lepton models
. It is found that the observed luminosities at 5 GHz ($L_{\rm
5GHz}$) and at 1 keV ($L_{\rm 1keV}$) are tightly correlated, and the two kinds
of sources can be roughly separated with a division of $L_{\rm 1keV}= L_{\rm
5GHz}$. Our SED fits show that the mechanisms of the X-rays are diverse. While the X-ray emission of a small fraction of the sources is a
simple extrapolation of the synchrotron radiation for the radio-to-optical
emission, an inverse Compton (IC) scattering component is necessary to model the X-rays
for most of the sources. Considering the sources at rest (the Doppler
factor $\delta=1$), the synchrotron-self-Compton (SSC) scattering would
dominate the IC process. This model can interpret the X-rays of some hot spots with a magnetic field strength ($B_{\rm ssc}^{\delta=1}$) being
consistent with the equipartition magnetic field ($B_{\rm eq}^{\delta=1}$)
in one order of magnitude, but an
unreasonably low $B_{\rm ssc}^{\delta=1}$
is required to model the X-rays for all knots. Measuring the deviation between $B_{\rm ssc}^{\delta=1}$ and $B_{\rm
eq}^{\delta=1}$ with ratio $R_B\equiv B_{\rm eq}^{\delta=1}/B_{\rm
ssc}^{\delta=1}$, we find that $R_B$ is greater than 1 and it is tightly
anti-correlated with ratio $R_L\equiv L_{\rm 1keV}/L_{\rm 5GHz}$ for both
the knots and the hot spots. We propose that the deviation may be due to the neglect of the relativistic bulk motion for these
sources. Considering this effect, the IC/CMB component would dominate the IC process. We show that the IC/CMB model well
explains the X-ray emission for most sources under the equipartition condition.
Although the derived beaming factor ($\delta$) and co-moving equipartition magnetic field ($B_{\rm eq}^{'}$)
of some hot spots are comparable to the knots, the $\delta$ values of the hot spots tend to be smaller and their $B_{\rm
eq}^{'}$ values tend to be larger than that of the knots, favoring the idea that the hot spots are jet
termination and knots are a part of a well-collimated jet. Both
$B_{eq}^{'}$ and $\delta$ are tentatively correlated with $R_L$. Corrected
by the beaming effect, the $L^{'}_{\rm 5GHz}-L^{'}_{\rm 1keV}$ relations for
the two kinds of sources are even tighter than the observed ones. These facts
suggest that, under the equipartition condition, the observational differences of the X-rays from the knots and hot spots may be mainly due the differences on the Doppler boosting effect and the co-moving magnetic field of the two kinds of sources. Our IC scattering models predict a prominent GeV-TeV
component in the SEDs for some sources, which are detectable with H.E.S.S. and
\emph{Fermi}/LAT.

\end{abstract}


\section{Introduction}           
\label{sect:intro} Hot spots and knots in large-scale
jets have been observed in many active galactic nuclei (AGNs). Hot
spots are often found near the outmost boundaries of radio lobes. They are
regarded as jet termination (Fnanaroff \& Riley 1974; Blandford \& Rees 1974;
Begelman et al. 1984; Bicknell 1985; Meisenheimer et al. 1989). Knots are
usually thought to be a part of a well-collimated jet (e.g., Harris \&
Krawczynski 2006).  Some of them are also detected in the optical and X-ray bands. With high spatial resolution and sensitivity, the \emph{Chandra} X-ray
Observatory opened a new era to study the X-ray emission of hot spots and
knots. It revealed that bright radio knots and hot spots in radio galaxies and
quasars are also often detected in the X-ray band (Wilson et al. 2001;
Hardcastle et al. 2004; Kataoka \& Stawarz 2005; Tavecchio et al. 2005).

The X-ray radiation mechanisms of hot spots and knots are highly debated. It is
generally believed that synchrotron radiation of relativistic electrons is
responsible for the radio emission. The detection of high polarization in the optical emission indicates that the
optical emission is also from synchrotron radiation (Roser \& Meisenheimer
1987; L\"{a}hteenm\"{a}ki \& Valtaoja 1999). It is uncertain whether the X-ray
emission is a simple extrapolation of the synchrotron radiation for the
radio-to-optical emission. Indeed, the X-rays of some hot spots (such as hot
spots N1 and N2 in 3C 33; Kraft et al. 2007) and knots (such as the knots in 3C
371 and M87; Sambruna et al. 2007; Liu \& Shen 2007) can be interpreted as
synchrotron radiation from the same electron population
responsible for the radio and optical emission. However, the X-ray emission of some hot spots and knots is apparently not a
simple extrapolation of the radio-to-optical component (Kataoka \& Stawarz
2005). An inverse Compton scattering (IC) component is necessary to model
the X-rays. As shown by Stawarz et al. (2007), the X-ray emission of hot spots in Cygnus A
is well interpreted with the synchrotron-self-Compton (SSC) scattering model. The
IC scattering of cosmic microwave background (IC/CMB) may also significantly
contribute to the observed X-ray emission, if the bulk flow of the material in
hot spots and knots is relativistic (Georganopoulos \& Kazanas 2003).

Broadband spectral energy distribution (SED) places strong constraints on the
radiation mechanism models. Systematical analysis on the SEDs in the radio and
X-ray bands of hot spots, knots and lobes were present by Hardcastle et al.
(2004) and Kataoka \& Stawarz (2005). Note that the optical data are
critical to characterize the SEDs, and may give more constraints on the
models. In this paper, we compile a large sample of the observed SEDs in the
radio-optical-X-ray band of hot spots and knots from literature, and fit them
with various models in order to study the radiation mechanisms of the X-rays
and to reveal the differences of the two kinds of sources.

The magnetic field and Doppler boosting effect are crucial ingredients in SED
modeling. The energy equipartition assumption between magnetic field and
radiating electrons is usually accepted in discussion of the energetics and
dynamics of radio sources. Hardcastle et al. (2004) reported that most
radio-luminous hot spots can be explained with the SSC model under this
assumption. With the equipartition condition, the derived magnetic field of
lobes is $\sim10^{-6}$G (Kataoka \& Stawarz 2005), being comparable to the
strength of the magnetic field in the interstellar medium. Since hot spots are
believed to be the terminal of a relativistic jet, the Doppler boosting effect
would be less prominent than that in knots, but their magnetic fields may be
magnified up to $\sim10^{-4}$G by strong external shocks produced by
interaction of a relativistic jet with circum medium (Kataoka \& Stawarz 2005).
Therefore, the Doppler boosting effect and co-moving magnetic field are
essential to discriminate the two kinds of sources, if they are physically
different. With our detailed SED fits, we compare their Doppler factor ($\delta$)
and co-moving magnetic field ($B^{'}$) between the knots and hot spots in our sample under the equipartition assumption.

The observed SEDs and our data analysis are present in \S 2. Models and our SED
fits are shown in \S 3. Conclusions and discussion are present in \S4.
Throughout, $H_{0}$=70 km s$^{-1}$ Mpc$^{-1}$, $\Omega_{\Lambda}=0.7$ and
$\Omega_{m}=0.3$ are adopted.

\section{Sample and Data Analysis}
\label{sect:data} Twenty-two hot spots and 45 knots from 35 AGNs (15 radio
galaxies, 16 quasars, 3 BL Lac objects, and 1 Seyfert galaxy; see Table 1) are
included in our sample. Most of them are taken from XJET Home
Page\footnote{Http://hea-www.harvard.edu/XJET/.}. Observations for these hot
spots and knots are summarized in Table 2, and their SEDs are displayed in
Figure 1. We show the distributions and the correlations of the spectral indices in the
radio and X-ray bands ($\alpha_r$ and $\alpha_X$) in Figure 2. It is found that
$\alpha_r$ is not correlated with $\alpha_X$. The $\alpha_r$ of both the knots
and hot spots are smaller than 1. The $\alpha_X$ narrowly clusters at $0.7\sim
1.2$ for the hot spots, but it ranges in $0.2\sim 2.4$ for the knots. We test
whether $\alpha_r$ and $\alpha_X$ distributions show statistical differences
between the two kinds of sources with the Kolmogorov-Smirnov test (K-S test),
which yields a chance probability $p_{\rm KS}$. A K-S test probability larger
than 0.1 would strongly suggest no statistical difference between two distributions. We
get $p_{\rm KS}=0.37$ and $p_{\rm KS}=0.41$ for the radio spectral indices and
the X-ray spectral indices between the two samples, respectively, indicating
that no statistical difference is found for the $\alpha_r$ and $\alpha_X$
distributions of two kinds of sources.

Figure \ref{Fig:3}(a) shows the correlations between the observed luminosities
at 5 GHz and 1 keV ($L_{\rm 5GHz}$ and $L_{\rm 1keV}$) for the hot spots and
the knots. It is found that $L_{\rm 1keV}$ is tightly correlated with $L_{\rm
5GHz}$. We measure the correlations with the Spearman correlation analysis,
which yields $\log L_{\rm 1keV}=(6.2\pm 4.7)+(0.84\pm 0.11)\log L_{\rm 5GHz}$
with a correlation coefficient $r=0.86$ and a chance probability $p<10^{-4}$
for the hot spots and $\log L_{\rm 1keV}=(6.9\pm 2.2)+(0.85\pm 0.05)\log L_{\rm
5GHz}$ with $r=0.92$ and $p<10^{-4}$ for the knots. The slopes of the two
correlations are the same in the error scopes, but averagely speaking, the
X-ray luminosity of the knots is larger than that of the hot spots with $\sim
0.7$ order of magnitude, indicating systematical difference between the two
kinds of sources. As seen in \S 3.3, this correlation is much tighter by
correcting with the Doppler boosting effect (see Figure \ref{Fig:3}(b) and details
in \S 3.3). The knots and the hot spots in the $L_{\rm 5GHz}$-$L_{\rm 1keV}$
plane are roughly separated with a division line of $L_{\rm 1keV}=L_{\rm
5GHz}$. Therefore, the ratio of $R_L\equiv L_{\rm 5GHz}/L_{\rm 1keV}$ is a
characteristic to distinguish the hot spots and the knots. This ratio should be
an intrinsic parameter independent of the Doppler boosting effect and the
cosmological effect. It may reflect the properties of the radiation regions.
\section{modeling the Observed SEDs}
The tight $L_{\rm 5GHz}$-$L_{\rm 1keV}$ correlation indicates that the
radiations in the two energy bands may be produced by the same electron
population. For few cases, such as hot spots N1 and N2 in 3C 33 and knots in
M87 (Kraft et al. 2007; Liu \& Shen 2007), their X-ray spectra are soft with
$\alpha_X>1$, and smoothly connect to the spectrum in the radio and optical
band. The X-rays of these sources may be the high energy tail of the
synchrotron radiation by the same electron population for the radio and optical
emission. The synchrotron radiation model is preferred to fit the X-rays of
these sources.

Some well-sampled SEDs in Figure 1 roughly show two bumps similar to that
observed in blazars. The two-bump feature is generally interpreted with the
synchrotron radiation and IC scattering by the relativistic electrons. Therefore, we fit these SEDs with
single-zone
synchrotron + IC scatting models. The IC seed photons may be originated from the synchrotron radiation itself
(SSC) or from the CMB. The photon field energy density of synchrotron radiation
in the co-moving frame is given by $U^{'}_{\rm syn}=L_{\rm syn}/(4\pi
R^{2}c\delta^{4})\approx 2.65\times10^{-12}L_{\rm
syn,40}R^{-2}_{20}\delta^{-4}$ erg cm$^{-3}$, where $Q_{n}=Q/10^{n}$ in cgs
units, $\delta$ the beaming factor, $R$ the radius of the radiation region, and
$c$ the speed of light. The energy density of the CMB is $U_{\rm
CMB}\approx4\times10^{-13}(1+z)^{4}\Gamma^{2}$ erg cm$^{-3}$, which
dramatically increases with the redshift $z$ of the sources and the bulk
Lorentz factor $\Gamma$ (taking $\Gamma\simeq\delta$) of the radiation site.
Without considering the Doppler boosting effect, the IC component should be
dominated by the SSC process, but it may be dominated by the IC/CMB, if the
source is relativistic motion. We take the two scenarios into account.

In our models, the radiation region is assumed to be a homogeneous sphere with
radius $R$. The radius is derived from the angular radius $\theta$ (see table
2), which is obtained from the optical or the X-ray observations. Considering
the beaming effect, \emph{R} and volume \emph{V} of the emitting region are
needed to take a relativistic transformation. We simply assume that the
emitting region still is a sphere with $V^{'}=V/\delta$ and derive the radius
of emitting region in co-moving frame by $\emph{R}^{'}=(3V^{'}/4\pi)^{1/3}$.
The electron distribution as a function of electron energy ($\gamma$) is taken
as a single power-law or a broken power law,
\begin{equation}
N(\gamma )= N_{0}\left\{ \begin{array}{ll}
                   \gamma ^{-p_1}  &  \mbox{ $\gamma \leq \gamma _b$}, \\
          \gamma _b^{p_2-p_1} \gamma ^{-p_2}  &  \mbox{ $\gamma > \gamma _b$,}
           \end{array}
       \right.
\end{equation}
where $p_{1,2}=2\alpha_{1,2}+1$ are the energy
indices of electrons below and above the break energy $\gamma_{b}$, and $\alpha_{1,2}$ are the observed spectral indices.

In our calculations, the Klein-Nishina effect for the radiation in the GeV-TeV
band is considered, but the absorption in the GeV-TeV band by the infrared
background light and by CMB photon during the gamma-ray photons propagating to
the Earth (Stecker et al. 2006) is not taken into account.

\subsection{Equipartition Magnetic Field and the Synchrotron Radiation model}
As mentioned in \S 1, the equipartition condition, which assumes that the
magnetic field energy density $U_{B}$ is equal to the electron energy density
$U_{e}$, is usually adopted in discussion of the X-ray origin. We first derive
the magnetic field strength $B_{\rm eq}^{\delta=1}$ (see Appendix A) under this
condition for the hot spots and knots in our sample without considering the
beaming effect ($\delta=1$). The calculation of $B_{\rm eq}^{\delta=1}$ depends
on $\gamma_{\min}$ (see Eqs. A5 and A10). The $\gamma_{\min}$ is quite
uncertain (e.g., Harris \& Krawczynski 2006). The $\gamma_{\min}$ values of 12
hot spots and all the knots in our sample are constrained with the observed
SEDs via a method reported by Tavecchio et al. (2000). The average of
$\gamma_{\min}$ for the 12 hot spots is $\sim 200$. For those hot spots that
their $\gamma_{\min}$ lost constraints from the observed SEDs, we take
$\gamma_{\min}=200$ in our calculation.

We fit the observed SEDs in the radio-optical band with the synchrotron
radiation model to derive $B_{\rm eq}^{\delta=1}$. Our results are reported in
Table 2. They are roughly consistent with the results derived from the formulae
given by Brunetti et al. (1997). The distributions of $B_{\rm eq}^{\delta=1}$
for both the hot spots and knots are shown in Figure \ref{Fig:4}(a). They range
in $10\sim 700$ $\mu$G. No systematical difference of $B_{\rm eq}^{\delta=1}$
is found between the two kinds of sources.

The observed X-ray spectra of the knots in  4C 73.18 (K-A), PKS 0521 (K), M87
(K-A, B, C1, D, E, F), 3C 31 (K), 3C 66B (K-B), and 3C 120 (K-K7) smoothly
connect to the spectra in the radio and optical band. They are well fit with
the synchrotron radiation model (the thin solid line in Figure \ref{Fig:1}), indicating
that the X-rays of these sources are the high energy tail of the synchrotron
radiation by the same electron population for the radio and optical emission.  We do not take these
sources into account in our following analysis.

\subsection{Representing the X-rays with SSC model for the Sources at Rest}
\begin{figure*}
\includegraphics[angle=0,scale=0.80]{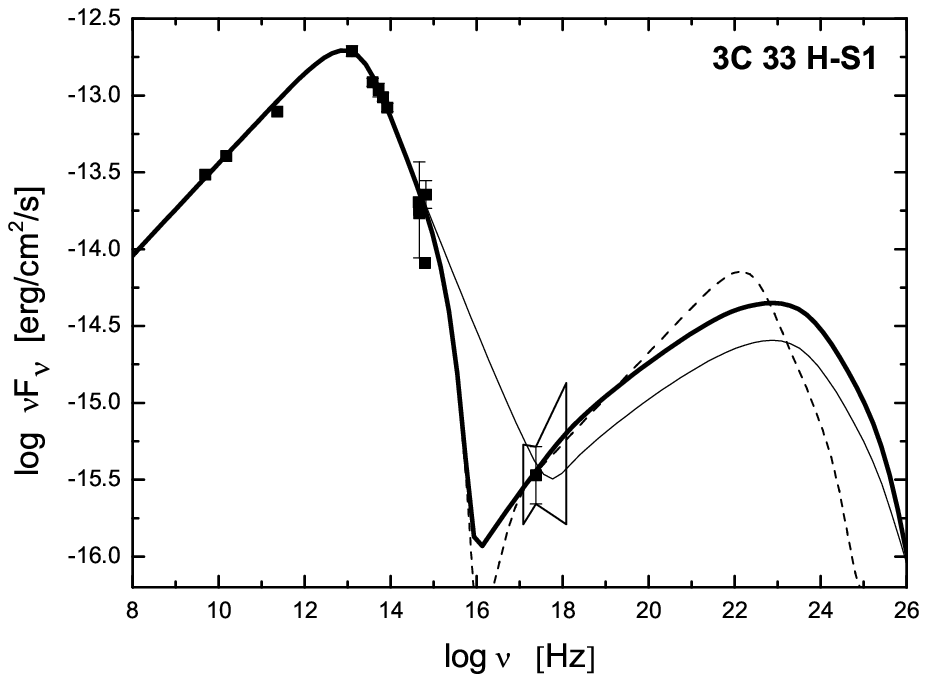}
\includegraphics[angle=0,scale=0.80]{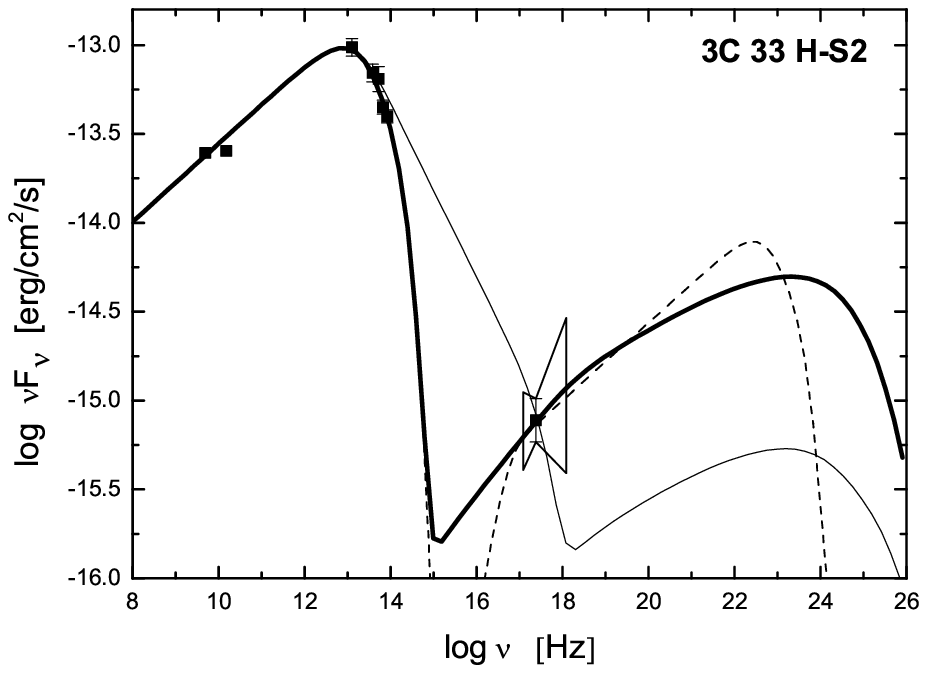}
\includegraphics[angle=0,scale=0.80]{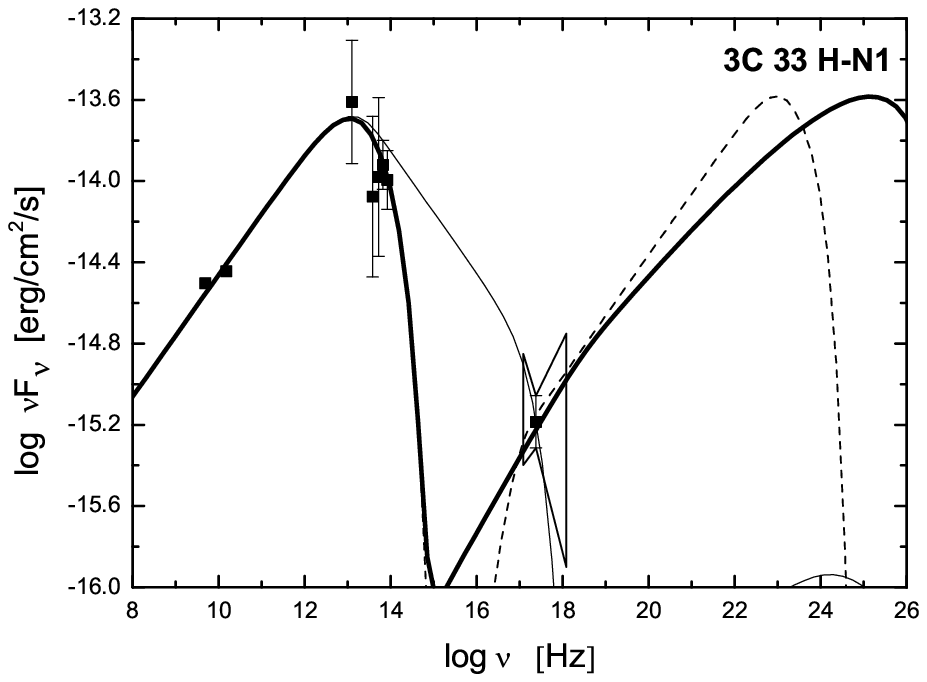}
\includegraphics[angle=0,scale=0.80]{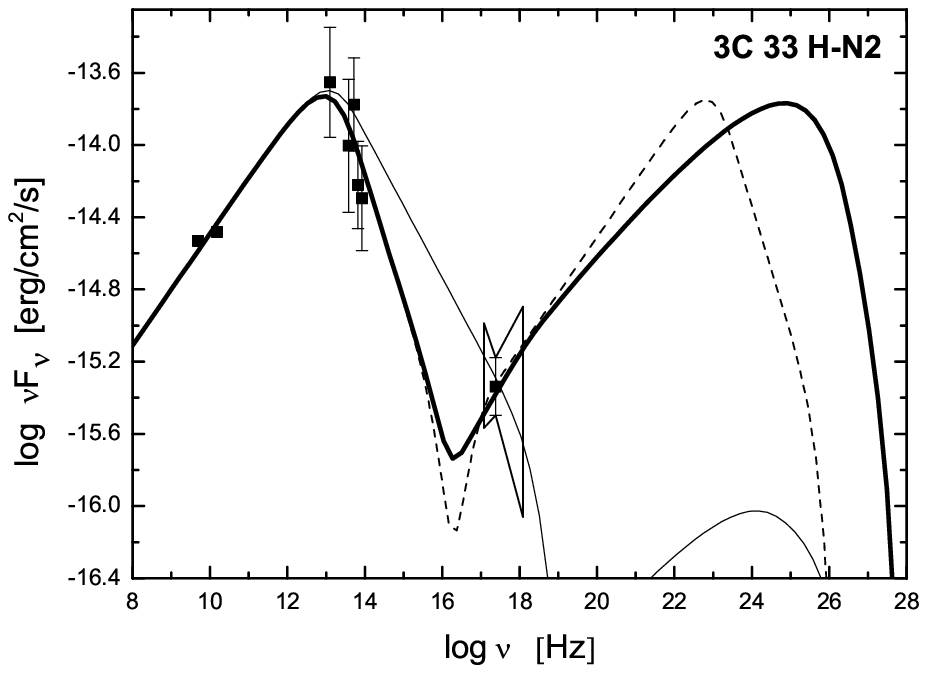}
\includegraphics[angle=0,scale=0.80]{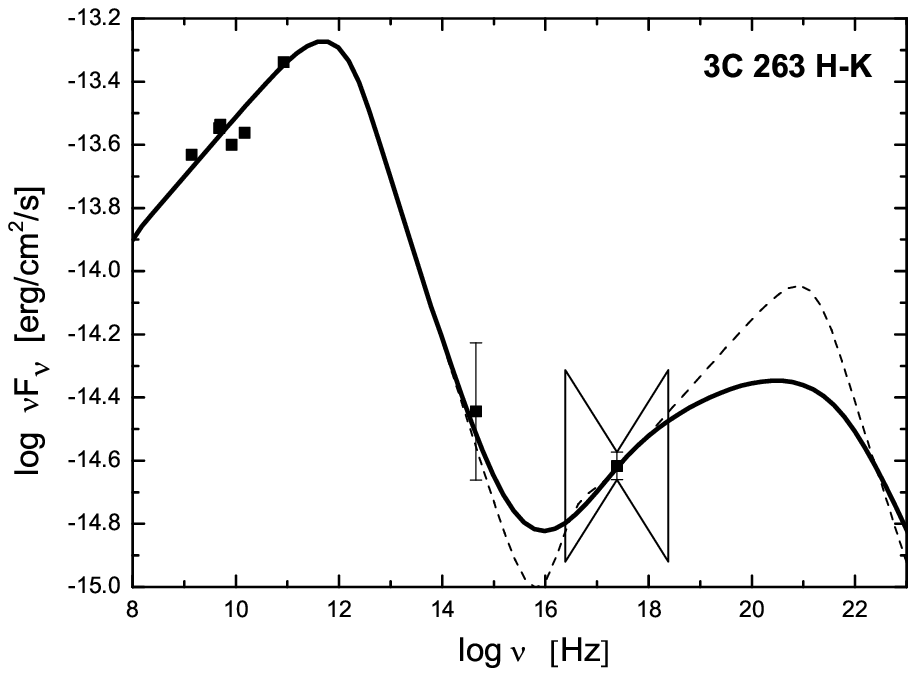}
\hfill
\includegraphics[angle=0,scale=0.80]{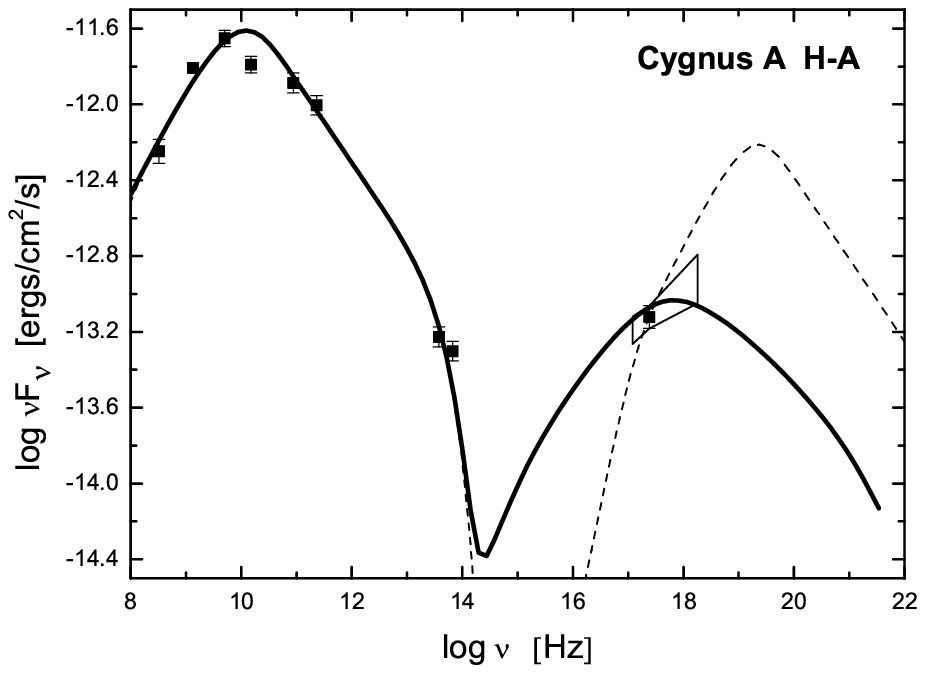}
\caption{Observed SEDs ({\em squares}) with our model fits: {\em thick solid}
line---SSC model with $\delta=1$; {\em dashed} line ---IC/CMB by
considering the beaming effect; {\em thin solid} line
---synchrotron radiation. The uncertainty of the X-ray flux is shown as a bow-tie symbol. The thresholds of \emph{Fermi}/LAT and H.E.S.S. are also marked for the sources with predicted GeV-TeV flux over the thresholds.} \label{Fig:1}
\end{figure*}

The observed X-ray spectral indices and SEDs shown in Table 2 and Figure
\ref{Fig:1} indicate that the X-rays of most hot spots and knots should be
contributed by IC scattering. We model the SEDs with the synchrotron + IC model
assuming that the sources are at rest. In this scenario, the SSC process should
dominate the IC process, as mentioned above. Although the contribution of
IC/CMB to the X-ray emission is not negligible for some sources at high
redshift, we only consider the SSC component in this section.

We first calculate the X-ray flux density at 1 keV ($F_{\rm 1keV}^{\rm eq}$)
with the synchrotron + SSC model under the equipartition assumption, i.e., $B=B_{\rm
eq}^{\delta=1}$. Our results are reported in Table 2. We measure the
consistency between $F_{\rm 1keV}^{\rm eq}$ and $F_{\rm 1keV}^{\rm obs}$ with
ratio $R_{F}\equiv F_{\rm 1keV}^{\rm obs}/F_{\rm 1keV}^{\rm eq}$, where $F_{\rm
1keV}^{\rm obs}$ is the observed flux density at 1 keV. It is found that $R_F$
is much larger than 1 for almost all the hot spots and knots in our sample,
indicating that the observed X-ray flux density is much larger than the model
prediction. The distributions of $R_F$, and $R_F$ as a function of $L_{\rm
5GHz}$, $L_{\rm 1keV}$, and $R_L$ for the hot spots and knots are shown in
Figure \ref{Fig:5}.  A tentative correlation presented in the $R_F-L_{\rm
5GHz}$ plane shows that the brighter sources in the radio band tend to be more
consistent with the equipartition condition (see Figure \ref{Fig:5}(b)). However,
no similar feature is seen for the X-ray bright sources (see Figure
\ref{Fig:5}(c)). It is interesting that $R_F$ is anti-correlated with $R_L$, and
both the hot spots and knots shape a well sequence (see Figure \ref{Fig:5}(d)).
The hot spots are at the lower end of the sequence and they tend to be closer
to the equipartition condition than the knots.

As shown above, the derived X-ray flux densities from the SSC model under the
equipartition condition significantly deviate the observations, especially for
the knots. In order to model the observed SEDs with the synchrotron + SSC model, we have
to get rid of this assumption. Keeping the model parameters the same as that
used above, we fit the SEDs with this model and derive the magnetic field
strengths ($B^{\delta=1}_{\rm ssc}$). The fits are shown in Figure \ref{Fig:1}
(the thick solid line). Although the SSC model can represent the observed flux
at 1 keV for the knots in 4C 73.18 (K-A), PKS
0521, M87 (K-A, B, C1, D, E, F), 3C 31, 3C 66B (K-B), 3C 120 (K-K7), 3C 273 (K-C1, C2, D1, D2H3), the bow ties defined by the errors of the X-ray spectral indices
clearly rule out this model for these knots. As discussed in Section 3.1, the X-rays of these knots are well fitted by the
synchrotron radiation model except knots in 3C 273. We do not include these knots in our following statistics.

The derived $B^{\delta=1}_{\rm ssc}$ are listed in Table 2, and their
distributions are shown in Figure \ref{Fig:4}(b). It is found that the
$B^{\delta=1}_{\rm ssc}$ of the knots are much smaller than the hot spots, with
medians of $1\mu$G and $\sim 30$ $\mu$G for the knots and hot spots,
respectively. Comparing  $B^{\delta=1}_{\rm ssc}$ with $B^{\delta=1}_{\rm eq}$, it is found that they are roughly consistent for the hot spots, indicating that the X-rays of the hot spots can be roughly fitted with the SSC model under the
equipartition condition. However, the $B^{\delta=1}_{\rm ssc}$ of the knots are
much smaller than $B^{\delta=1}_{\rm eq}$, even unreasonably smaller than the
magnetic field strength of the interstellar medium for some knots. Therefore, the X-rays
of these knots may not be dominated by the SSC component.

To investigate the deviation of $B^{\delta=1}_{\rm ssc}$ to $B^{\delta=1}_{\rm
eq}$ for individual source, we define ratio $R_B\equiv B^{\delta=1}_{\rm
eq}/B^{\delta=1}_{\rm ssc}$, which is physically the same as $R_F$. Similar to
$R_F$, $R_B$ is much larger than 1 for almost all the sources, especially for
the knots. The distributions of $R_B$ and $R_B$ as a function of $L_{\rm
5GHz}$, $L_{\rm 1keV}$, and $R_L$ are shown in Figure \ref{Fig:6}. We find the
same features as shown in Figure \ref{Fig:5}.

\subsection{Modeling the X-Rays by Considering Relativistic Bulk Motion}
It is generally believed that the knots should have relativistic motion. The
hot spots may also be relativistic (Dennett-Thorpe et al. 1997;
Tavecchio et al. 2005; Harris \& Krawczynski 2006). In this section, we fit the
SEDs with the synchrotron + IC model by considering the beaming effect under the equipartition condition. In this scenario, the IC/CMB process should dominate the IC component.
Although the contribution of SSC is negligible comparing with IC/CMB in this case, we still take SSC process into account in our calculation.

Our fits are shown in Figure \ref{Fig:1} (the dashed line). The distributions
of $\delta$ for the knots and hot spots are shown in Figure \ref{Fig:7}. It is
found that, averagely,  $\delta\sim 10$ for most of the knots and $\delta\sim
5$ for most of the hot spots. Some sources in our sample are included in
Kataoka \& Stawarz (2005). We compare our results of $\delta$ for these sources
with that reported by Kataoka \& Stawarz (2005) ($\delta^{\rm KS05}$) in Figure
\ref{Fig:8}. They are roughly consistent.

The $B_{\rm eq}^{'}$ distributions with comparison to $B_{\rm eq}^{\delta=1}$
and $B_{\rm SSC}^{\delta=1}$ are shown in Figure \ref{Fig:4}(c). The $B_{\rm
eq}^{'}$ distributions are more consistent with $B_{\rm eq}^{\delta=1}$ than
$B_{\rm SSC}^{\delta=1}$. Both $B_{\rm eq}^{'}$ and $B_{\rm eq}^{\delta=1}$
distributions approximately span an order of magnitude, much narrower than that
of $B_{\rm SSC}^{\delta=1}$, especially for the knots. All $B_{\rm eq}^{'}$ are
larger than the magnetic filed strength of the interstellar medium, implying
that the magnetic filed of the interstellar medium would be amplified in
the knots and hot spots by the turbulence of the relativistic shocks. The
$B_{\rm eq}^{'}$ of the knots are smaller than that of the hot spots, with
typical values of 10 $\mu$G and 40 $\mu$G for the knots and the hot spots,
respectively, favoring the idea of different origins of the shocks (internal
vs. external) for the two kinds of sources (e.g., Harris \& Krawczynski 2006).

As our discussed in \S 2, $R_L$ is an intrinsic characteristic for the sources.
It may be a representative of the co-moving magnetic field and the Doppler
boosting effect of the sources. We show $R_L$ as a function of $B_{\rm eq}^{'}$
and $\delta$ in Figure \ref{Fig:9}. It is found that $R_L$ is
correlated with $B_{\rm eq}^{'}$ for the knots, with a linear coefficient of
$r=0.77$ and chance probability $p<10^{-4}$. We obtain $\log R_L=(-2.23\pm
0.24)+(1.45\pm 0.23)\log B_{\rm eq}^{'}$. No significant correlation between
$R_L$ and $B_{\rm eq}^{'}$ is found for the hot spots. However, both the knots
and hot spots form a sequence in the $R_L-B_{\rm eq}^{'}$ plane, with a best
linear fit $\log R_L=(-2.55\pm 0.23)+(1.92\pm 0.17)\log B_{\rm eq}^{'}$. The
hot spots locate at the higher end of the sequence. Similar feature is also
observed in the $R_L-\delta$ plane, as shown in Figure 9(b). These results
imply that $R_L$ would be determined by both $B_{\rm eq}^{'}$ and $\delta$. The
strong anti-correlation of $R_F-R_L$ (or $R_B-R_L$) shown in Figure \ref{Fig:5}
(or Figure \ref{Fig:6}) may be due to neglect of the beaming effect. This effect plays
important role on the observed flux since the observed flux is proportional to
$\delta^4$. Correcting by the Doppler boosting effect, we show the $L_{\rm
5GHz}^{'}-L_{\rm 1keV}^{'}$ relations in Figure 3(b). The relations are tighter
than the observed ones, with a linear coefficient of $r=0.98$ and $r=0.90$ for
the knots and hot spots, respectively.

\section{Conclusions and Discussion}
We have present extensive analysis and SED fits for 22 hot spots and 45 knots
in 35 AGN jets. We find that $L_{\rm 5GHz}$ and $L_{\rm 1keV}$ are tightly
correlated. The two kinds of sources can be roughly separated with a division
of $L_{\rm 1keV}\sim L_{\rm 5GHz}$. Our SED fits show that the mechanisms of
the X-rays are diverse. While the X-ray emission of a small fraction of the
sources is a simple extrapolation of the synchrotron radiation for the
radio-to-optical emission, the IC component may dominate the observed X-rays
for most of the sources. Without considering the relativistic bulk motion, the SSC
model can explain the X-rays for some hot spots with $B_{\rm ssc}^{\delta=1}$
being consistent with the equipartition magnetic field $B_{\rm eq}^{\delta=1}$
in one order of magnitude, but an unreasonably low magnetic field strength is
required in modeling the X-rays for all knots with this model. Considering
relativistic bulk motion for the sources, the IC/CMB dominated model well
explains the X-ray emission for most sources under the equipartition condition.
Although the derived $B_{\rm eq}^{'}$ and $\delta$ for some hot spots are
comparable to that of the knots, the $B_{\rm eq}^{'}$ value for the knots tends
to be smaller than that of the hot spots and the $\delta$ tends to be larger,
favoring the idea that the hot spots are jet termination and knots are a part
of a well-collimated jet. Corrected by the beaming effect, the $L^{'}_{\rm
5GHz}-L^{'}_{\rm 1keV}$ relations for the two kinds of sources are even tighter
than the observed ones, indicating that the correlations are intrinsic.
The ratio $R_L$ is correlated with $B_{eq}^{'}$ and $\delta$. These facts suggest that, under the equipartition
condition, the differences on the X-ray observations for the knots and hot
spots would be mainly due to the differences of the Doppler boosting effect and
the co-moving magnetic field, although some hot spots have similar feature to the knots.

The $R_L$ may be an indicator of $B_{eq}^{'}$ and $\delta$. It is an intrinsic
parameter independent of the Doppler boosting and the cosmological effects. Our
results show that the X-rays of a hot spot with larger $R_L$ are better to be
fitted with the SSC model under equipartition condition without considering the
beaming effect. This is consistent with that reported by Hardcastle et al.
(2004), who found that the X-rays of the radio-bright hot spots can be
explained with the SSC model under equipartition condition. The strong
anti-correlation between $R_B$ (or $R_F$) and $R_L$ may also offer a tool to
discriminate the two kinds of sources. We find in the Figures \ref{Fig:3},
\ref{Fig:5} and \ref{Fig:6} that the hot spot H-A and the knot
K-B\footnote{This source is not included in our discussion, only marked in the
Figures \ref{Fig:3}a, \ref{Fig:5} and \ref{Fig:6}} in PKS B1421-490 are
significant outliers. They were reported as knots by Gelbord et al. (2005). The
knot K-B is very peculiar for its extreme optical output, with a ratio of
knot/core optical flux $\sim 300$. Gelbord et al. (2005) suspected that it is a
core between components A and C. We find that K-B looks is a significant
outlier in the figures and K-A resembles a hot spot. Most recently, Godfrey et
al. (2009) confirmed that the K-B is a core and the K-A is a hot spot
with VLBI observations.

The $B_{\rm eq}^{'}$ value for the knots tends to be smaller than that of the
hot spots and the $\delta$ tends to be larger, favoring the idea that the hot
spots are jet termination and knots are a part of a well-collimated jet.
However, the $B_{\rm eq}^{'}$ and $\delta$ for some hot spots are comparable to
that of the knots, making uncertainty on identifying a component as a knot or a
hot spot. For example, the north-east double hot spots in 3C 351, which locate
at the outer boundary of a lobe, have relativistic
motion feature, hence may be identified as knots of the jet (Harris \& Krawczynski 2006).

The synchrotron radiation in the optical band indicates that there are
relativistic electrons existed in these extended regions (Roser \& Meisenheimer
1987; L$\ddot{a}$hteenm\"{a}ki \& Valtaoja 1999). Moreover, the X-ray emission
of some sources may be also synchrotron radiation as mentioned in \S 3.1.
Assuming a magnetic field strength $B\sim 10^{-5}$ G, one can estimate the
energy of relativistic electrons is $\gamma \sim 10^{6}$, which contribute to
the optical emission by synchrotron process. These electrons may interact with
the synchrotron photons and external field photons to produce very high energy
$\gamma$-ray photons by IC scattering. As shown in Figure 1, both the SSC and
IC/CMB models predict a prominent GeV-TeV component in the SEDs of some
sources. We check if the predicted GeV-TeV emission can be detectable with
H.E.S.S. and \emph{Fermi}/LAT, and also show the sensitivity curves of H.E.S.S.
and \emph{Fermi}/LAT in Figure 1 for these sources\footnote{The absorption by
the infrared background light and CMB during the GeV-TeV photons propagating to
the Earth is not taken into account (Stecker et al. 2006)}. The detections of
these high energy emission would place much stronger constraints on the
radiation mechanisms and on the physical parameters of these sources. The origin of the high energy TeV gamma-ray emission is also a debating issue,
and detections of these high energy emission would drastically improved our
view of the universe (see Cui 2009 for a review).

Note that our one-single zone lepton models cannot explain the observed SEDs
for the four knots in 3C 273 (K-C1, K-C2, K-D1, and K-D2H3). Jester et al.
(2006) had reported that the X-ray spectra rule out the single-zone model of
X-ray emission for some jet knots in 3C 273. It is possible that these sources
may have a complex structure as the western hot spot in Pictor A (Zhang et al.
2009).

\acknowledgments

We thank the anonymous referee for his/her valuable suggestions. This
work was supported by the National Natural Science Foundation of China (grants 10778702, 10533050, 10873002), the National Basic
Research Program (``973" Program) of China (2009CB824800),
and the West PhD project of the training Programme for the
Talents of West Light Foundation of the CAS. JMB thanks
supports of the Bai-Ren-Ji-Hua and the Zhong-Yao-Fang-Xiang (grant KJCX2-YW-T21) projects of the CAS.

\begin{deluxetable}{lllll}
\tabletypesize{\footnotesize} \tablecaption{List of 35 AGNs with jet knots and hot spots included in our sample} \tablenum{1}
\tablehead{ \colhead{Name}& \colhead{\emph{z}\tablenotemark{a}}&
\colhead{$D_{L}$\tablenotemark{b} (Mpc)}&
\colhead{Class\tablenotemark{c}}& \colhead{Reference}}
\startdata
3C 15 & 0.073 & 329.9 & RG (FR I/II) & 1 \\
3C 31 & 0.0169 & 73.3 & RG (FR I) & 2 \\
3C 33 & 0.0597 & 267.3 & RG (FR II) & 3 \\
3C 228 & 0.5524 & 3194 & RG & 4 \\
3C 245 & 1.029 & 6845.4 & Q & 5 \\
3C 263 & 0.6563 & 3937 & LDQ & 6 \\
3C 275.1 & 0.557 & 3226.2 & LDQ & 4 \\
3C 280 & 0.996 & 6575 & RG (FR II) & 7 \\
3C 351 & 0.371 & 1988 & LDQ & 6 \\
3C 390.3 & 0.0561 & 250.5 & RG (FR II) & 8 \\
3C 295 & 0.461 & 2570.6 & RG (FR II) & 9 \\
3C 303 & 0.141 & 666.6 & RG (FR II) & 10 \\
3C 66B & 0.0215 & 93.6 & RG (FRI) & 2 \\
3C 120 & 0.033 & 144.9 & Sy I & 12 \\
3C 273 & 0.1583 & 756.5 & CDQ & 13 \\
3C 454.3 & 0.86 & 5485.1 & CDQ & 14 \\
3C 207 & 0.684 & 4141 & LDQ & 5 \\
3C 345 & 0.594 & 3487.4 & CDQ & 5 \\
3C 346 & 0.161 & 770.7 & RG (FR I) & 15 \\
3C 371 & 0.051 & 226.9 & BL & 16 \\
3C 403 & 0.059 & 264 & RG (FR II) & 17 \\
M87 & 0.0043 & 18.5 & RG (FR I) & 20 \\
Cygnus A & 0.0562 & 251 & RG (FR II) & 11 \\
Pictor A & 0.035 & 153.9 & RG (FR II) & 21 \\
PKS 0405-123 & 0.574 & 3345.6 & Q & 5 \\
PKS 0521-365 & 0.0554 & 247.3 & BL & 22 \\
PKS 0637-752 & 0.653 & 3913.1 & CDQ & 2 \\
PKS 1136-135 & 0.554 & 3205.2 & LDQ & 18 \\
PKS 1229-021 & 1.045 & 6977.3 & CDQ & 14 \\
PKS 1421-490 & 0.663 & 3986.3 & Q & 19 \\
PKS 2201+044 ((4C 04.77)) & 0.027 & 118 & BL & 16 \\
PKS 1928+738 (4C +73.18) & 0.302 & 1564.7 & CDQ & 5 \\
PKS 1354+195 (4C +19.44) & 0.72 & 4409.1 & CDQ & 5 \\
PKS 1150+497 (4C +49.22) & 0.334 & 1758.4 & CDQ & 5 \\
PKS 0836+299 (4C +29.30)  & 0.064 & 287.4 & RG (FR I) & 5 \\
\enddata
\tablenotetext{a}{z: redshift;} \tablenotetext{b}{$D_{L}$:
luminosity distance of the sources;} \tablenotetext{c}{RG: radio
galaxy of either Fanaroff-Riley class I (FR I) or class II (FR
II); Q: quasar, either core-dominated (CD) or lobe-dominated (LD);
Sy: Seyfert galaxy; BL: BL Lac objects.}

\tablerefs{ (1) Kataoka et al. 2003a; (2) Kataoka \& Stawarz 2005;
(3) Kraft et al. 2007; (4) Hardcastle et al. 2004; (5) Sambruna et
al. 2004; (6) Hardcastle et al. 2002; (7) Donahue et al. 2003; (8)
Harris et al. 1998; (9) Harris et al. 2000; (10) Meisenheimer et
al. 1997; Kataoka et al. 2003b; (11) Stawarz et al. 2007; (12)
Harris et al. 2004; (13) Jester et al. 2007; (14) Tavecchio et al.
2007; (15) Worrall \& Birkinshaw 2005; (16) Sambruna et al. 2007;
(17) Kraft et al. 2005; (18) Sambruna et al. 2006; (19) Gelbord et
al. 2005; (20) Liu \& Shen 2007; Perlman et al. 2001; (21) Wilson
et al. 2001; (22) Falomo et al. (2009).}
\end{deluxetable}

\begin{deluxetable}{llllllllllllllllll}
\tabletypesize{\footnotesize} \rotate
\tablecolumns{18} \tablewidth{0pc} \tablecaption{Observations and SED fit results for the hot
spots and knots in our sample}\tablenum{2} \tablehead{ \colhead{} & \colhead{} &
\multicolumn{4}{c}{Observations} &   \colhead{}   &
\multicolumn{3}{c}{$\delta=1$, SSC} & \colhead{} & \multicolumn{2}{c}{$\delta>1$, IC/CMB} & \colhead{} & \multicolumn{4}{c}{Preferred model}\\
\cline{3-6} \cline{8-10} \cline{12-13} \cline{15-18}\\\colhead{Source} & \colhead{Comp\tablenotemark{a}}   &
\colhead{$\alpha_{r}$} & \colhead{$\alpha_{X}$}& \colhead{$F_{\rm 1keV}^{\rm obs}$} & \colhead{$\theta$} &
\colhead{$\gamma_{\min}$} & \colhead{$B^{\delta=1}_{\rm eq}$} &\colhead{$F_{\rm 1keV}^{\rm eq}$}
&\colhead{$B^{\delta=1}_{\rm ssc}$ } & \colhead{} & \colhead{$\delta$} & \colhead{$B^{'}_{\rm eq}$} & \colhead{} & \colhead{Model} & \colhead{$\alpha_{X}^{\rm mod}$} &
\colhead{$p_{1}$} & \colhead{$p_{2}$} \\\colhead{}& \colhead{} & \colhead{} & \colhead{}& \colhead{(nJy)} & \colhead{(arcsec)} & \colhead{} &
\colhead{($\mu$G)} &\colhead{(nJy)} &\colhead{ ($\mu$G)} & \colhead{} &
\colhead{} & \colhead{($\mu$G)}& \colhead{} & \colhead{} & \colhead{} & \colhead{} & \colhead{}\\\colhead{(1)}& \colhead{(2)} & \colhead{(3)} & \colhead{(4)}& \colhead{(5)} & \colhead{(6)} & \colhead{(7)} &
\colhead{(8)} &\colhead{(9)} &\colhead{ (10)} & \colhead{} &
\colhead{(11)} & \colhead{(12)}& \colhead{} & \colhead{(13)} & \colhead{(14)} & \colhead{(15)} & \colhead{(16)}}
\startdata
3C 33&H-S1&0.75&0.8$\pm$0.6&0.14$\pm$0.06&0.5&200&158&0.086&120&&3.3&66.1&&SSC&0.63&2.4&4.4\\
&H-S2&0.98&0.8$\pm$0.6&0.32$\pm$0.09&1.5&200&64.6&0.036&20&&2.5&32.9&&SSC&0.71&2.56&3.98\\
&H-N1&0.88&1.2$\pm$0.8&0.27$\pm$0.08&1.25&200&36.4&0.0016&2.1&&4&13.2&&SYN&1.85&2.4&3.6\\
&H-N2&0.9&1.2$\pm$0.8&0.19$\pm$0.07&1.25&200&38.7&0.0012&2.3&&3.8&14.6&&SYN&1.41&2.38&3.8\\
3C 263&H-K&0.84&1.0$\pm$0.3&1.0$\pm$0.1&0.39&300&144&0.71&118&&3.1&70&&SSC&0.8&2.62&4.04\\
Cygnus A&H-A&0.5&0.77$\pm$0.13&31.2$\pm$4.3&1&200&214&19.4&155&&7&52.4&&SSC&0.8&1.84&3.86\\
&H-D&0.38&0.8$\pm$0.11&47.9$\pm$5.9&1&100&164&47.8&160&&7&40.7&&SSC&0.77&1.5&3.24\\
&H-B&0.59&0.7$\pm$0.35&6.8$\pm$2.6&1&200&330&0.55&85&&7.5&78.9&&SSC&0.77&2.02&3.96\\
3C 351&H-L&0.93&0.85$\pm$0.1&3.4$\pm$0.4&0.8&300&72.4&0.13&12.5&&5&28&&SSC&0.78&2.36&3.3\\
&H-J&0.76&0.5$\pm$0.1&4.3$\pm$0.3&0.16&200&186&0.13&29&&10&33.1&&IC/CMB\tablenotemark{b}&0.34&2.44&3.1\\
3C 303&H-W&0.84&0.4$\pm$0.2&4&1&200&68.9&0.03&5&&5.9&18.6&&IC/CMB\tablenotemark{b}&0.5&2.64&3.7\\
3C 295&H-NW&0.94&0.9$\pm$0.5&3.8&0.1&500&460&0.63&200&&12&102.9&&SSC&0.94&1.9&4.4\\
3C 390.3&H-B&0.71&0.9$\pm$0.15&4.2$\pm$0.87&1&200&38.1&0.004&0.85&&8&9.45&&IC/CMB&0.78&2.56&3.1\\
3C 275.1&H-N&\nodata&\nodata&1.78&0.6&100&119&0.061&22&&4.3&40.3&&SSC&0.84&2.79&\nodata\\
3C 228&H-S&\nodata&\nodata&1.3&0.27&200&131&0.063&28&&5&40&&SSC&0.87&2.62&3.1\\
3C 245&H-D&\nodata&\nodata&0.7$\pm$0.3&0.8&600&67.7&0.064&18&&2.8&32&&SSC&0.66&1.96&3.92\\
3C 280&H-W&0.8&1.3$\pm$1.0&0.79&0.3&200&147&0.052&35&&4.2&51.2&&SYN&1.22&2.6&3.3\\
&H-E&0.8&1.2&0.34&0.3&200&124&0.014&23&&3.8&46.5&&SYN&1.13&2.6&3.1\\
PKS 0405&H-N&\nodata&\nodata&1.6$\pm$0.5&0.7&400&81.8&0.061&16.5&&3.5&37.2&&SSC&0.91&2.8&3.22\\
PKS 0836&H-B&\nodata&\nodata&2.2$\pm$0.6&0.9&100&33.1&6.9E-4&0.5&&7.5&7.45&&SYN&1.04&2.68&3.06\\
Pictor A&H-W&0.74&1.07$\pm$0.11&45&0.3&110&392&0.76&48&&19.3&45.3&&SYN&1.32&2.38&3.66\\
PKS 1421&H-A&0.67&0.31$\pm$0.32&13.3$\pm$1.6&0.24&200&404&13.7&400&&8.5&86.2&&SSC&0.6&1.9&4.06\\
M87&K-D&\nodata&1.43$\pm$0.09&51.5$\pm$4.2&0.4&150&296&0.015&2.8&&\nodata&\nodata&&SYN&1.34&2.36&3.68\\
&K-A&\nodata&1.61$\pm$0.07&156$\pm$8.8&0.9&150&266&0.15&5&&\nodata&\nodata&&SYN&1.55&2.28&4.1\\
&K-E&\nodata&1.48$\pm$0.12&32.2$\pm$6.5&0.9&150&101&0.0016&0.38&&\nodata&\nodata&&SYN&1.38&2.42&3.76\\
&K-F&\nodata&1.64$\pm$0.15&20.1$\pm$5.2&1.2&150&118&0.0049&0.9&&\nodata&\nodata&&SYN&1.53&2.3&4.06\\
&K-B&\nodata&1.59$\pm$0.12&30.3$\pm$5.5&1.3&150&184&0.066&5.5&&\nodata&\nodata&&SYN&1.77&2.32&4.54\\
&K-C1&\nodata&1.33$\pm$0.06&14.6$\pm$5.2&0.5&100&380&0.067&20&&\nodata&\nodata&&SYN&1.7&2.36&4.4\\
PKS 1136&K-A&0.67&1.1$\pm$0.6&1.7$\pm$0.2&0.85&50&25.7&3.6E-4&0.18&&8&5.68&&IC/CMB&0.66&2.32&\nodata\\
&K-B&0.81&1.1$\pm$0.3&3.5$\pm$0.2&0.6&100&57.6&9.5E-4&0.67&&9&11.4&&IC/CMB&0.8&2.61&\nodata\\
&K-$\alpha$&0.75&0.9$\pm$0.4&1.9$\pm$0.2&0.75&60&34&1.3E-4&0.16&&10&6.95&&IC/CMB&0.71&2.42&\nodata\\
&K- D&0.71&0.5$\pm$0.5&1.0$\pm$0.2&0.6&100&94.7&0.0061&7.2&&4.7&30&&IC/CMB&0.9&2.8&\nodata\\
PKS 1150&K- B&0.72&0.7$\pm$0.2&7.6$\pm$0.5&0.76&100&41.8&8.8E-4&0.28&&12.2&6.7&&IC/CMB&0.72&2.46&4.6\\
&K-C&0.71&0.5$\pm$0.3&2.9$\pm$0.4&0.91&300&27&5.1E-4&0.2&&8.9&6.18&&IC/CMB&0.72&2.42&3.82\\
&K-D&0.68&0.7$\pm$0.5&1.3$\pm$0.3&0.91&50&28.4&1.1E-4&0.15&&9&6.7&&IC/CMB&0.69&2.38&\nodata\\
&K-E&0.67&0.7$\pm$0.3&1.7$\pm$0.2&0.71&800&27.6&7E-4&0.35&&8&7.43&&IC/CMB&0.68&2.36&3.4\\
&K-IJ&0.81&1.1$\pm$0.6&0.6$\pm$0.1&0.71&200&50.6&0.0021&2.5&&5&17.2&&IC/CMB&0.8&2.6&4.28\\
PKS 2201&K-A&0.71&1.1$\pm$0.4&5.6&0.5&100&37.2&1.5E-4&0.06&&24&4.08&&SYN&1.1&2.2&3.12\\
&K-$\beta$&0.59&0.9$\pm$0.5&3.8&0.3&50&67.4&1.5E-5&0.031&&45&4.34&&IC/CMB&0.53&2.15&5\\
3C 371&K-A&0.69&1.1$\pm$0.4&7&0.7&50&32.5&7.5E-4&0.11&&17&4.14&&SYN&1.09&2.34&3.12\\
PKS 1928&K-A&\nodata&1.66$\pm$0.74&6.9$\pm$1.1&0.8&100&36.4&5.2E-4&0.18&&\nodata&\nodata&&SYN&0.88&2.6&4.06\\
PKS 1354&K-A&\nodata&0.6$\pm$0.32&16.1$\pm$8.2&1.9&300&27.2&0.0041&0.18&&8.5&5.92&&IC/CMB&0.41&2.12&4\\
&K-B&\nodata&\nodata&0.7$\pm$0.3&1.4&300&23.5&0.0027&0.9&&3.2&10.1&&IC/CMB&0.5&2.32&3.4\\
PKS 1229&K-A&\nodata&\nodata&8.5$\pm$3.2&1&250&60.3&0.038&3&&5.4&17.6&&IC/CMB&0.63&2.4&3.6\\
PKS 0637&K&0.8&0.9$\pm$0.1&6.2&0.4&100&108&0.012&3.8&&9.5&20.6&&IC/CMB&0.8&2.6&\nodata\\
PKS  0521&K&0.89&1.3$\pm$0.3&14&0.4&100&118&0.014&2.5&&\nodata&\nodata&&SYN&1.25&2.4&3.5\\
3C 454.3&K-A&\nodata&\nodata&6$\pm$1.4&1&300&51.8&0.008&1.5&&6&13.8&&IC/CMB&0.3&2.6&5\\
&K-B&\nodata&\nodata&6$\pm$1.4&1&40&95.1&0.079&8.5&&6.1&34.6&&IC/CMB&0.72&2.44&5\\
3C 403&K-F1&\nodata&0.75$\pm$0.4&0.9$\pm$0.2&0.75&150&56.3&5.2E-4&1.1&&8&13&&IC/CMB&0.79&2.58&\nodata\\
&K-F6&\nodata&0.7$\pm$0.3&2.3$\pm$0.2&0.75&100&60.8&0.0012&0.8&&11&10.6&&IC/CMB&0.68&2.4&\nodata\\
3C 346&K-C&\nodata&1.0$\pm$0.3&1.6$\pm$0.2&0.9&100&72.3&0.017&6.1&&5.4&20.9&&IC/CMB&0.8&2.6&\nodata\\
3C 345&K-A&\nodata&0.66$\pm$0.86&3.8$\pm$0.7&0.6&50&131&0.089&17&&6.4&36.6&&IC/CMB&0.76&2.52&\nodata\\
3C 207&K-A&\nodata&0.1$\pm$0.3&3.0$\pm$0.7&0.5&100&62.7&0.007&1&&11&11.2&&IC/CMB&0.4&1.8&3.6\\
3C 66B&K-A&0.75&0.97$\pm$0.34&4.0$\pm$0.3&0.7&60&45.5&2.2E-5&0.045&&24&4.47&&SYN&1.16&2.46&3.4\\
&K-B&0.6&1.17$\pm$0.14&6.1$\pm$0.4&0.6&50&73.2&8.5E-4&0.33&&\nodata&\nodata&&SYN&1.27&2.22&3.54\\
3C 31&K-K&0.55&1.1$\pm$0.2&7.3&0.6&150&88.2&5.3E-4&0.2&&\nodata&\nodata&&SYN&0.98&2.72&3.4\\
3C 15&K-C&0.9&0.71$\pm$0.4&0.934$\pm$0.2&0.4&30&162&0.0021&7.2&&9&31.6&&IC/CMB&0.9&2.8&\nodata\\
3C 273&K-A&0.85&0.83$\pm$0.02&46.5$\pm$0.54&0.8&20&91.8&0.0085&0.77&&22.5&12.2&&IC/CMB&0.75&2.5&5\\
&K-C1&0.73&1.07$\pm$0.06&4.85$\pm$0.16&0.6&20&110&0.016&4.5&&10.5&19.6&&IC/CMB&0.74&2.48&4\\
&K-C2&0.75&0.96$\pm$0.05&6.25$\pm$0.18&0.7&20&118&0.039&6.8&&9.5&22.7&&IC/CMB&0.75&2.5&3.82\\
&K-B1&0.82&0.8$\pm$0.03&10.9$\pm$0.25&0.6&20&126&0.0056&2.2&&17&21.8&&IC/CMB&0.82&2.64&6\\
&K-D1&0.77&1.02$\pm$0.05&5.16$\pm$0.17&0.7&20&149&0.057&13&&8.5&31&&IC/CMB&0.79&2.58&4.74\\
&K-DH&0.85&1.04$\pm$0.04&7.82$\pm$0.2&1&20&174&0.19&25.5&&6.8&41.9&&IC/CMB&0.86&2.72&4.4\\
3C 120&K-K4&0.74&0.9$\pm$0.2&10$\pm$2&0.7&100&76.2&0.0018&0.58&&18&9.24&&IC/CMB&0.58&2.4&\nodata\\
&K-S2&0.67&0.2$\pm$0.6&0.882&1.6&100&17.1&6.9E-5&0.085&&8&3.8&&IC/CMB&0.62&2.24&\nodata\\
&K-S3&0.69&\nodata&0.8$\pm$0.6&1.6&100&16.1&3.4E-5&0.055&&8&3.5&&SYN&1.16&2.36&3.28\\
&K-K7&0.68&2.4$\pm$0.6&6.3$\pm$1.6&1.5&100&35.3&\nodata&\nodata&&\nodata&\nodata&&SYN&1.2&2.6&7\\

\enddata
\tablenotetext{a}{The first capital represents the kind of the  structure,
``K" indicating ``knot" and ``H" indicating ``hot spot". The suffix denotes the
name of the extended region.}
\tablenotetext{b}{Both the IC/SSC and IC/CMB models can not match the X-ray spectra well, but the IC/CMB model is better to represent the data.}
\tablecomments{Columns: (3) Radio spectral
index $\alpha_{r}$; (4) X-ray spectral index $\alpha_{x}$ at 1 keV; (5) The observed X-ray flux density at 1 keV; (6) Size of the
emitting region in arcsec; (7) The minimum Lorentz factor of the
electrons $\gamma_{min}$; (8) The equipartition magnetic field
$B^{\delta=1}_{\rm eq}$; (9) The predicted flux density at 1 keV; (10) The fitting magnetic field $B^{\delta=1}_{\rm ssc}$ with SSC model; (11) The beaming factors $\delta$ considering the IC/CMB model; (12) The equipartition magnetic field $B^{'}_{\rm eq}$ by considering the
beaming effect; (13) The preferred model; (14) The
derived spectral index $\alpha_{X}^{\rm mod}$ at 1 keV by the preferred model; (15) (16) The energy indices $p_{1}, p_{2}$ of electrons below and above
the break.}
\end{deluxetable}

\clearpage
\clearpage \setlength{\voffset}{0mm}

\clearpage
\begin{figure*}
\includegraphics[angle=0,scale=0.80]{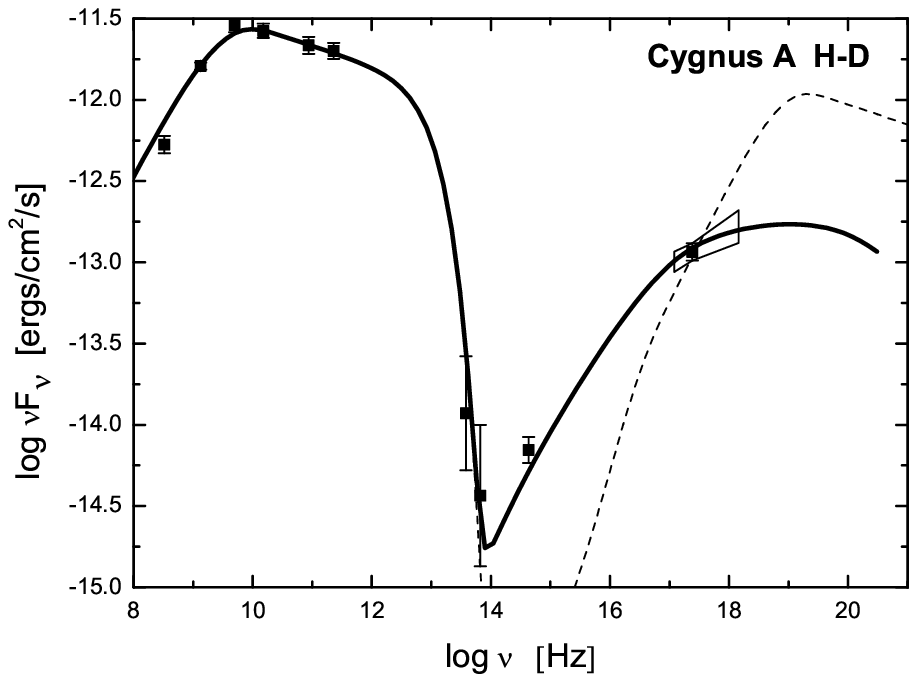}
\includegraphics[angle=0,scale=0.80]{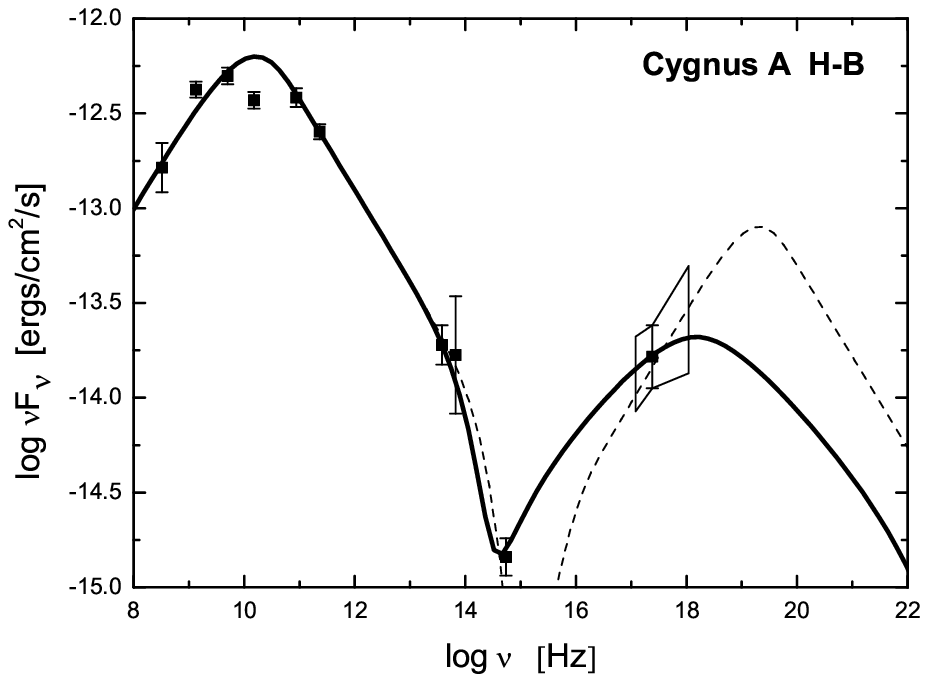}
\includegraphics[angle=0,scale=0.80]{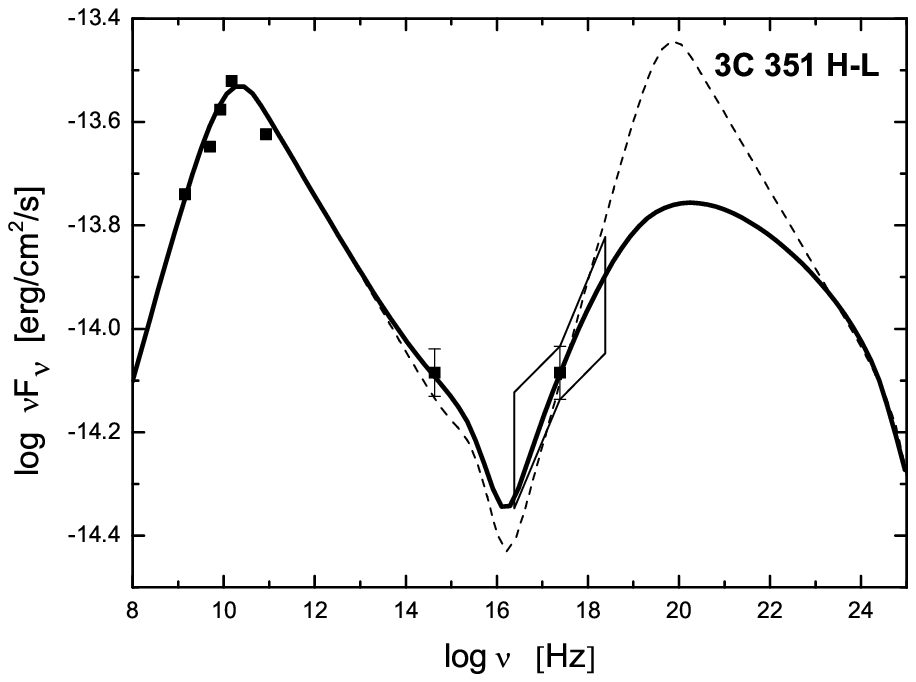}
\includegraphics[angle=0,scale=0.80]{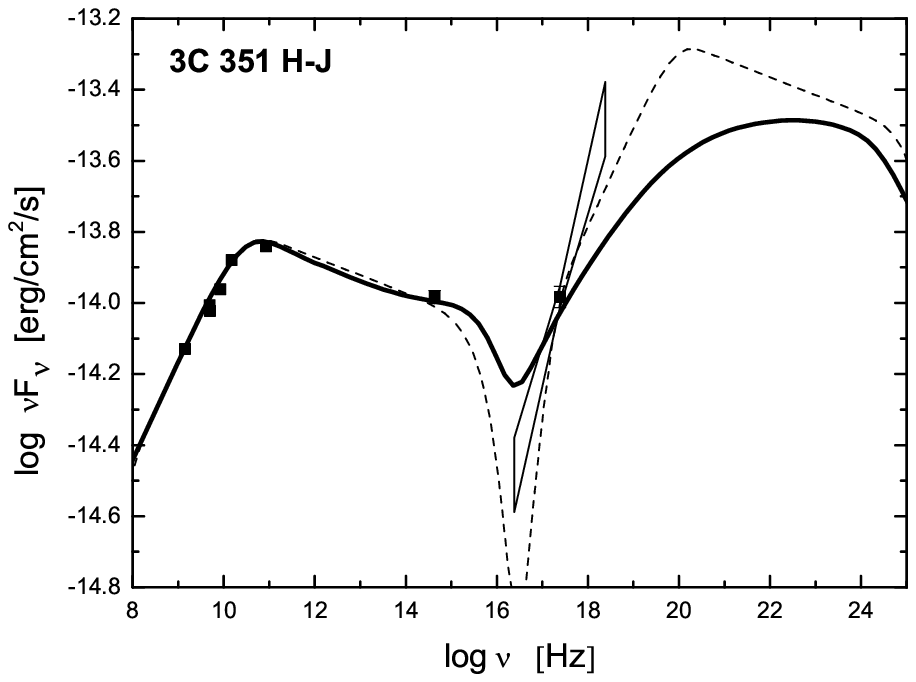}
\includegraphics[angle=0,scale=0.80]{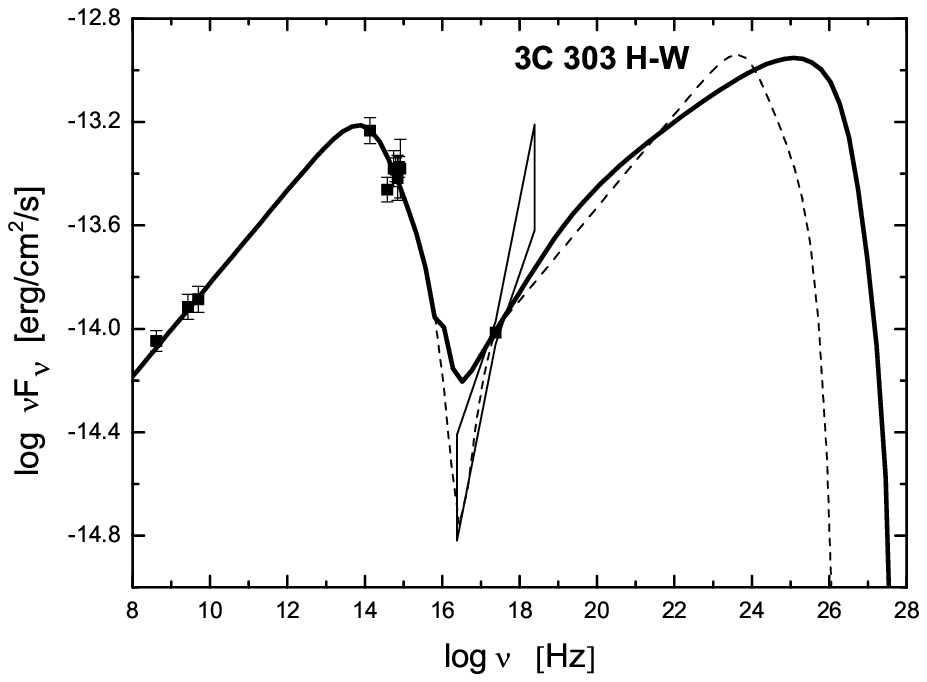}
\includegraphics[angle=0,scale=0.80]{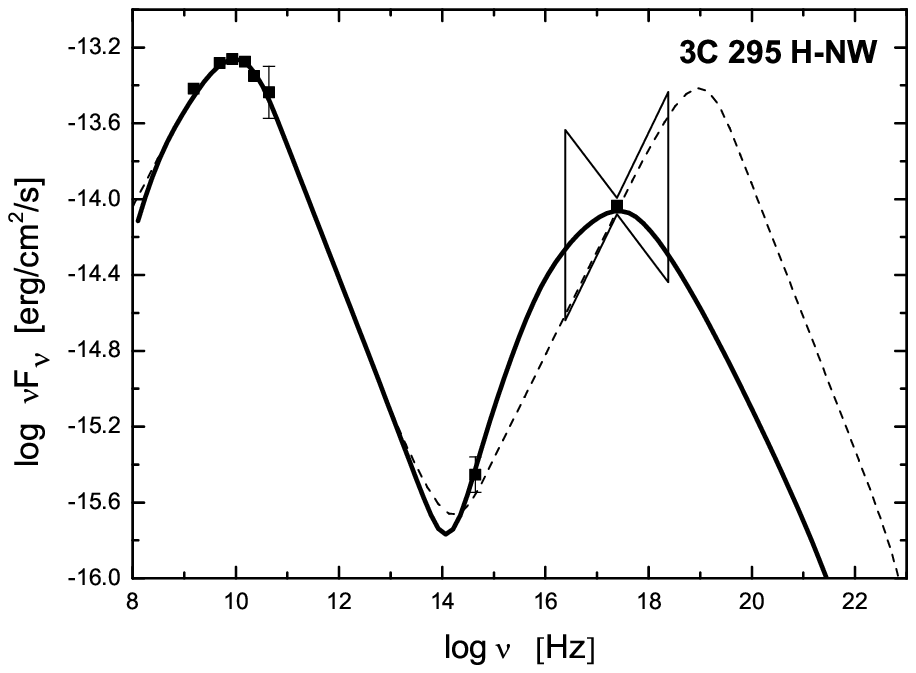}
\includegraphics[angle=0,scale=0.80]{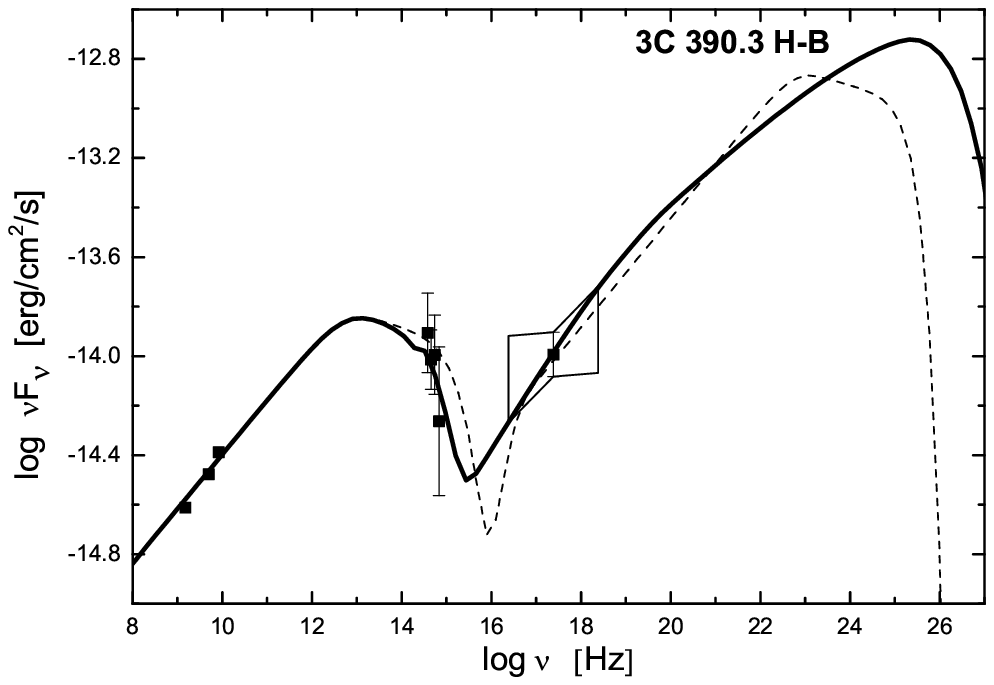}
\hfill
\includegraphics[angle=0,scale=0.80]{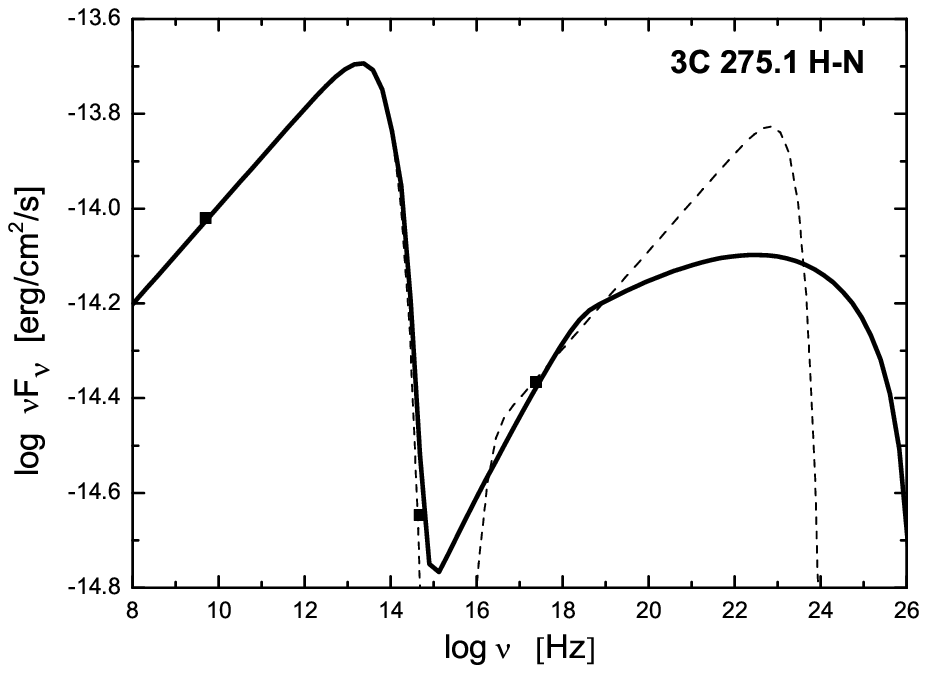}2
\hfill\center{Fig. 1---  continued}
\end{figure*}
\clearpage \setlength{\voffset}{0mm}

\begin{figure*}
\includegraphics[angle=0,scale=0.80]{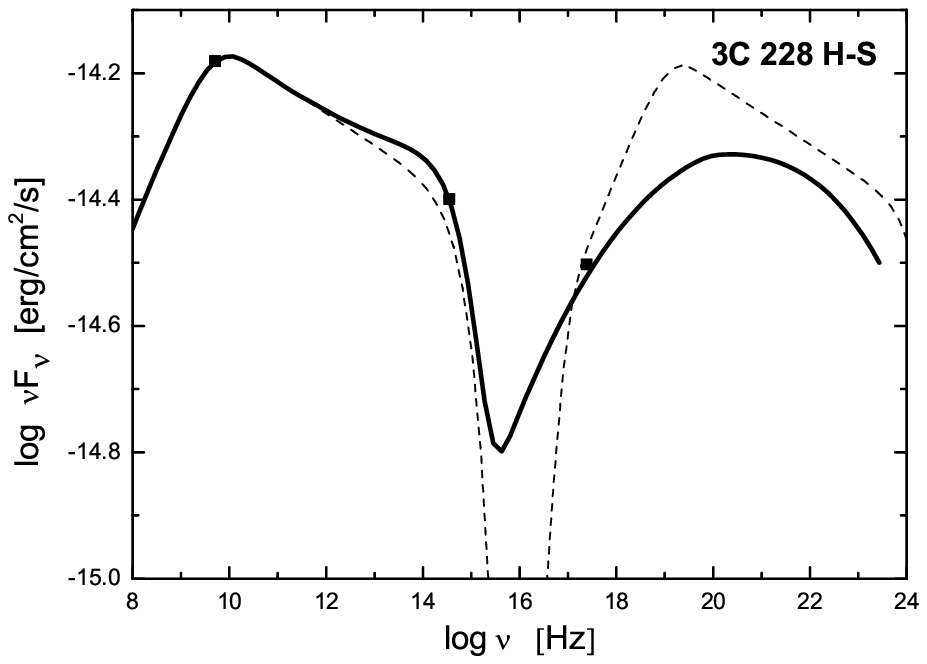}
\includegraphics[angle=0,scale=0.80]{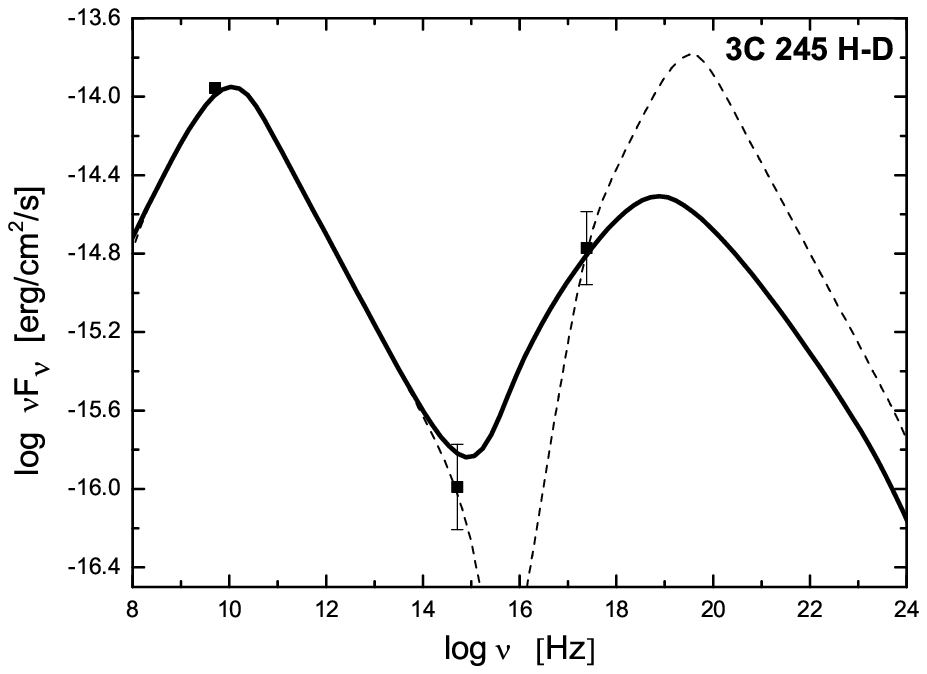}
\includegraphics[angle=0,scale=0.80]{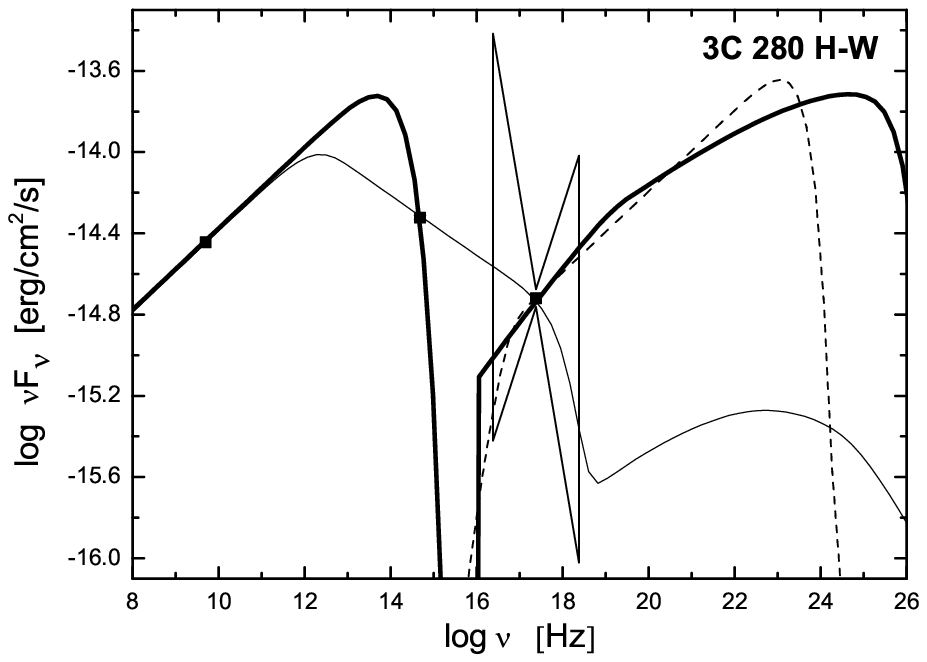}
\includegraphics[angle=0,scale=0.80]{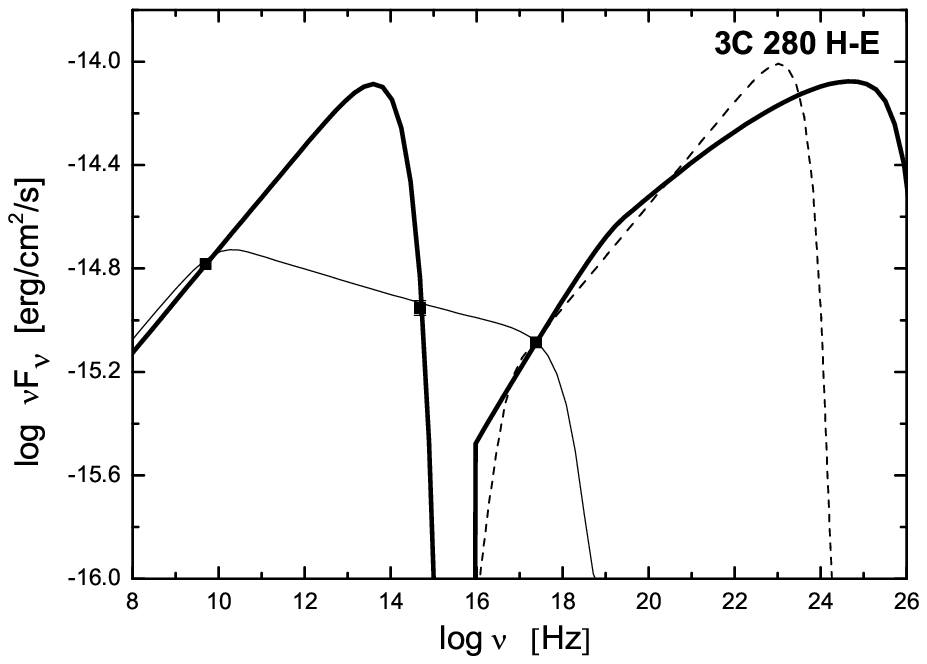}
\includegraphics[angle=0,scale=0.80]{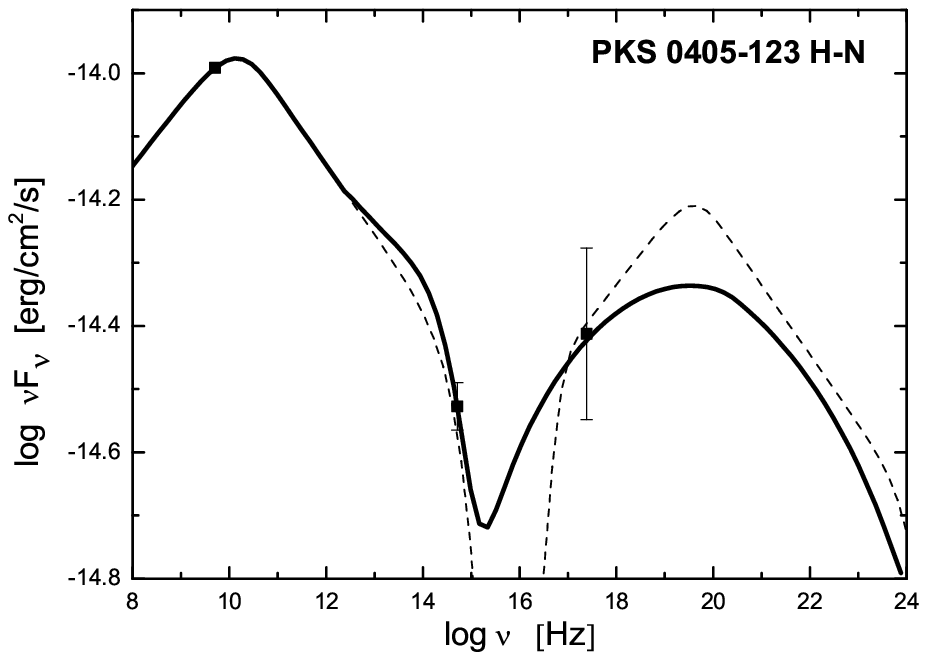}
\includegraphics[angle=0,scale=0.80]{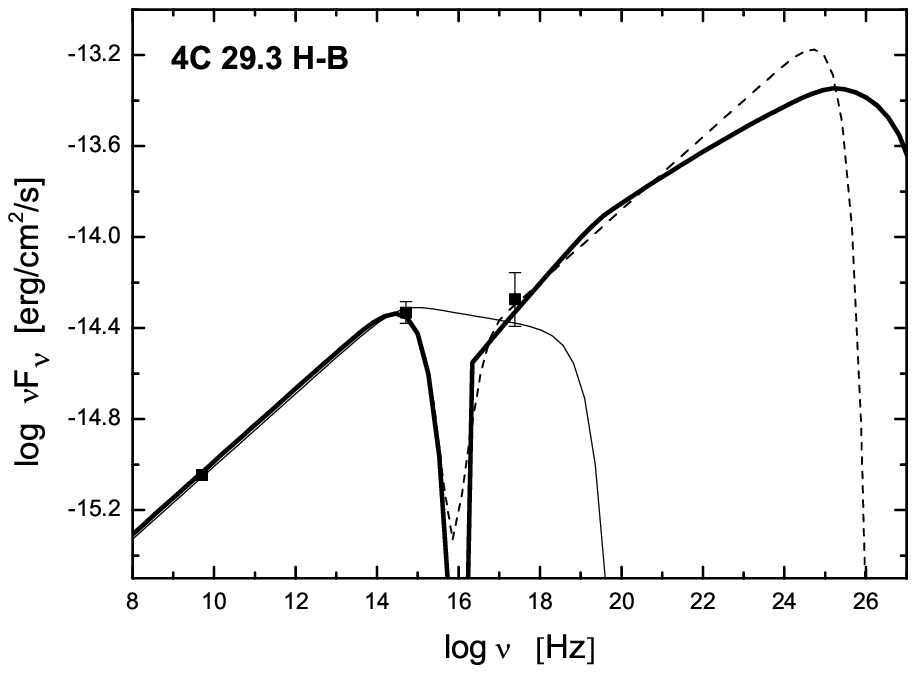}
\includegraphics[angle=0,scale=0.80]{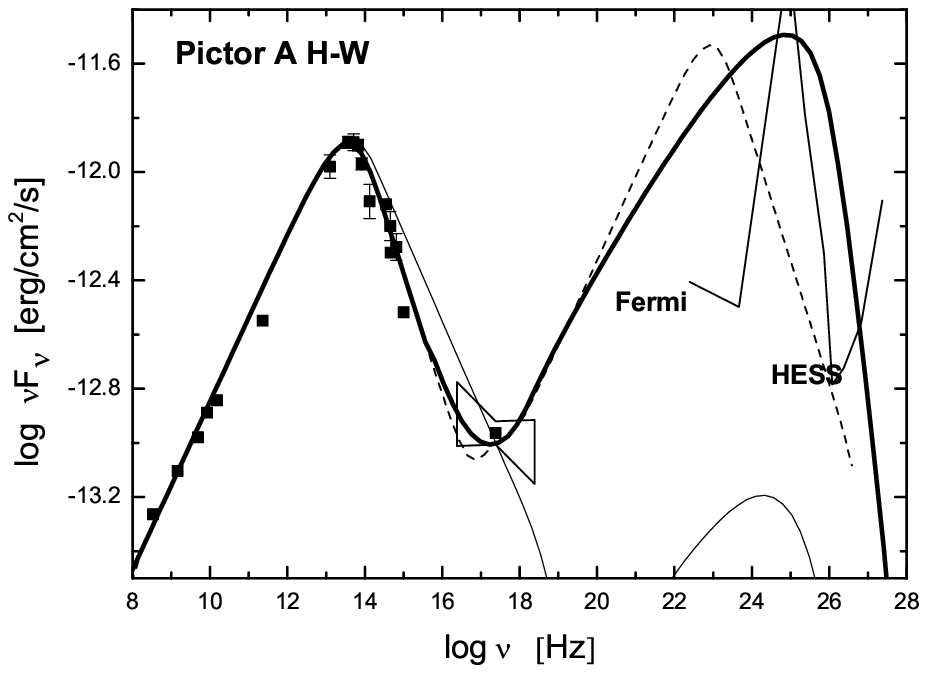}
\hfill
\includegraphics[angle=0,scale=0.80]{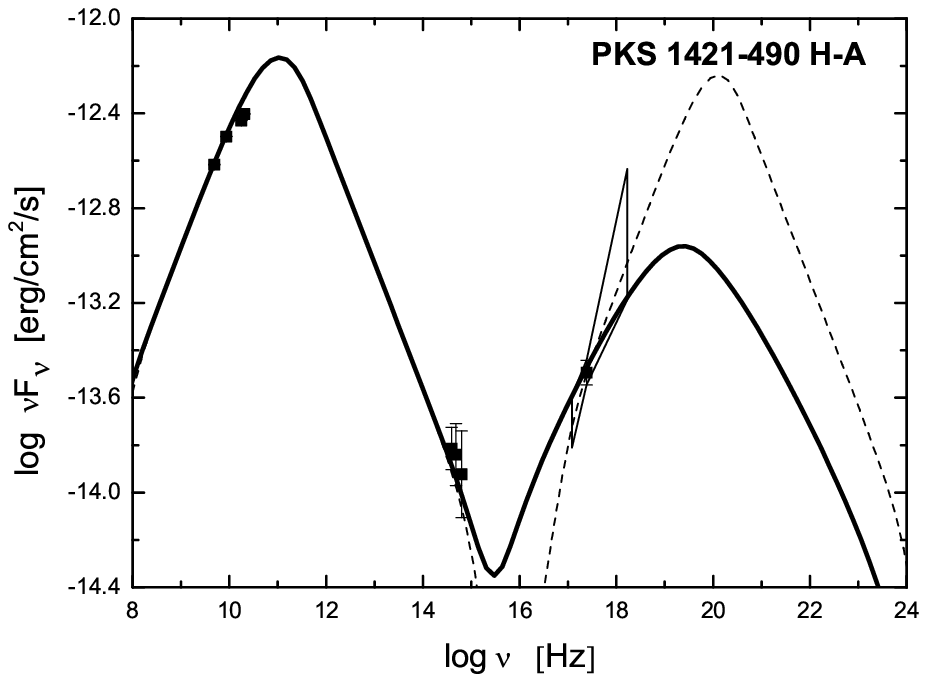}
\hfill\center{Fig. 1---  continued}
\end{figure*}
\clearpage \setlength{\voffset}{0mm}

\begin{figure*}
\includegraphics[angle=0,scale=0.80]{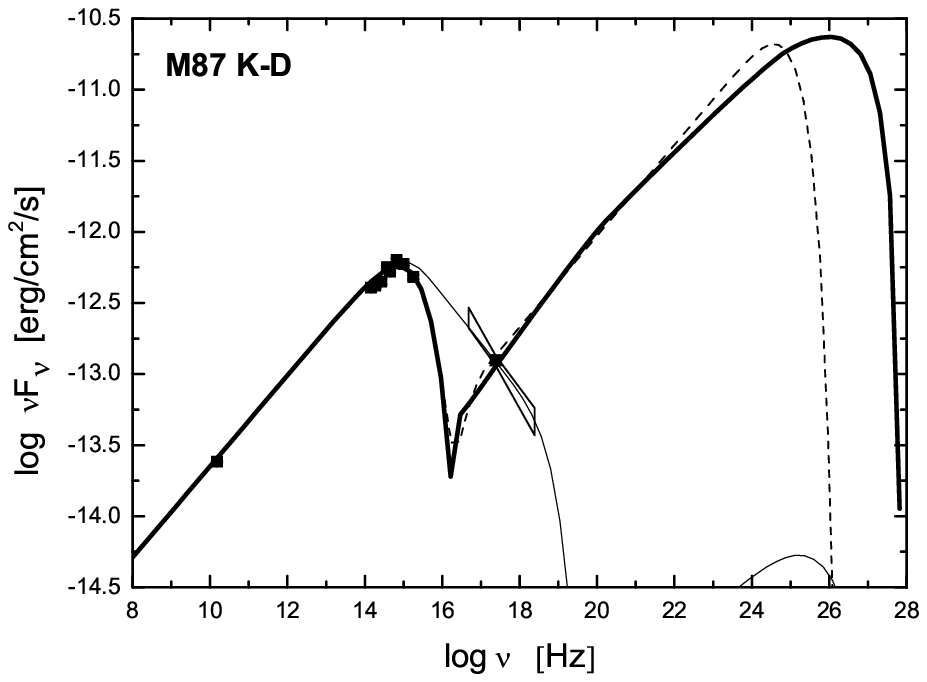}
\includegraphics[angle=0,scale=0.80]{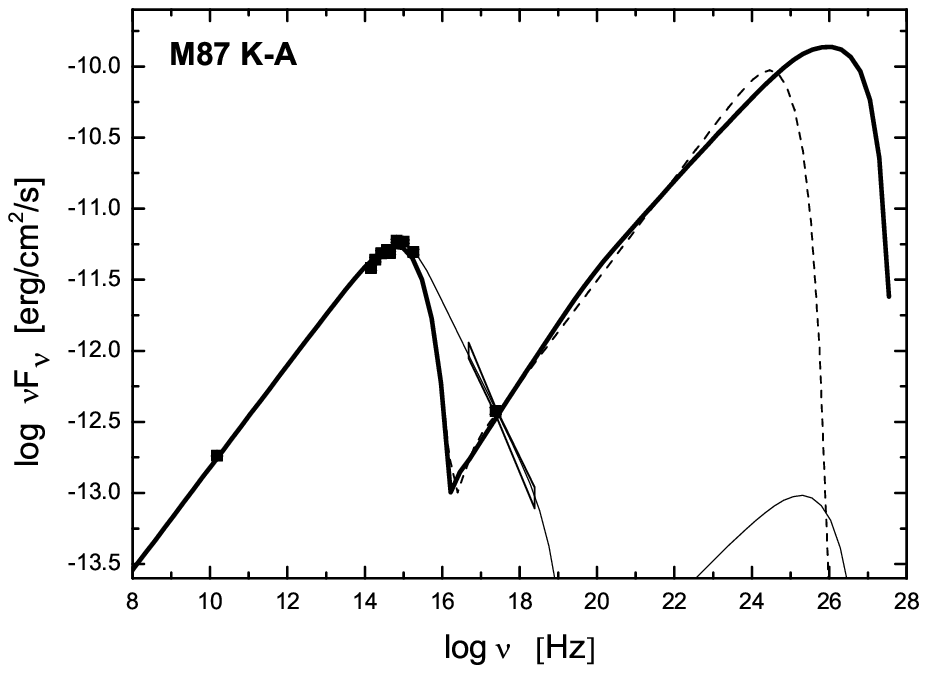}
\includegraphics[angle=0,scale=0.80]{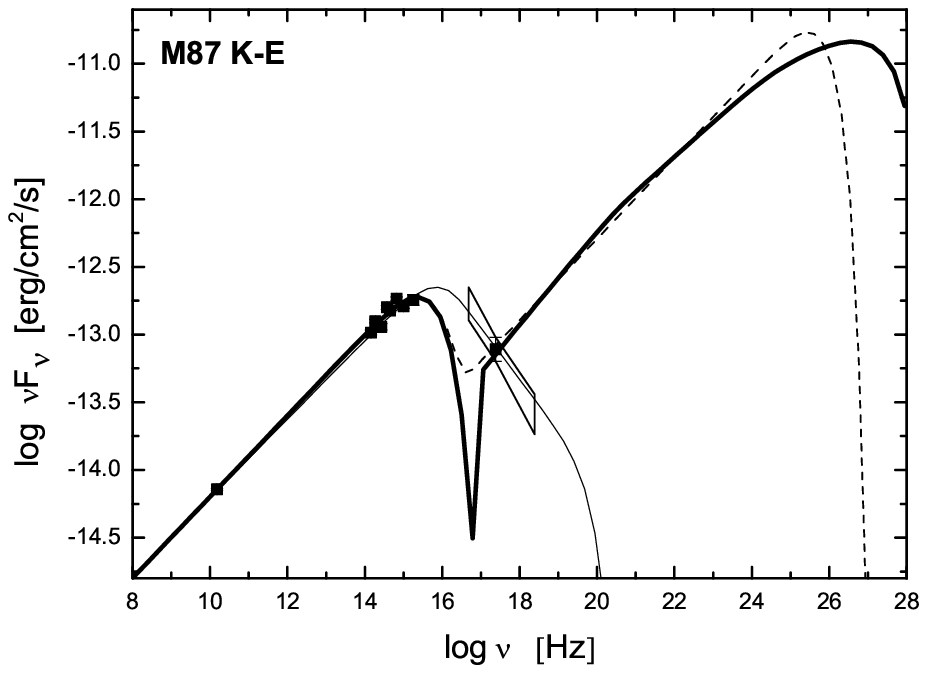}
\includegraphics[angle=0,scale=0.80]{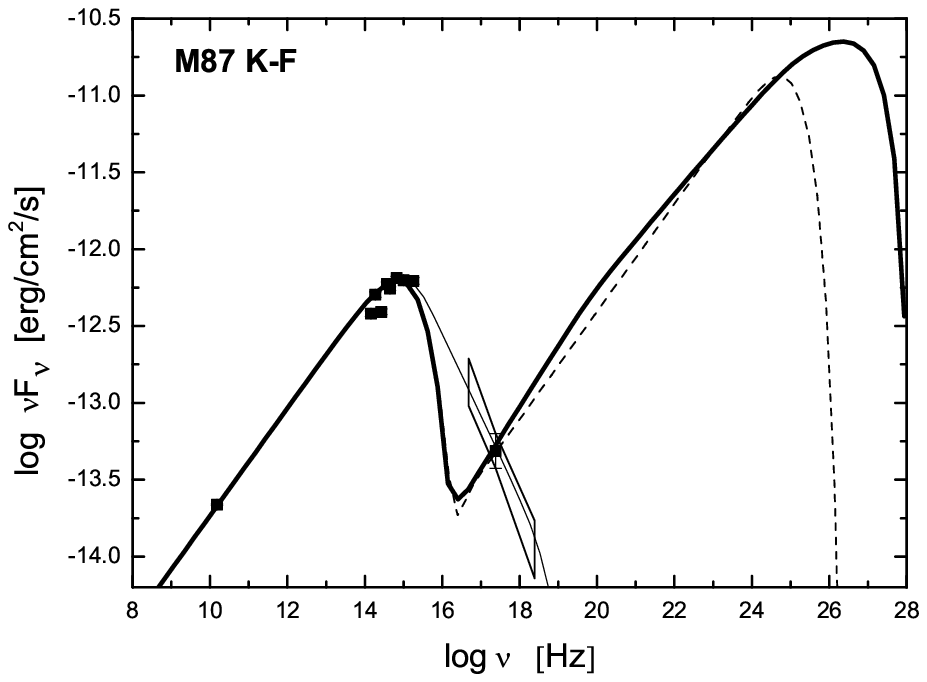}
\includegraphics[angle=0,scale=0.80]{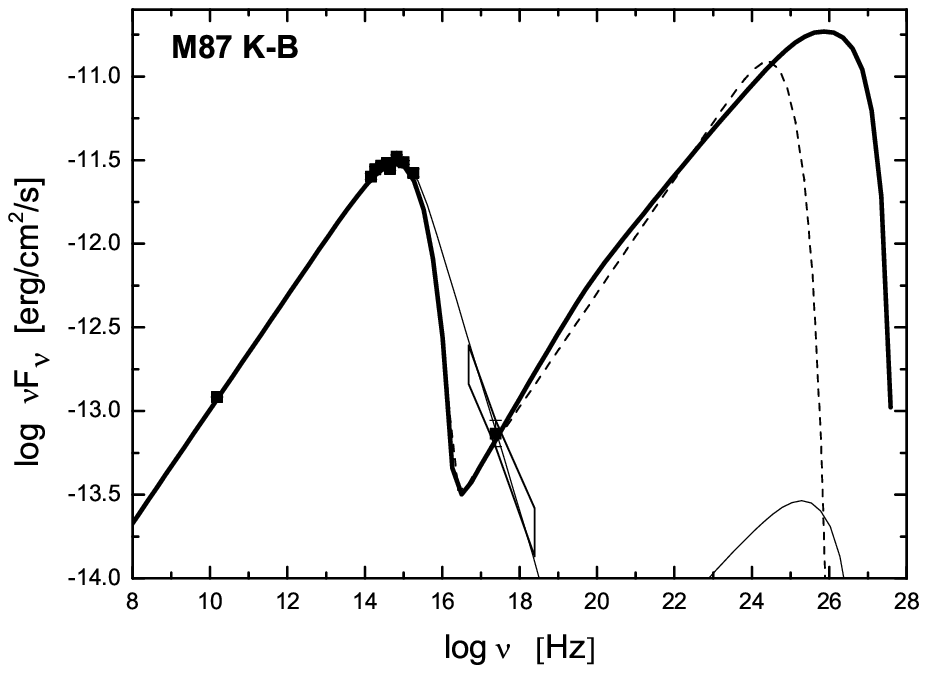}
\includegraphics[angle=0,scale=0.80]{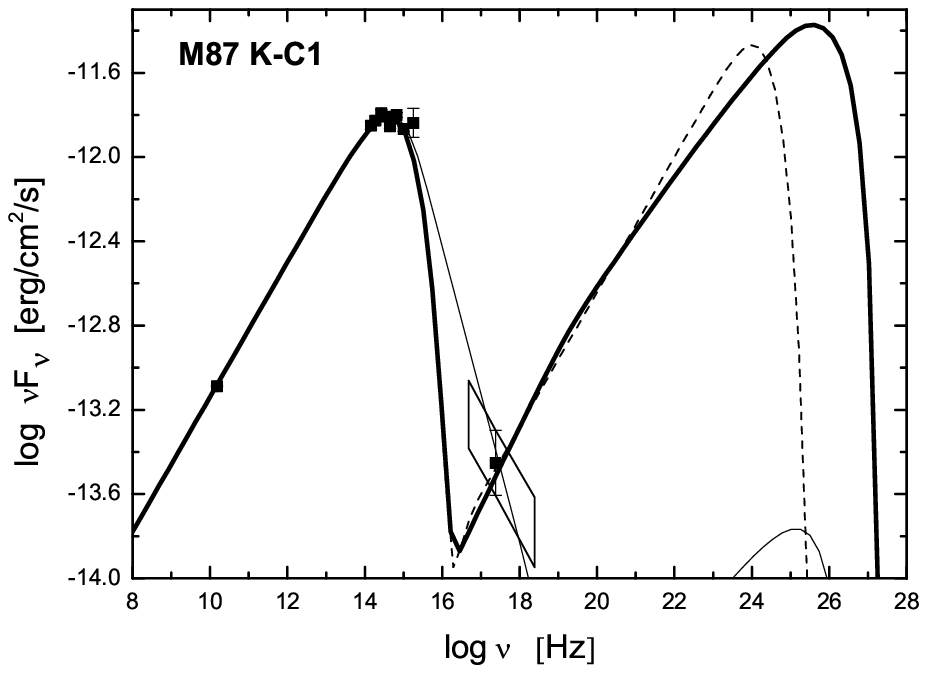}
\includegraphics[angle=0,scale=0.80]{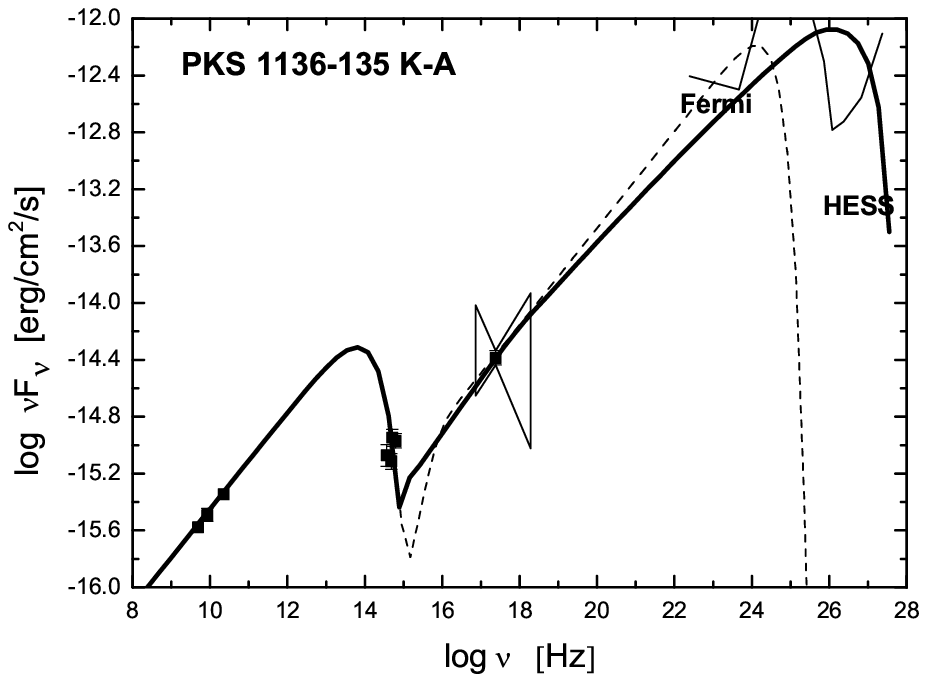}
\hfill
\includegraphics[angle=0,scale=0.80]{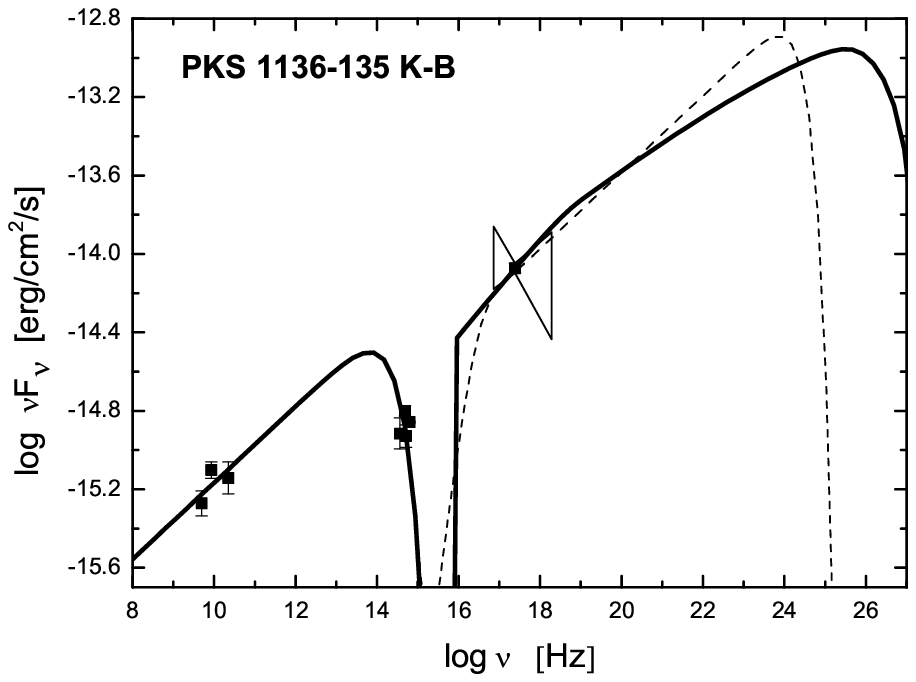}
\hfill\center{Fig. 1---  continued}
\end{figure*}
\clearpage \setlength{\voffset}{0mm}

\begin{figure*}
\includegraphics[angle=0,scale=0.80]{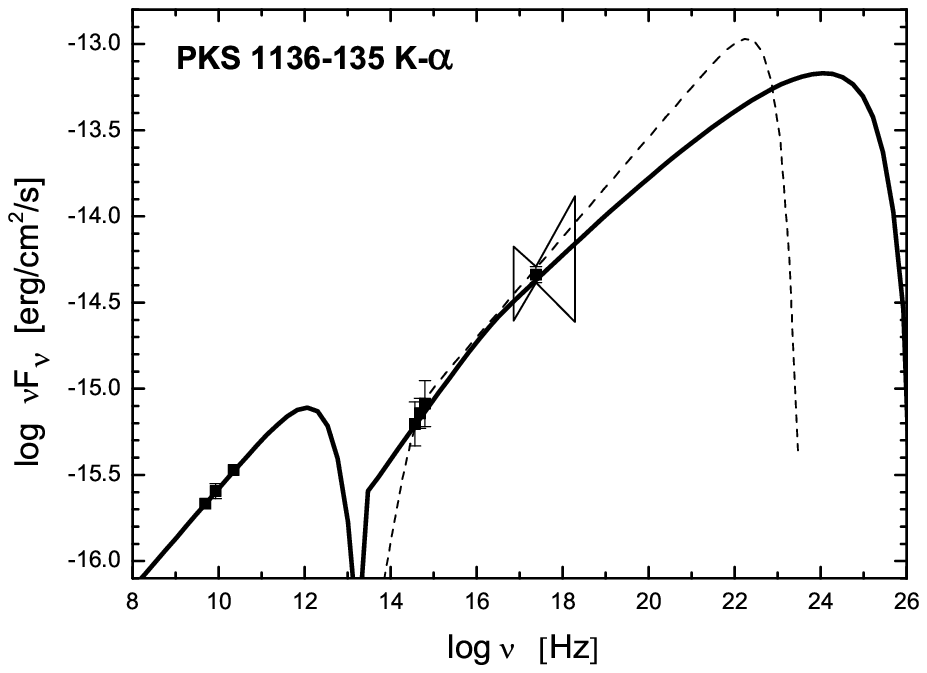}
\includegraphics[angle=0,scale=0.80]{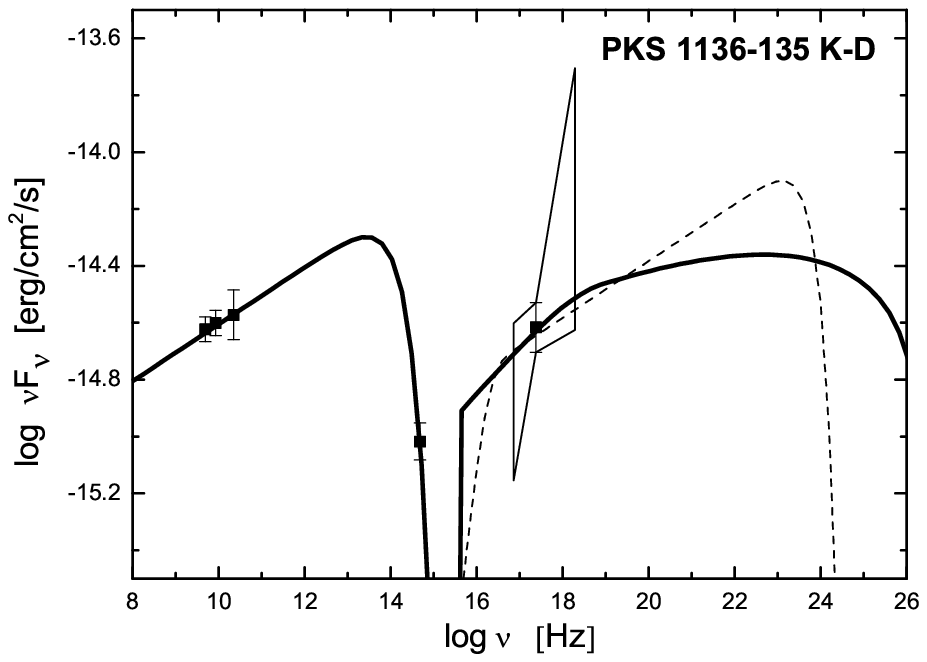}
\includegraphics[angle=0,scale=0.80]{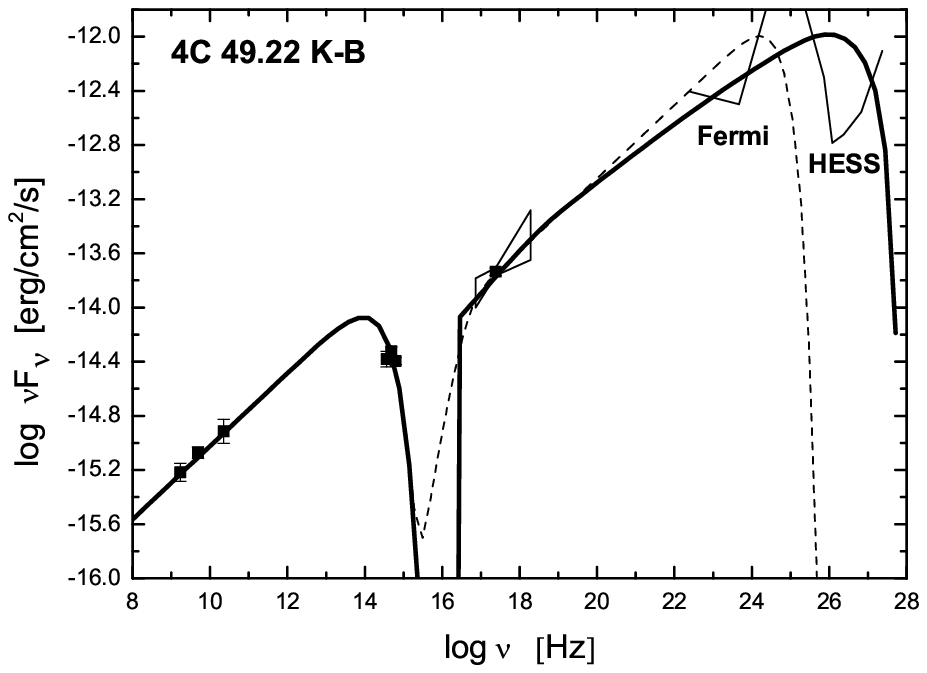}
\includegraphics[angle=0,scale=0.80]{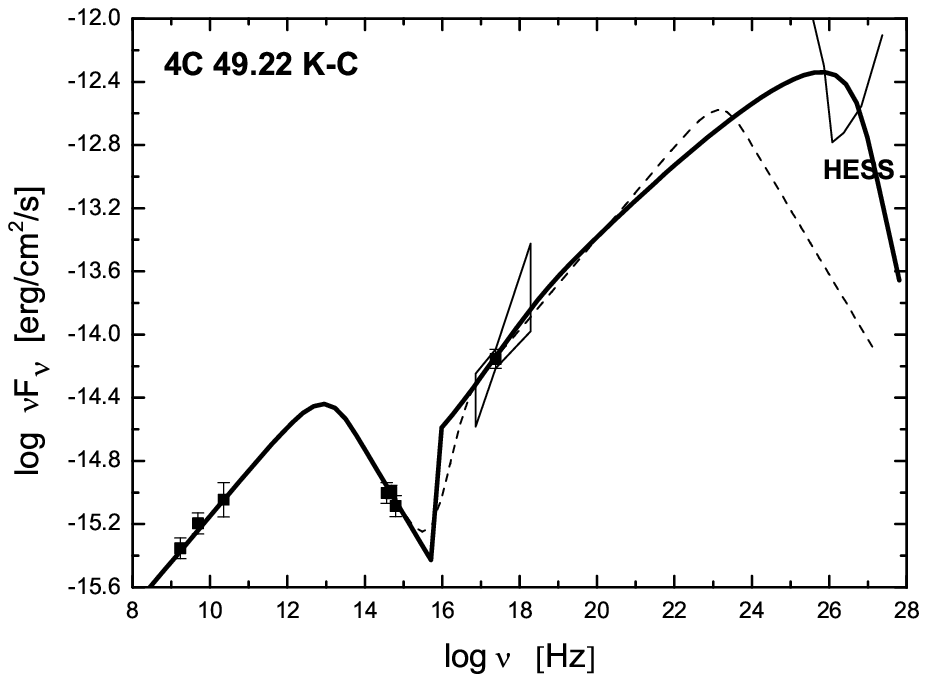}
\includegraphics[angle=0,scale=0.80]{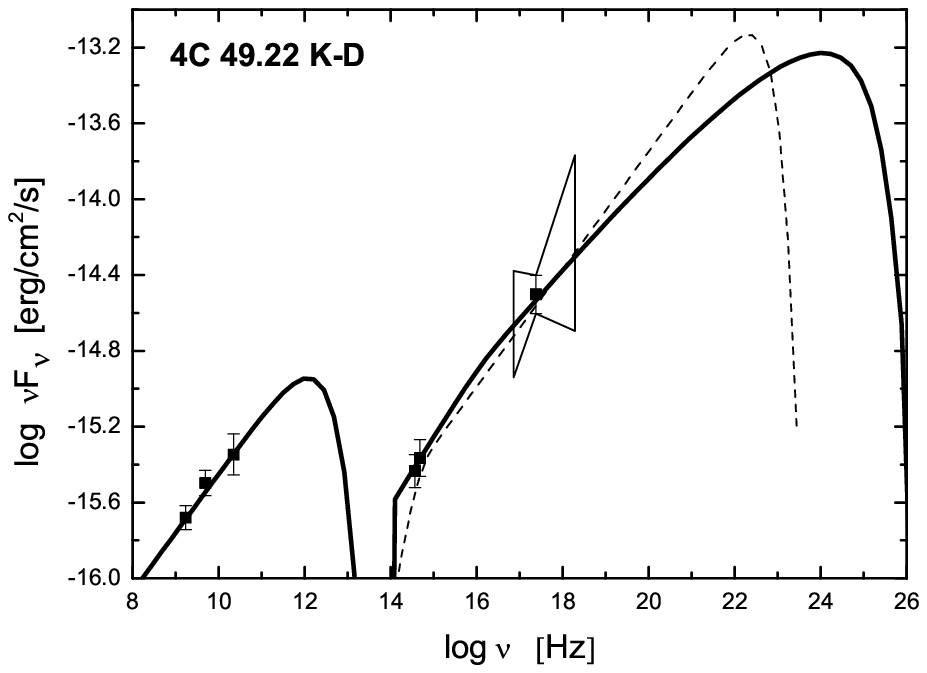}
\includegraphics[angle=0,scale=0.80]{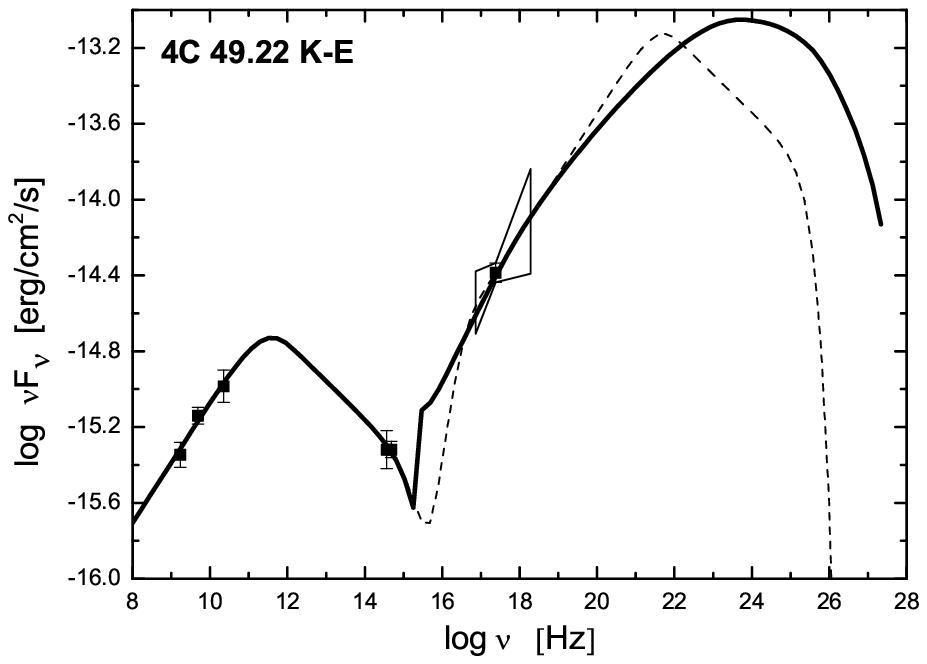}
\includegraphics[angle=0,scale=0.80]{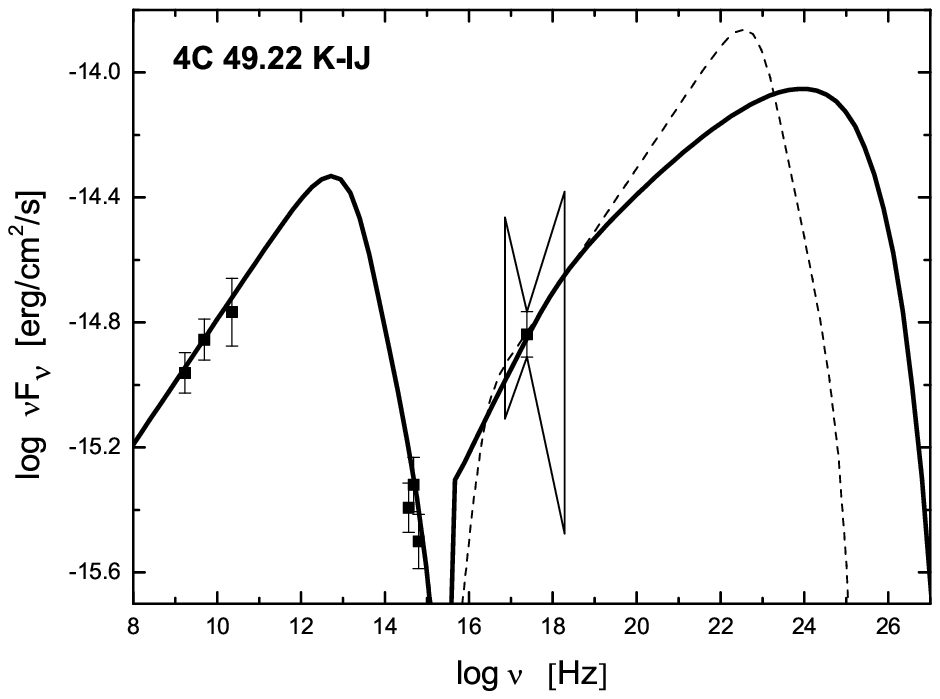}
\hfill
\includegraphics[angle=0,scale=0.80]{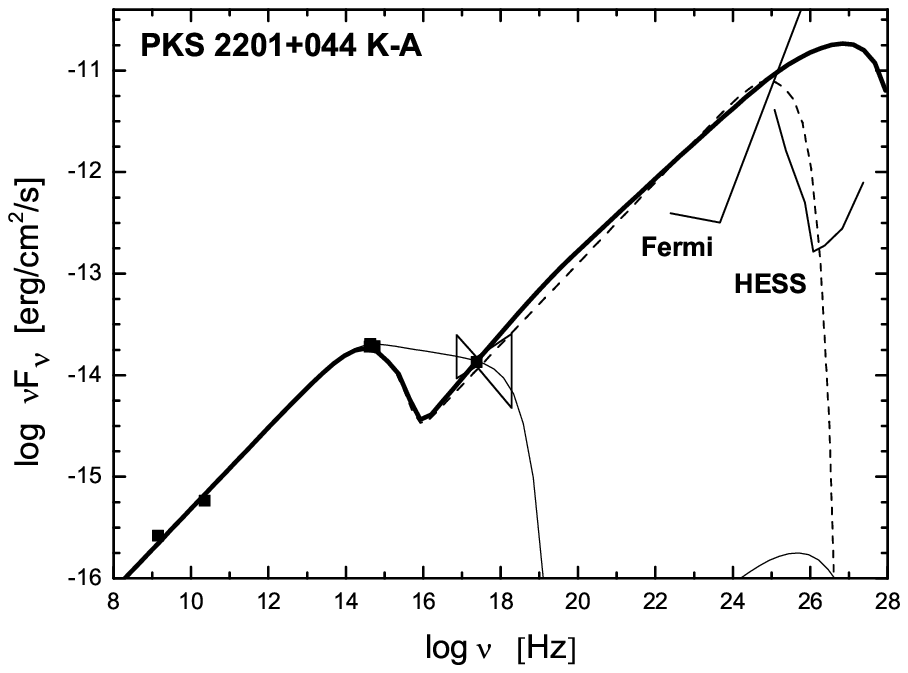}
\hfill\center{Fig. 1---  continued}
\end{figure*}
\clearpage \setlength{\voffset}{0mm}

\begin{figure*}
\includegraphics[angle=0,scale=0.80]{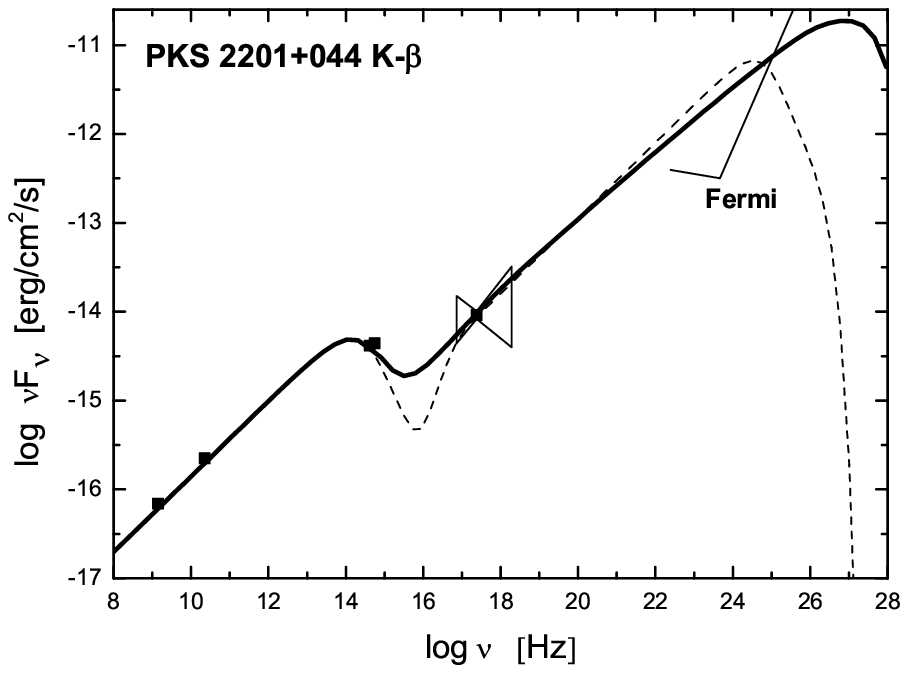}
\includegraphics[angle=0,scale=0.80]{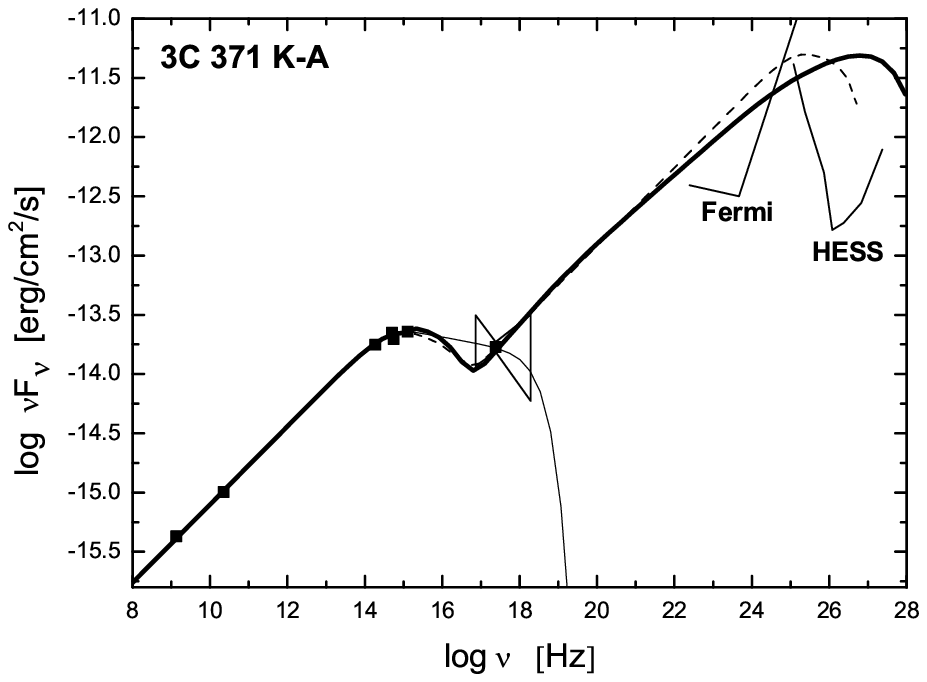}
\includegraphics[angle=0,scale=0.80]{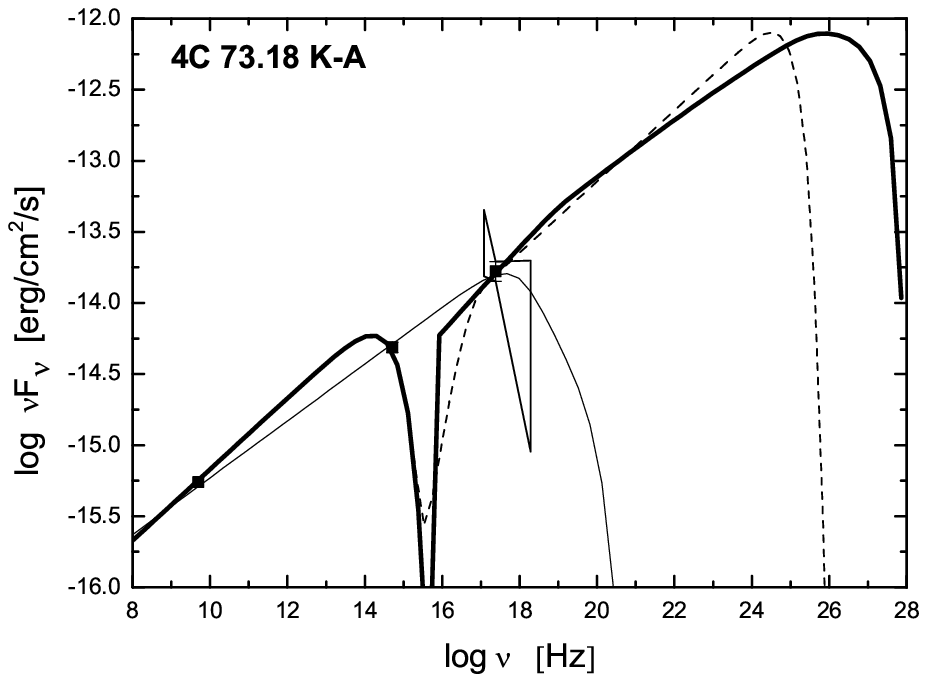}
\includegraphics[angle=0,scale=0.80]{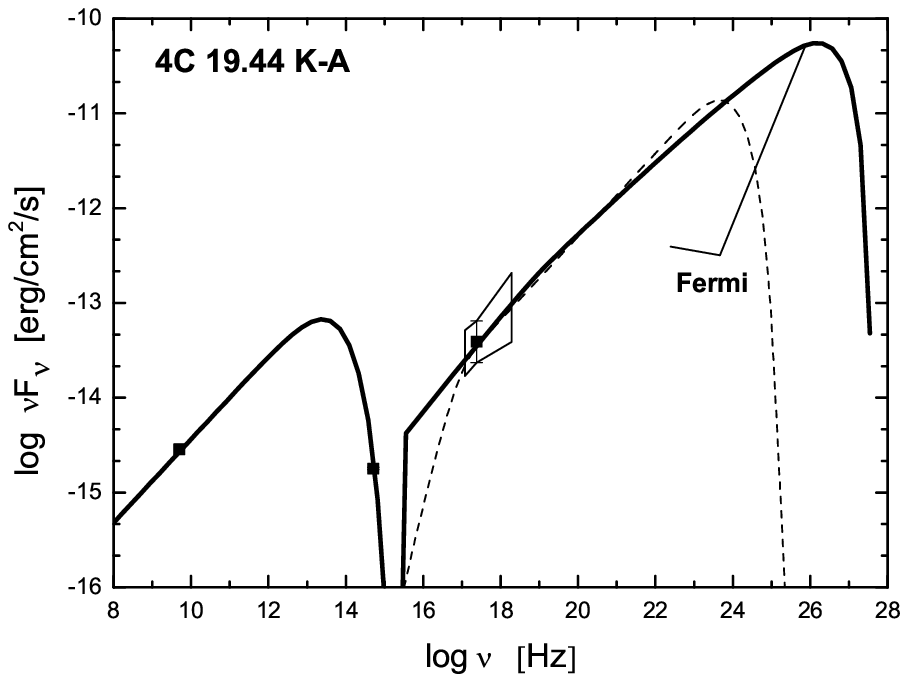}
\includegraphics[angle=0,scale=0.80]{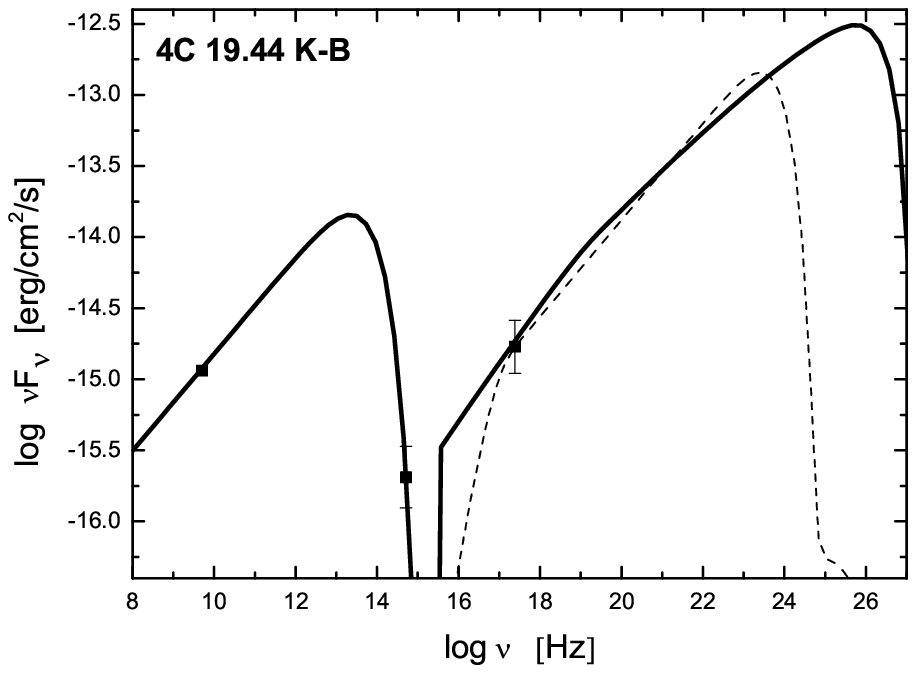}
\includegraphics[angle=0,scale=0.80]{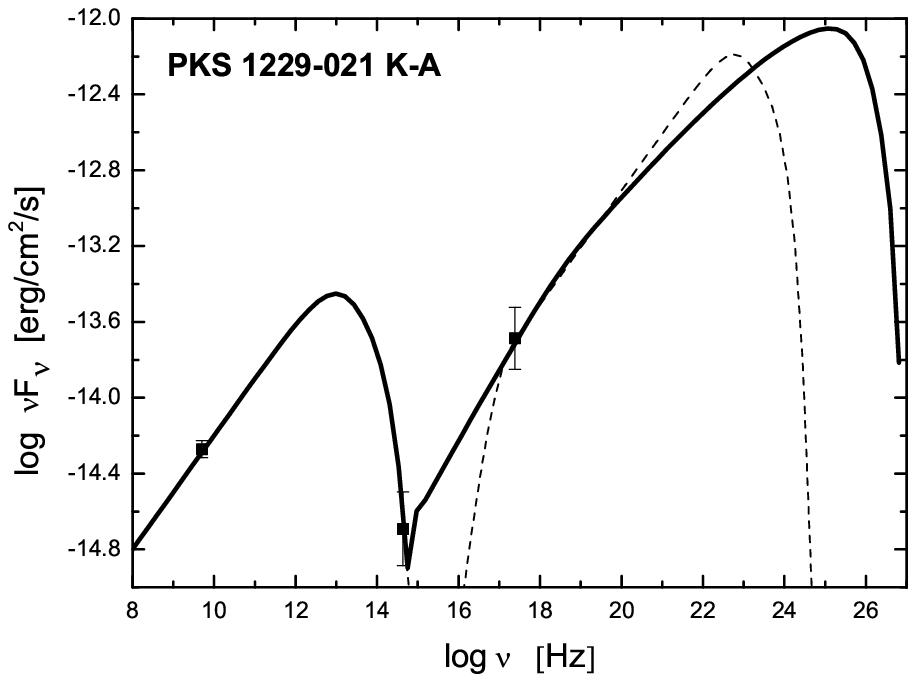}
\includegraphics[angle=0,scale=0.80]{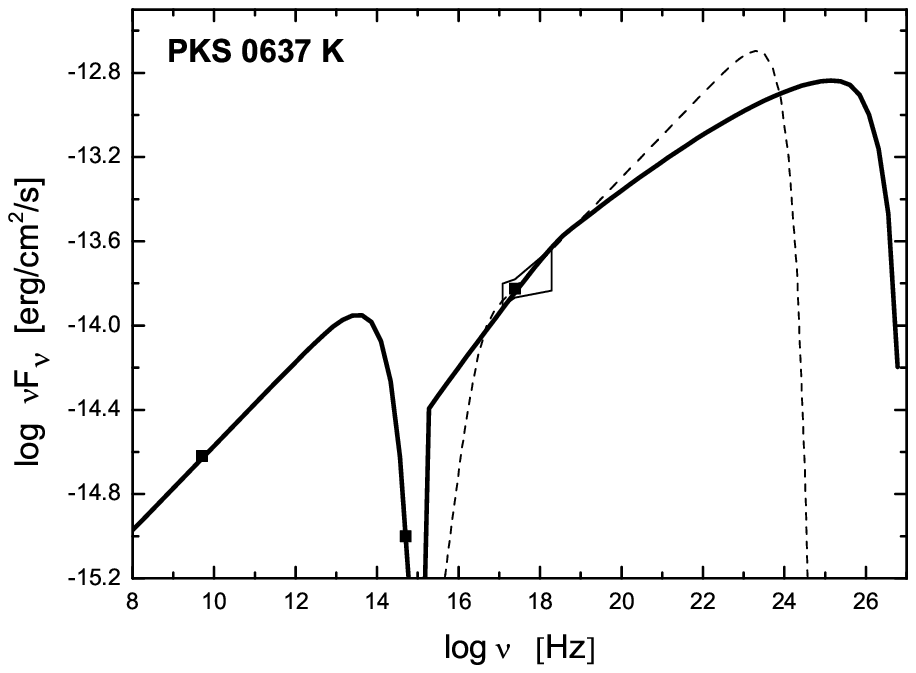}
\hfill
\includegraphics[angle=0,scale=0.80]{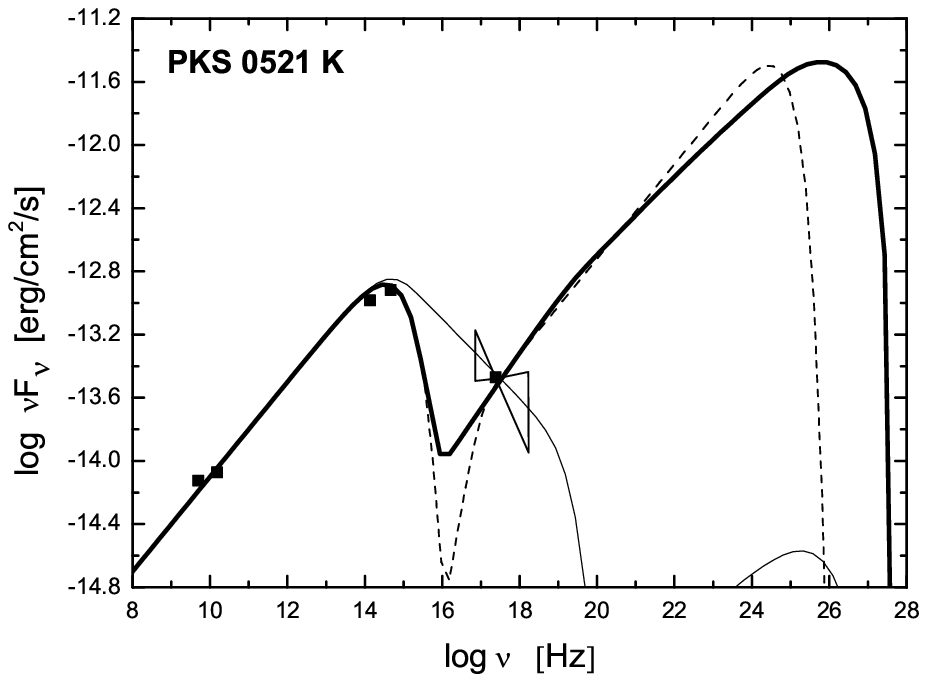}
\hfill\center{Fig. 1---  continued}
\end{figure*}
\clearpage \setlength{\voffset}{0mm}

\begin{figure*}
\includegraphics[angle=0,scale=0.80]{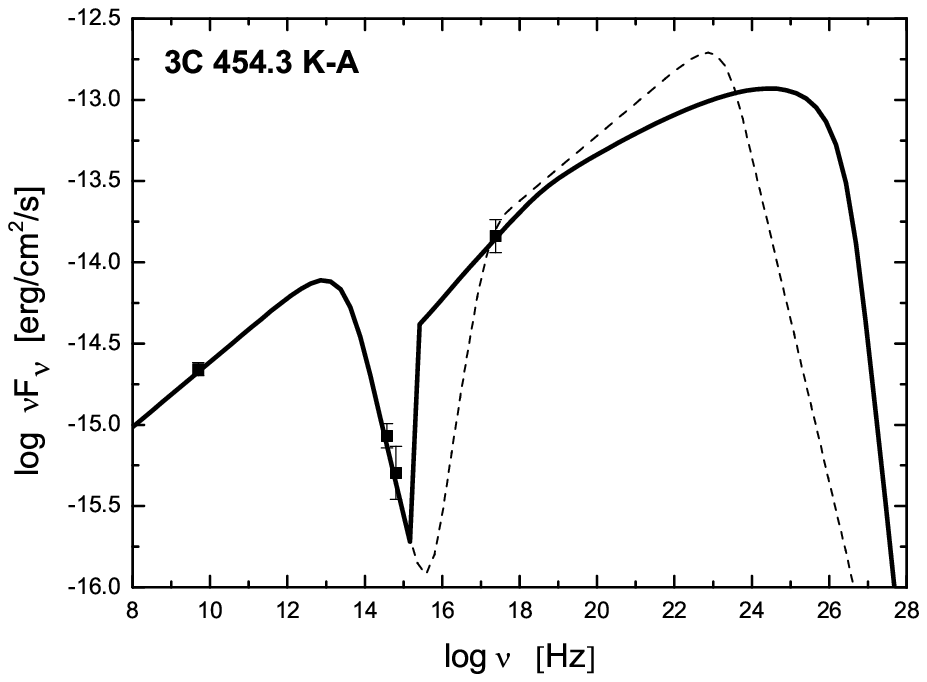}
\includegraphics[angle=0,scale=0.80]{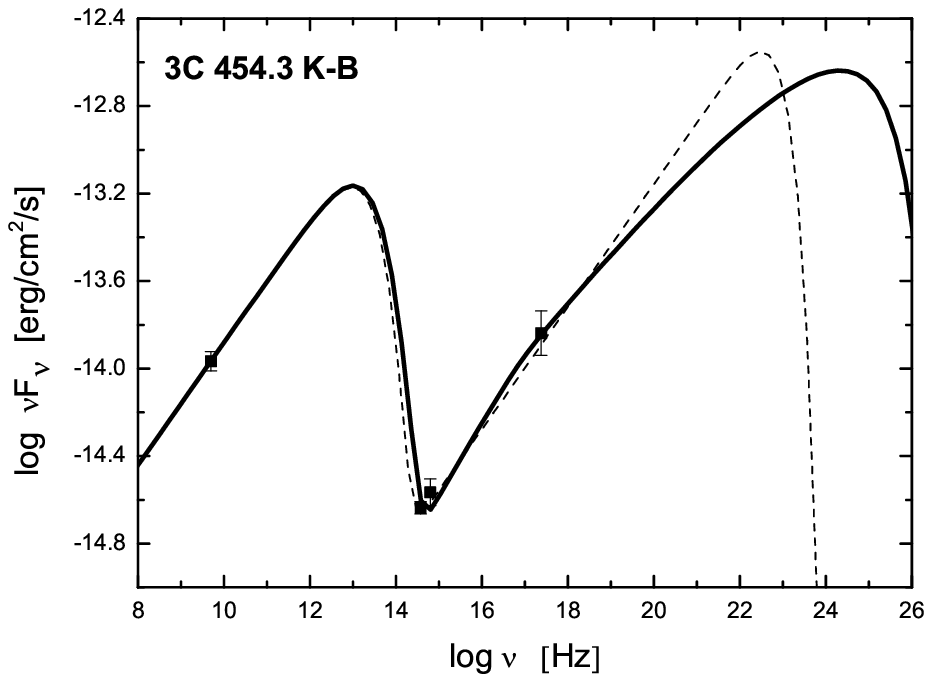}
\includegraphics[angle=0,scale=0.80]{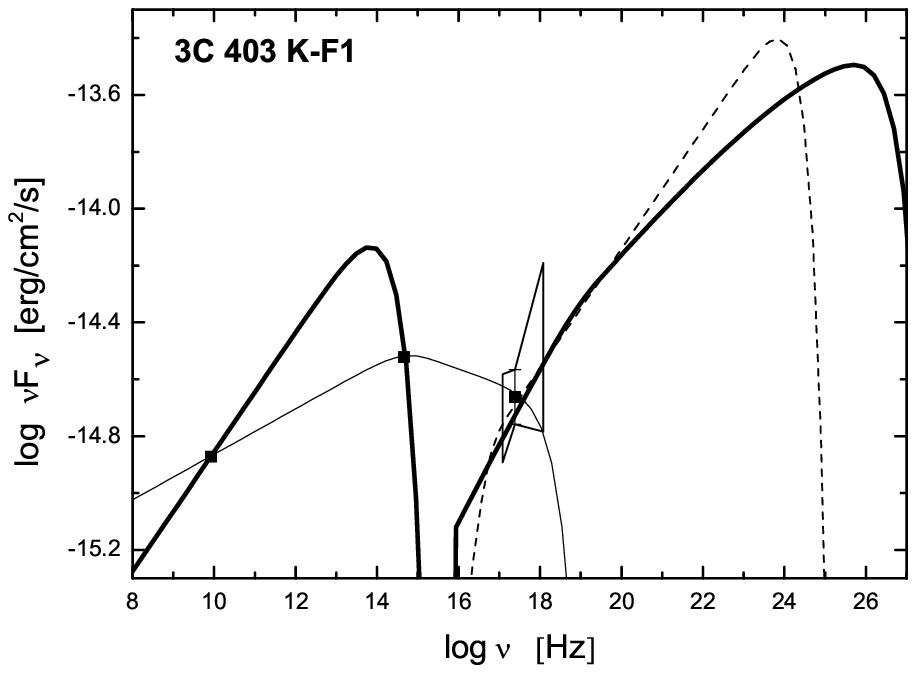}
\includegraphics[angle=0,scale=0.80]{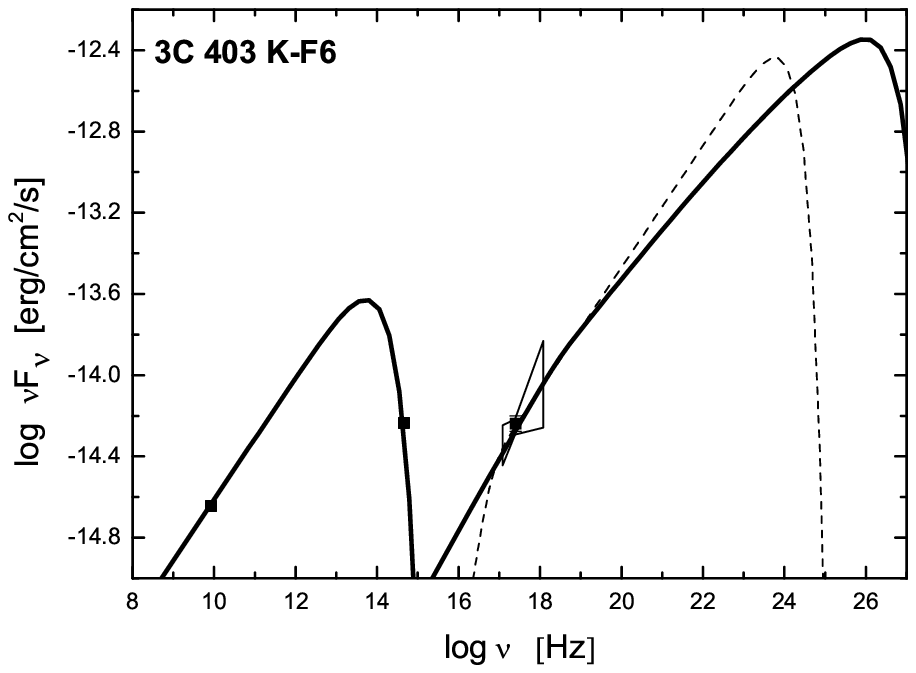}
\includegraphics[angle=0,scale=0.80]{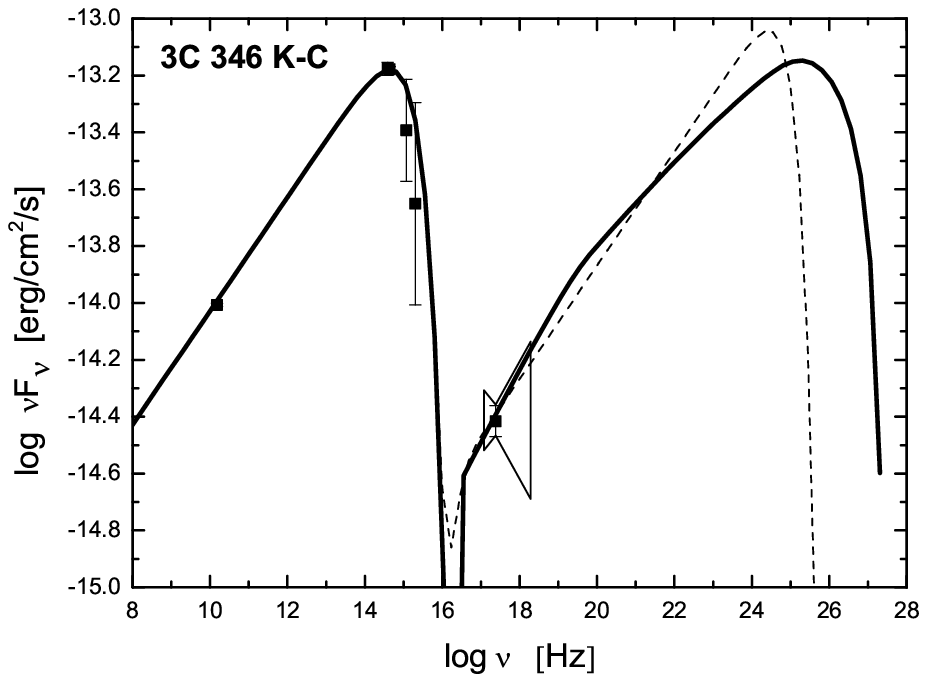}
\includegraphics[angle=0,scale=0.80]{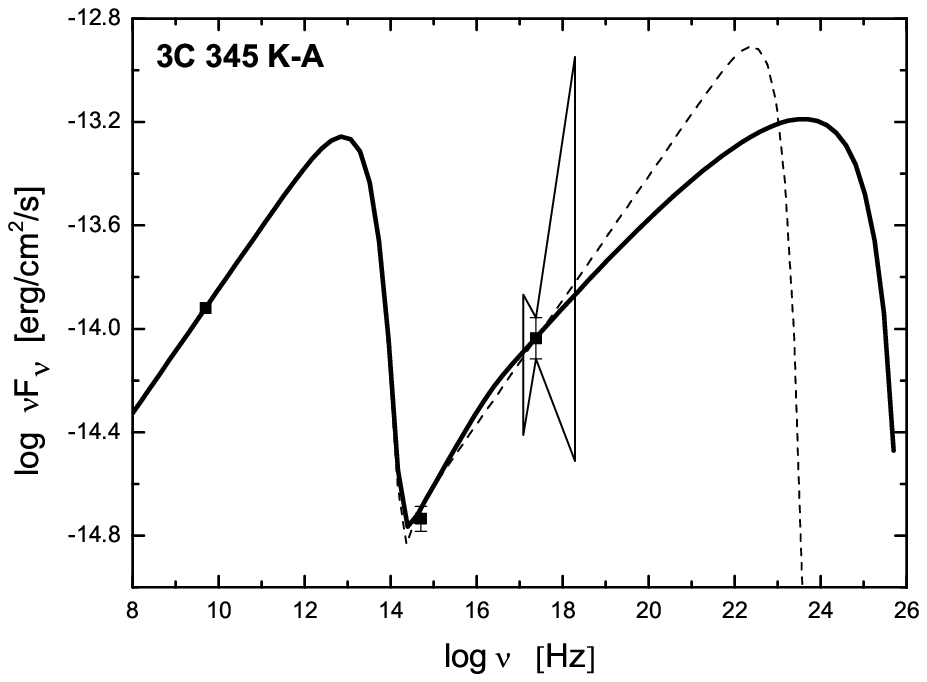}
\includegraphics[angle=0,scale=0.80]{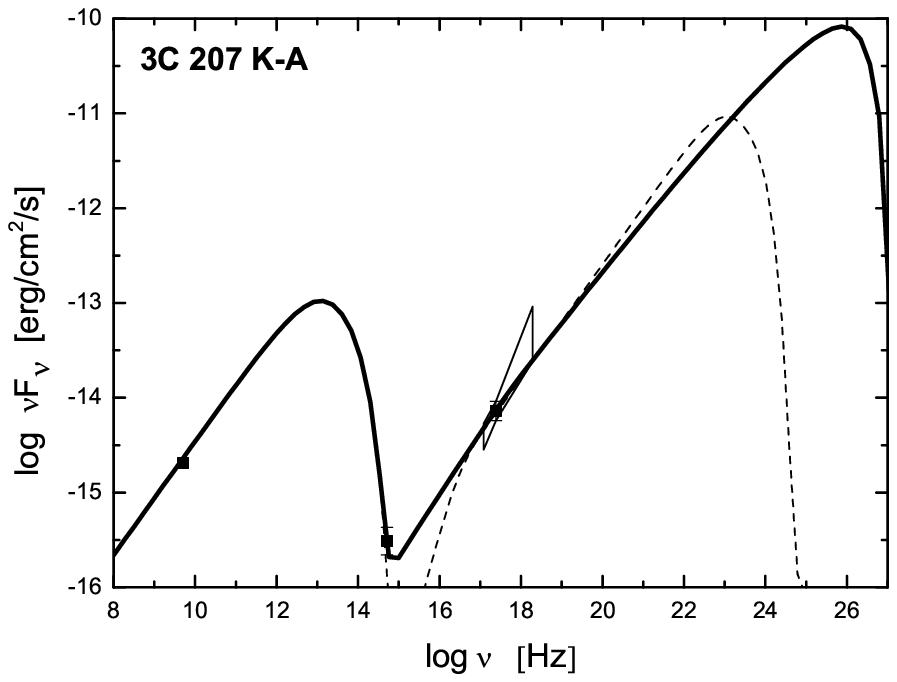}
\hfill
\includegraphics[angle=0,scale=0.80]{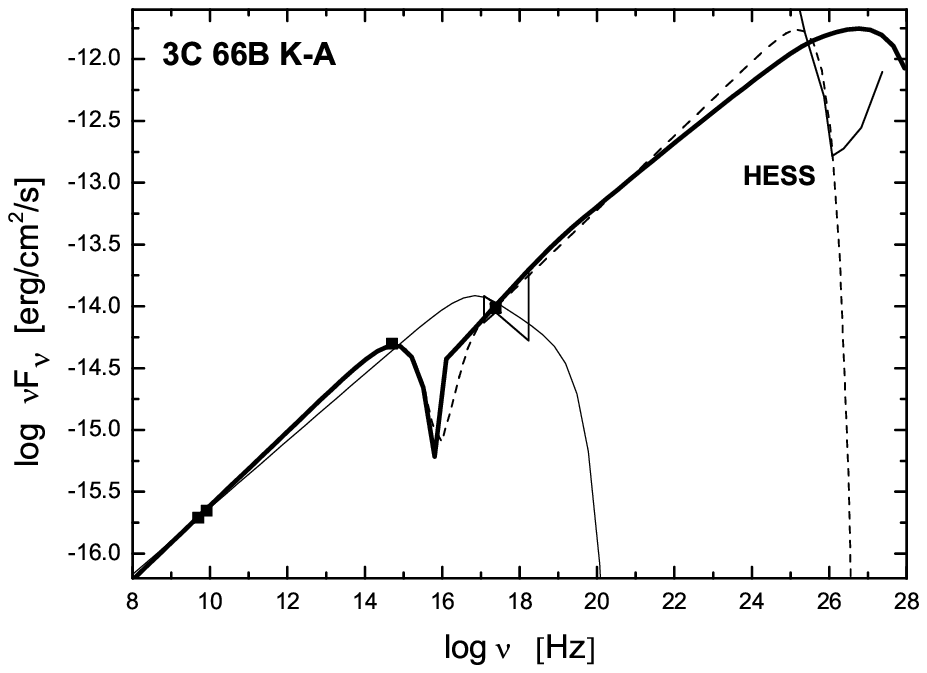}
\hfill\center{Fig. 1---  continued}
\end{figure*}
\clearpage \setlength{\voffset}{0mm}

\begin{figure*}
\includegraphics[angle=0,scale=0.80]{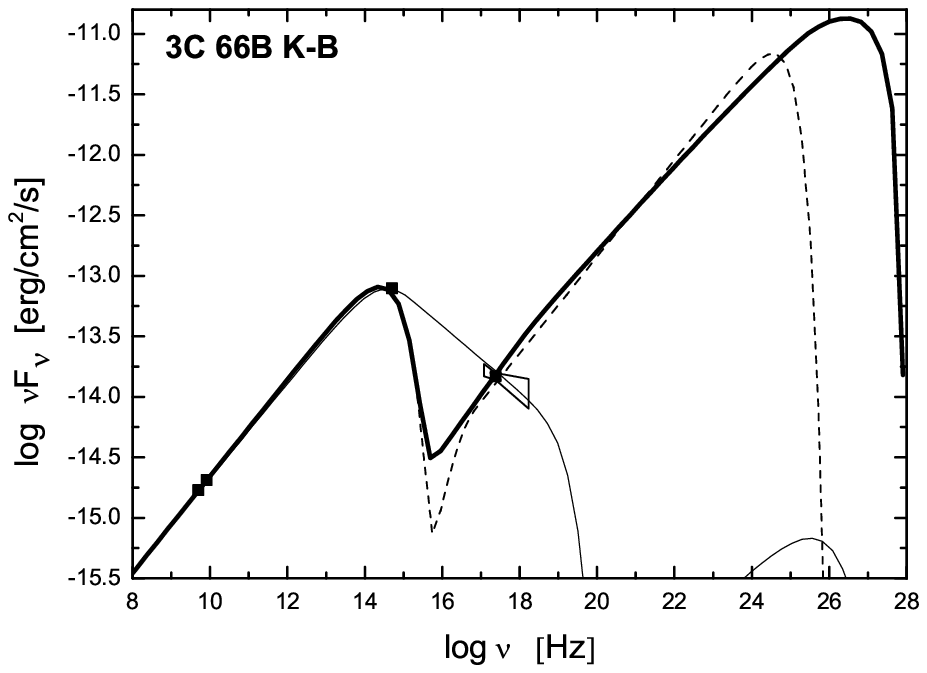}
\includegraphics[angle=0,scale=0.80]{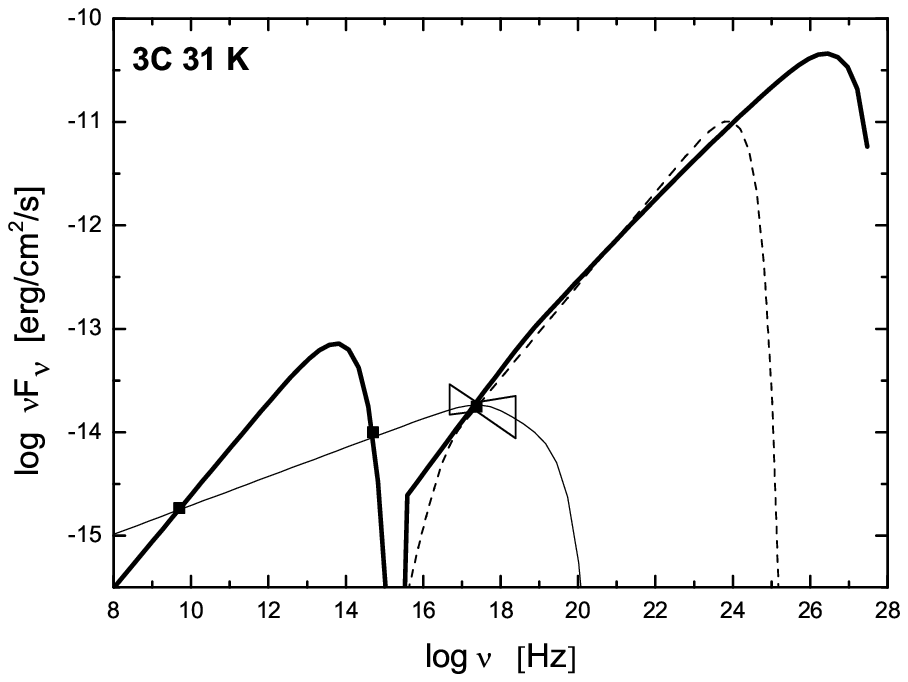}
\includegraphics[angle=0,scale=0.80]{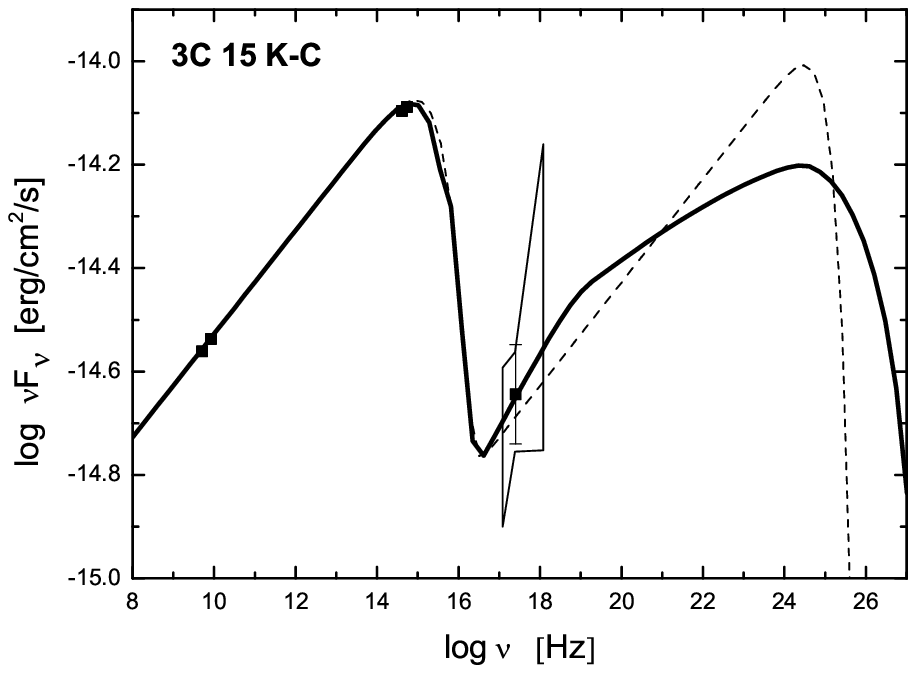}
\includegraphics[angle=0,scale=0.80]{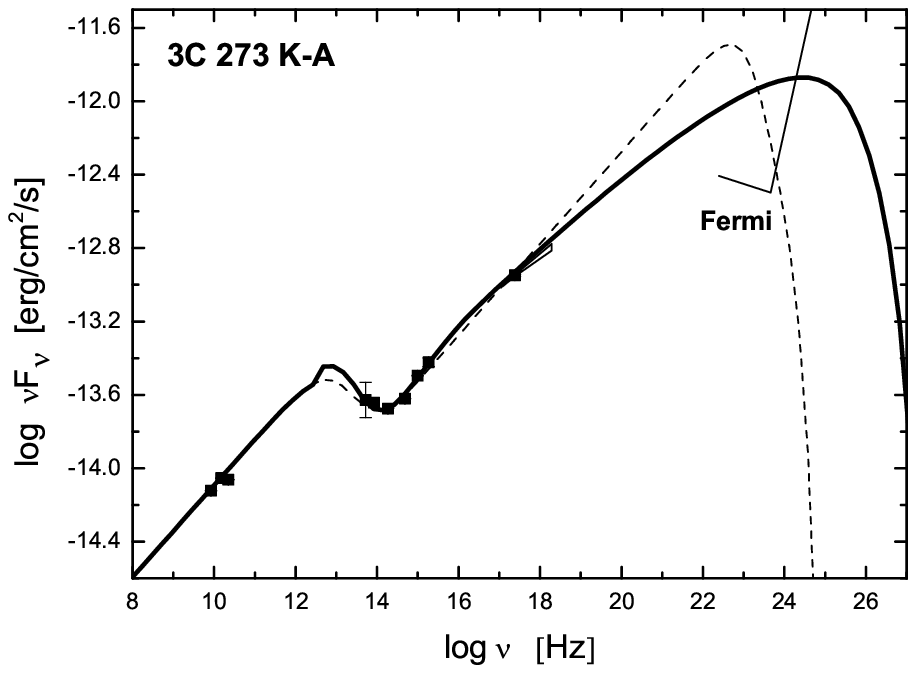}
\includegraphics[angle=0,scale=0.80]{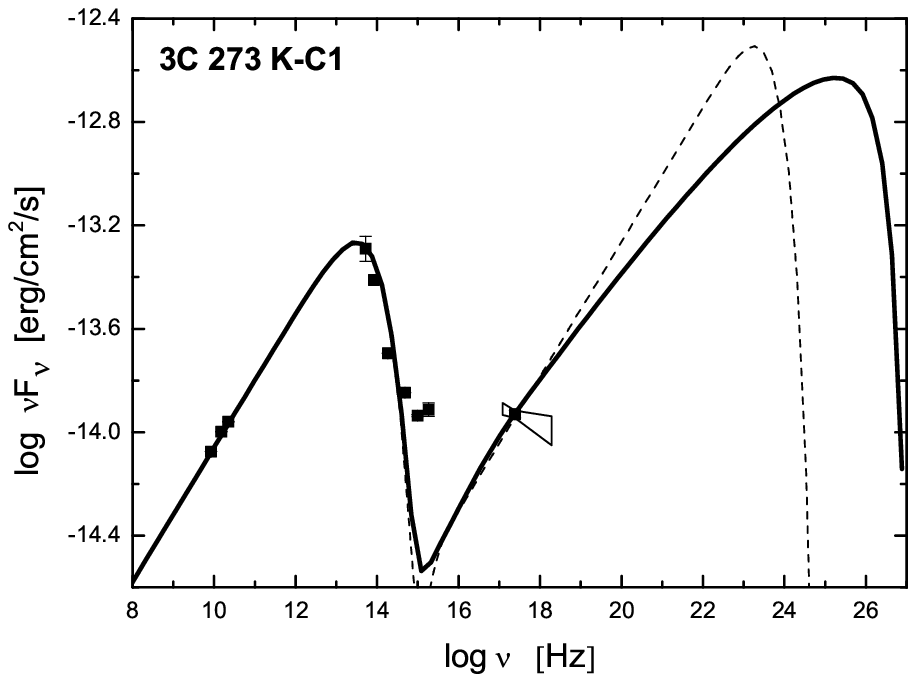}
\includegraphics[angle=0,scale=0.80]{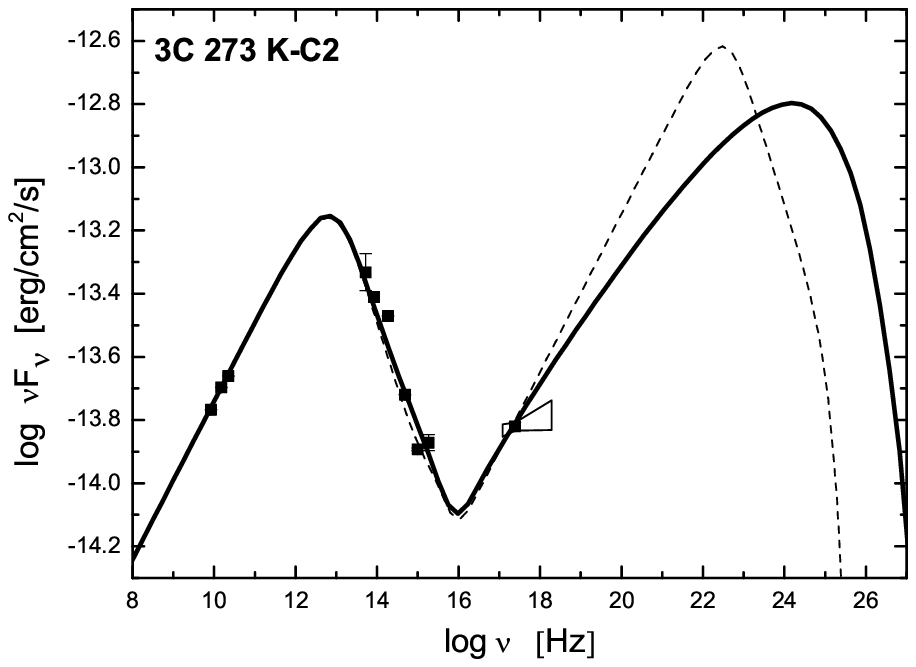}
\includegraphics[angle=0,scale=0.80]{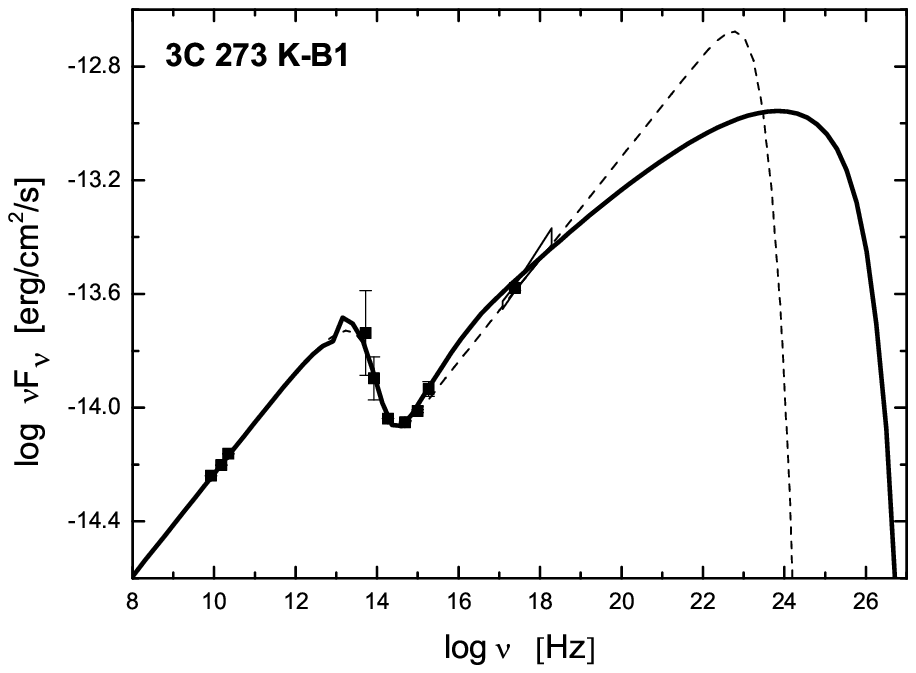}
\hfill
\includegraphics[angle=0,scale=0.80]{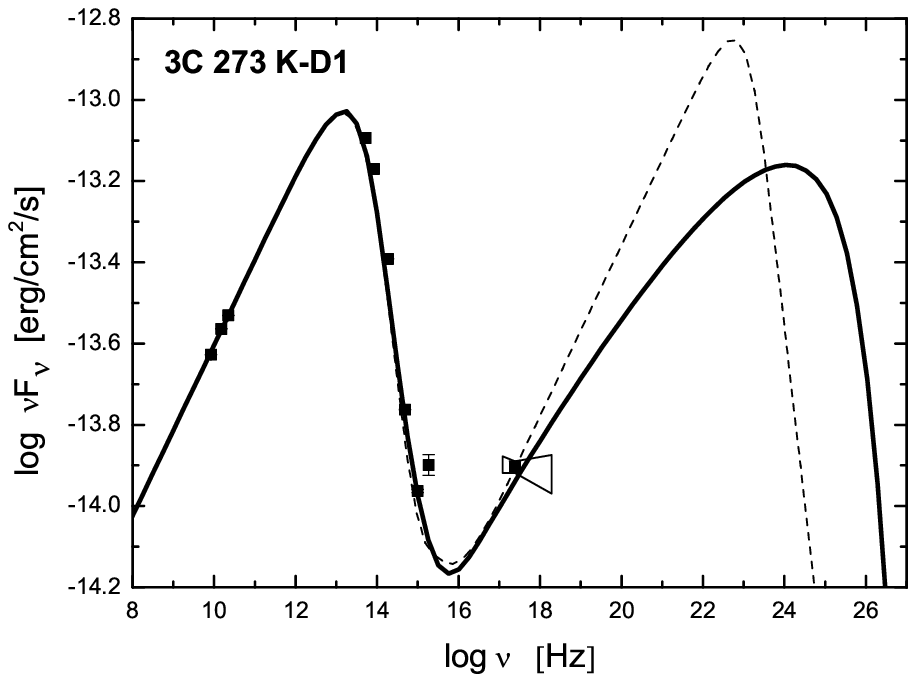}
\hfill\center{Fig. 1---  continued}
\end{figure*}
\clearpage \setlength{\voffset}{0mm}

\begin{figure*}
\includegraphics[angle=0,scale=0.80]{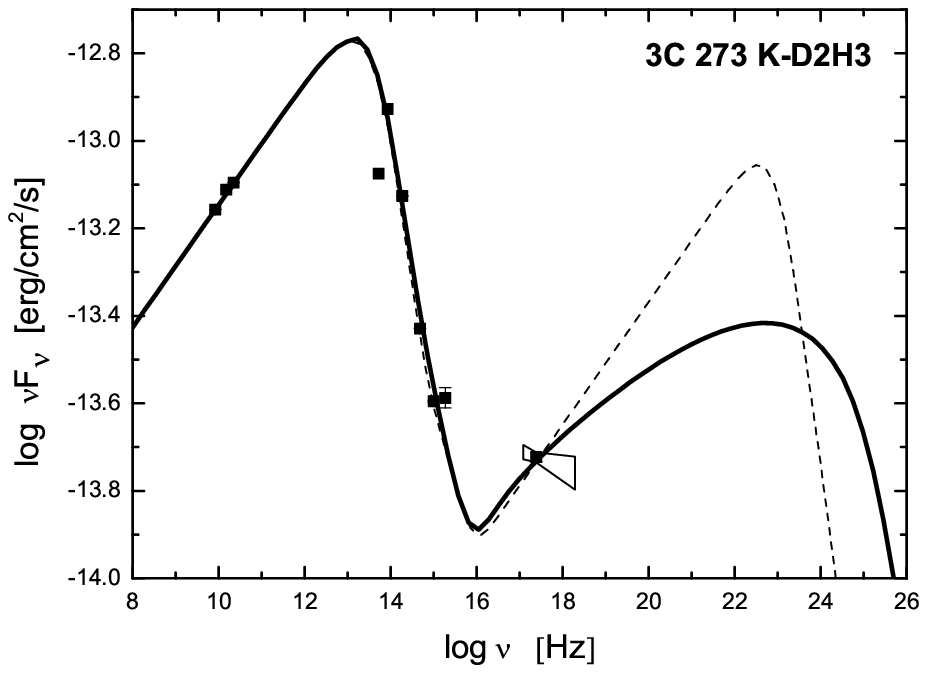}
\includegraphics[angle=0,scale=0.80]{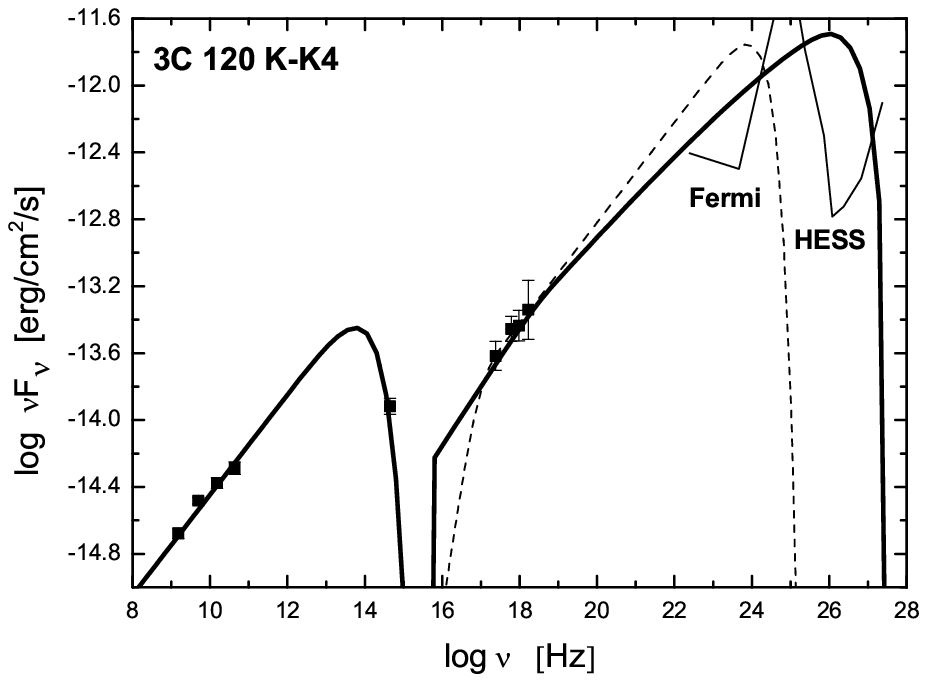}
\includegraphics[angle=0,scale=0.80]{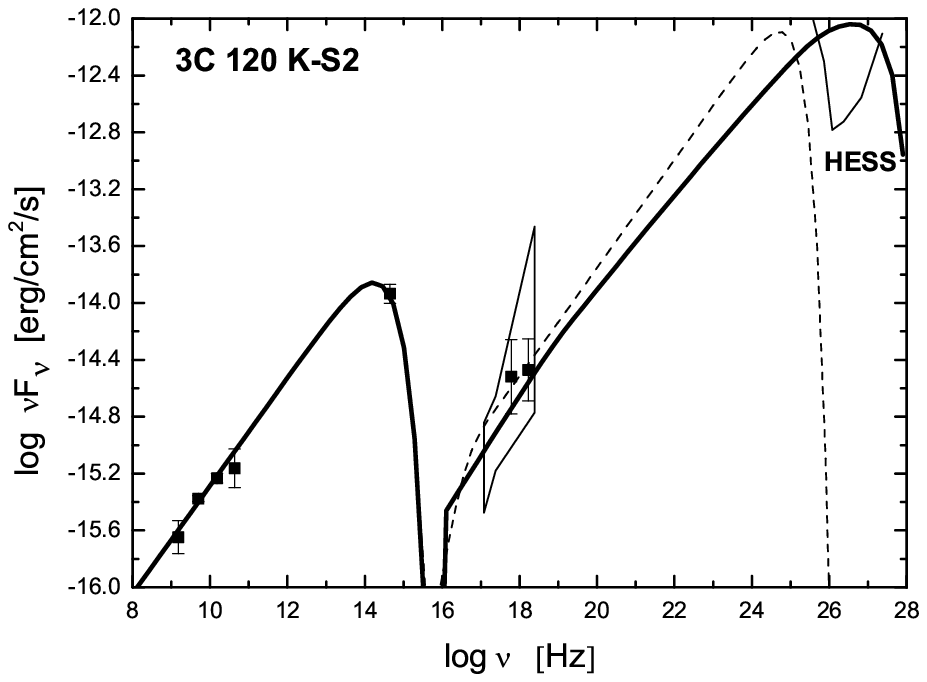}
\includegraphics[angle=0,scale=0.80]{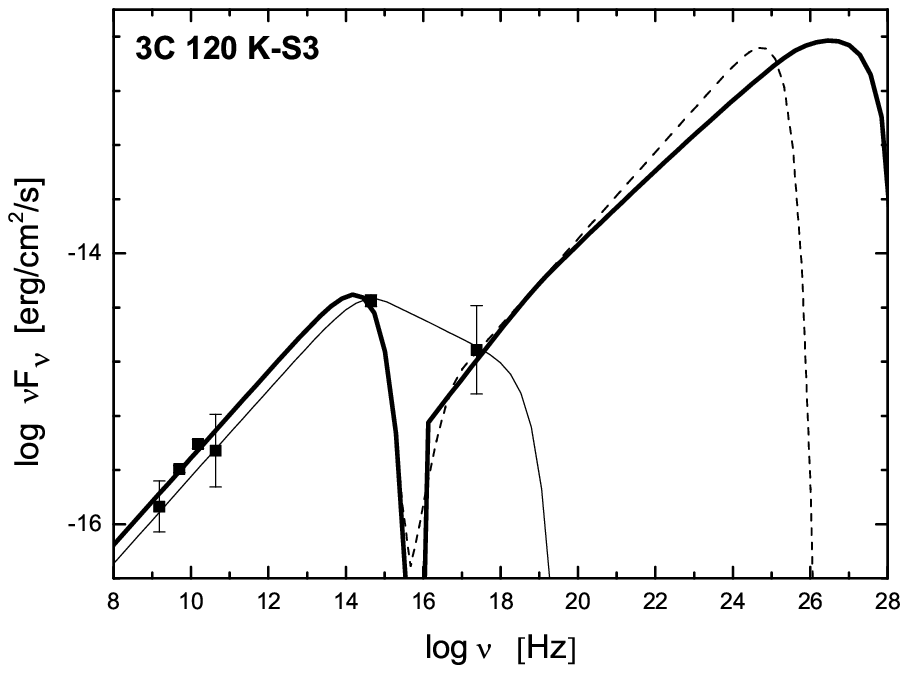}
\includegraphics[angle=0,scale=0.80]{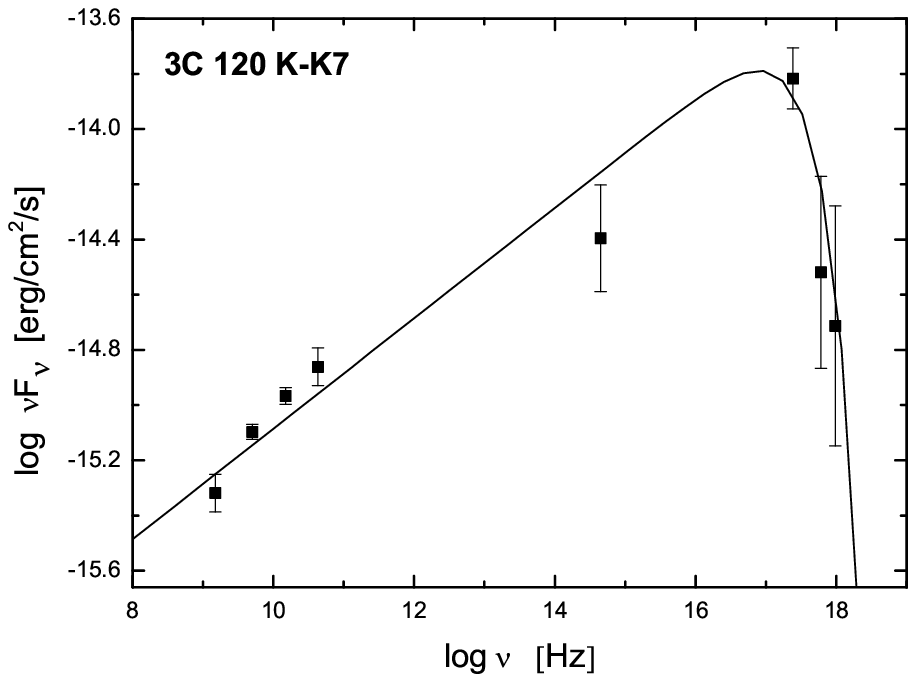}
\end{figure*}

\clearpage
\begin{figure}
\plotone{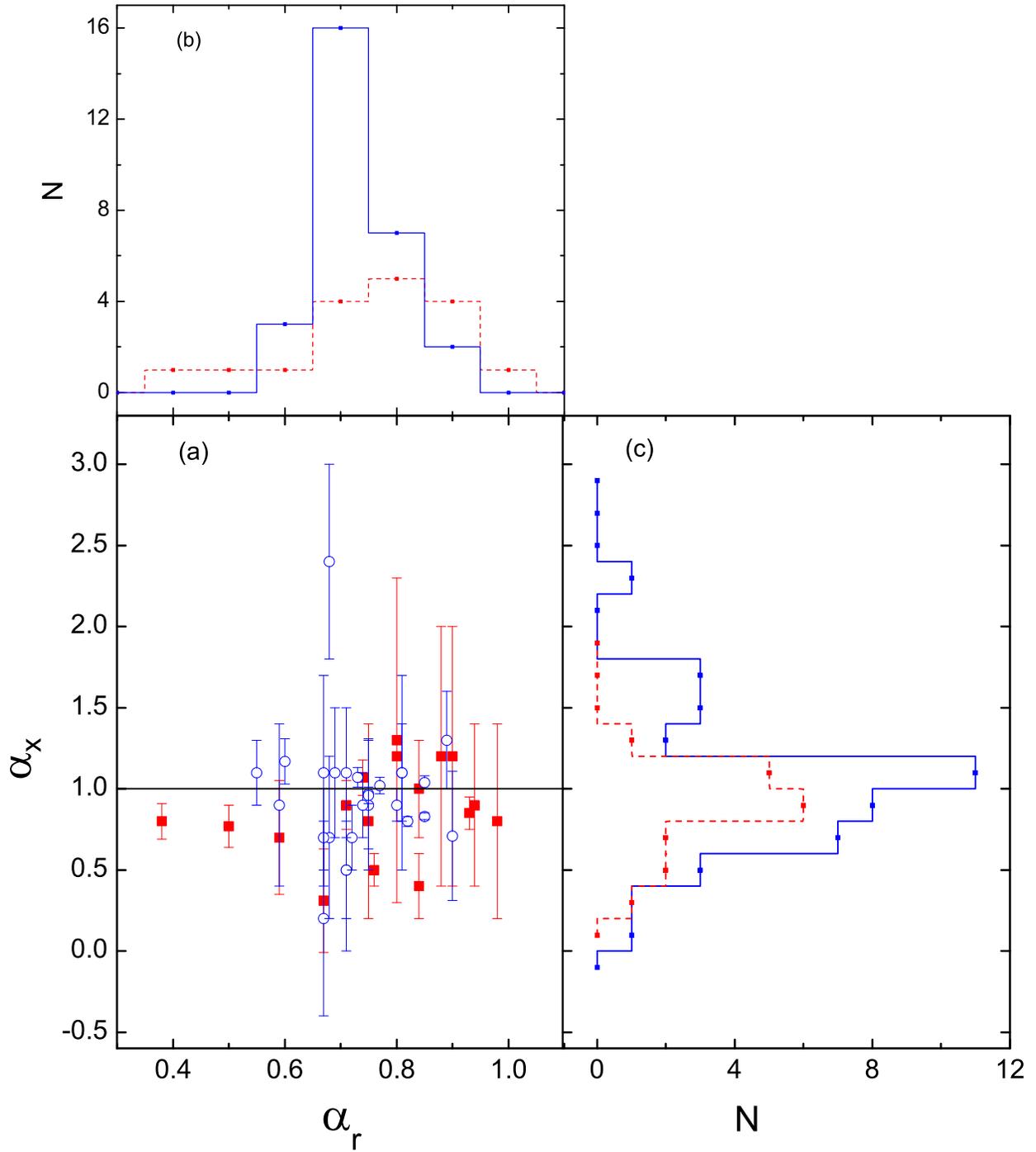} \caption{Distributions of the observed spectral
indices in the radio and X-ray bands for the knots ({\em
circles/solid lines}) and hot spots ({\em squares/dashed lines}).
 \label{Fig:2}}
\end{figure}

\clearpage
\begin{figure*}
\includegraphics[angle=0,scale=0.9]{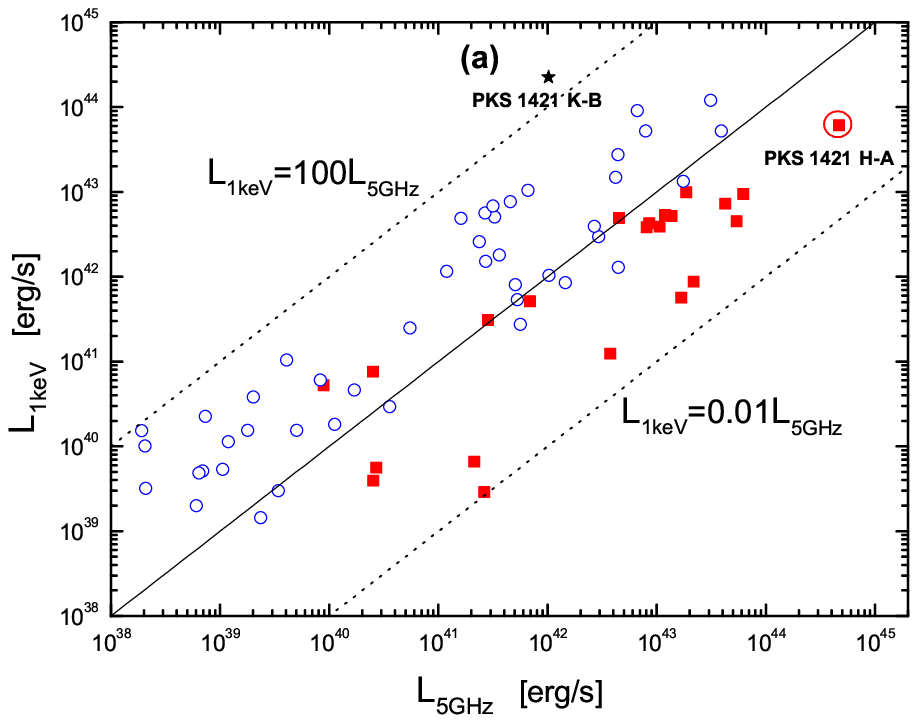}
\includegraphics[angle=0,scale=0.9]{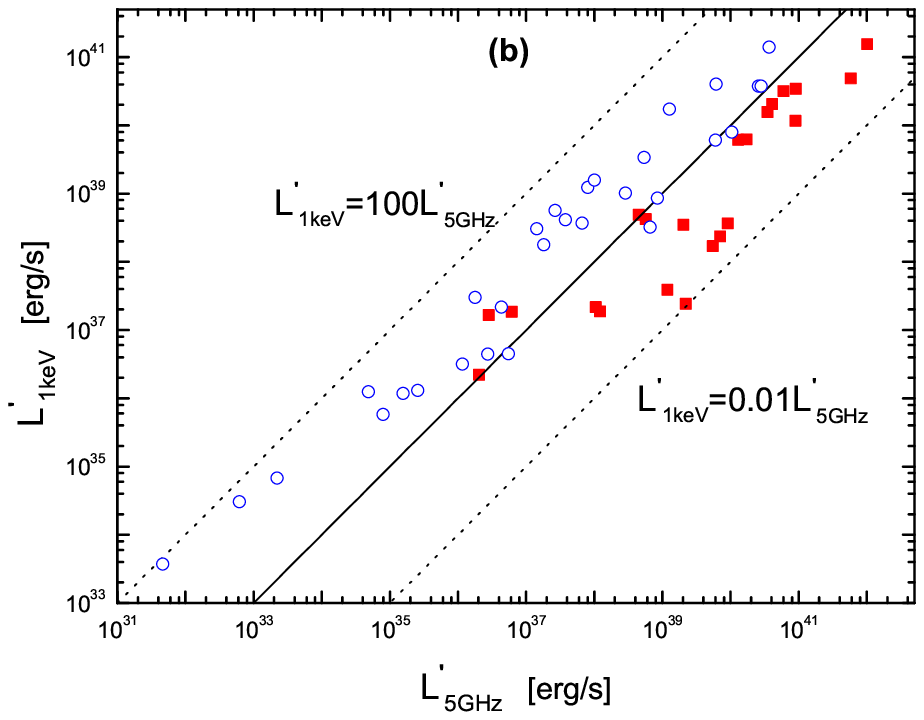}
\caption{Luminosity at 1 keV as a function of
that at 5 GHz for the knots and hot spots without ({\em Panel a}) and with ({\em Panel b}) corrected by the beaming factors. The symbols are the same as in Fig. 2.\label{Fig:3}}
\end{figure*}

\clearpage
\begin{figure*}
\includegraphics[angle=0,scale=1.7]{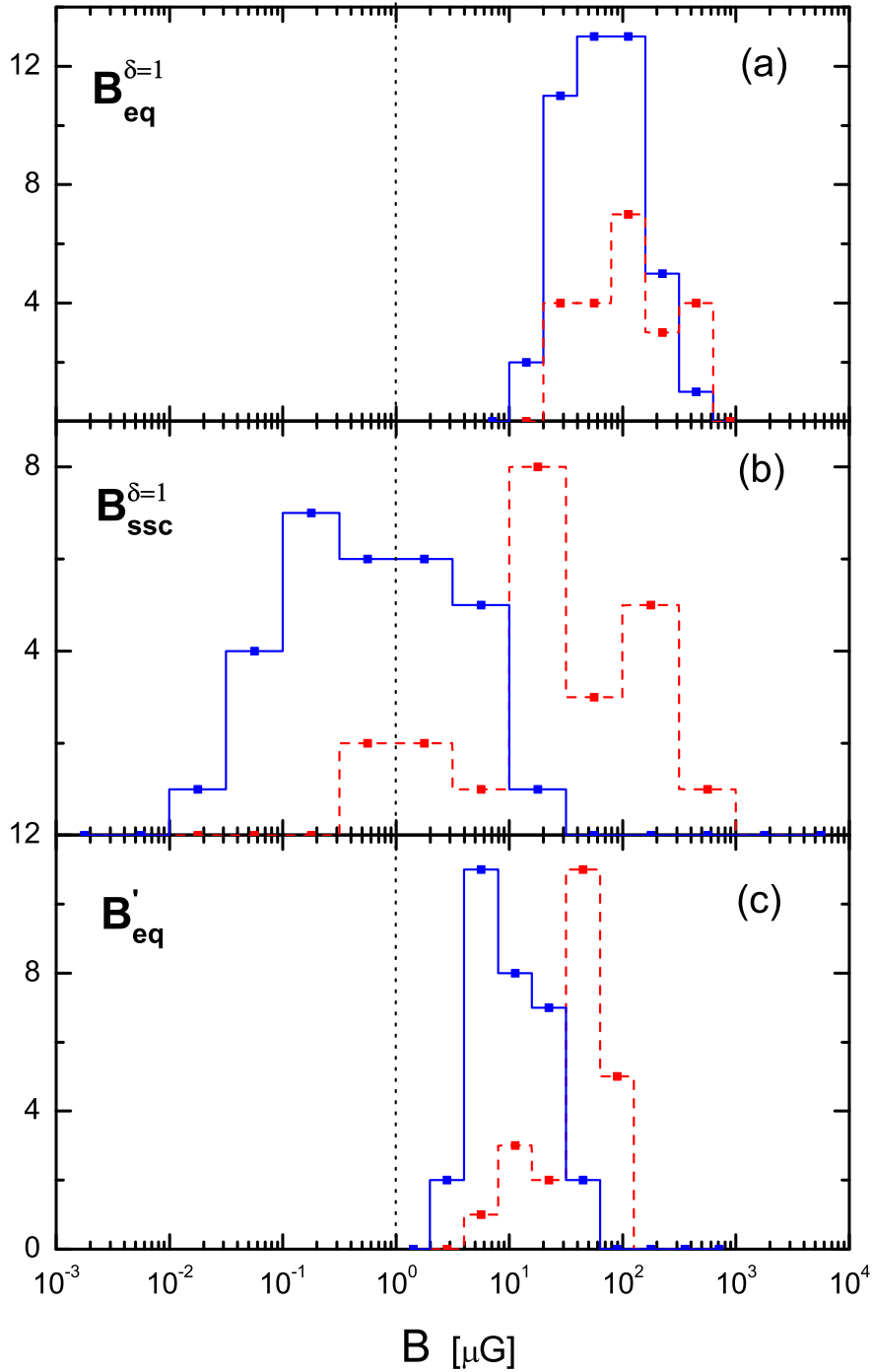}
\caption{Distributions of the magnetic field strength for the knots and hot spots in cases of (a) assuming equipartition condition and $\delta=1$, (b) derived from the SSC model by assuming $\delta=1$, and (c) considering the beaming effect. The vertical {\em dotted} line is the magnetic field strength for the interstellar medium, i.e. $B=1 \mu$G. The symbol styles are the same as in Fig. 2.
 \label{Fig:4}}
\end{figure*}

\clearpage
\begin{figure*}
\includegraphics[angle=0,scale=0.80]{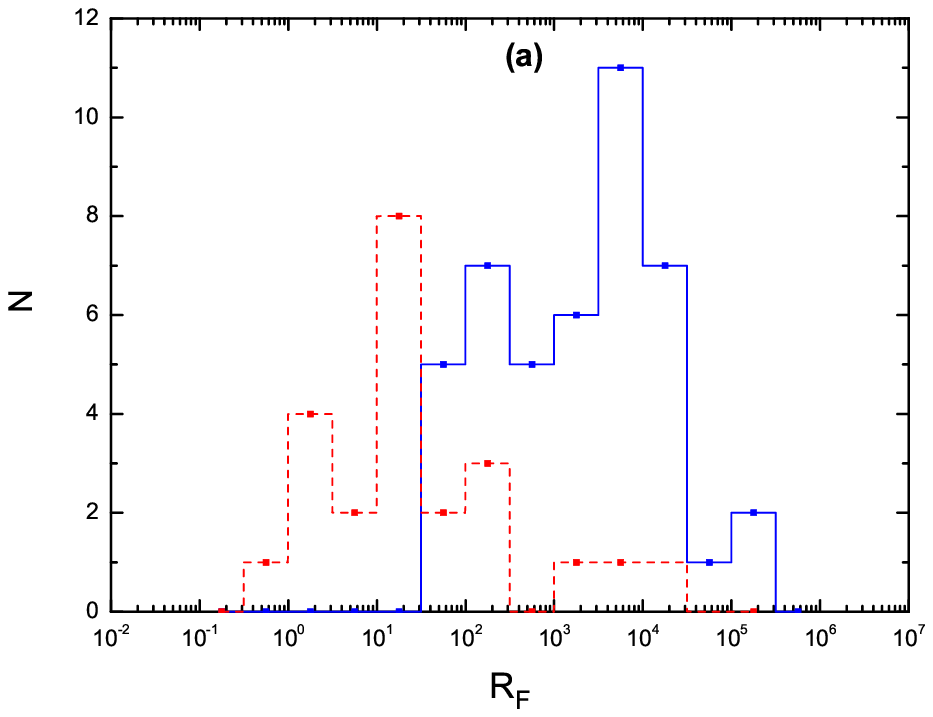}
\includegraphics[angle=0,scale=0.80]{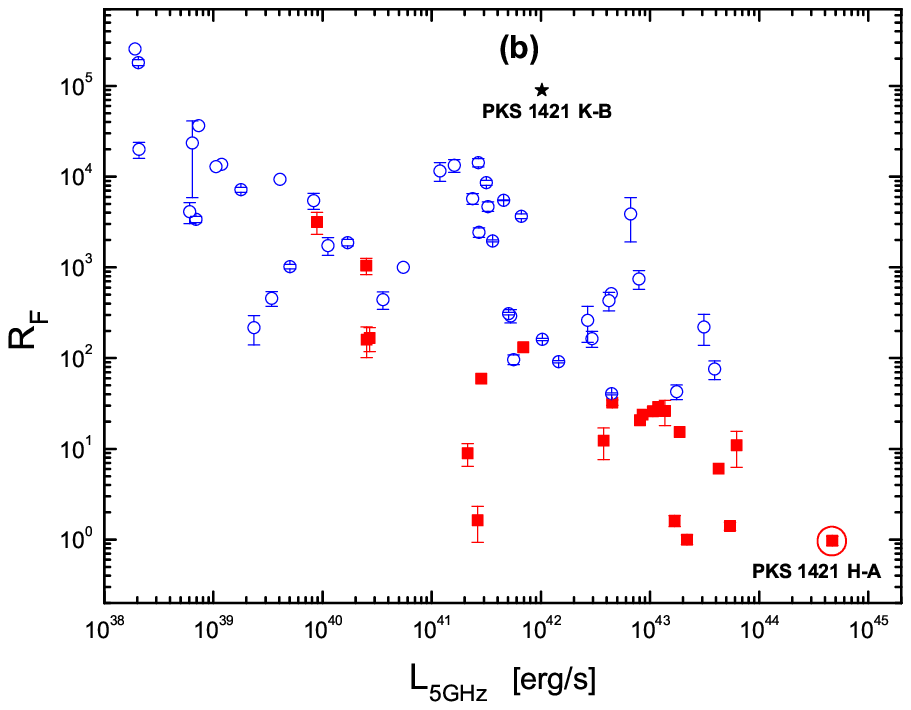}
\includegraphics[angle=0,scale=0.80]{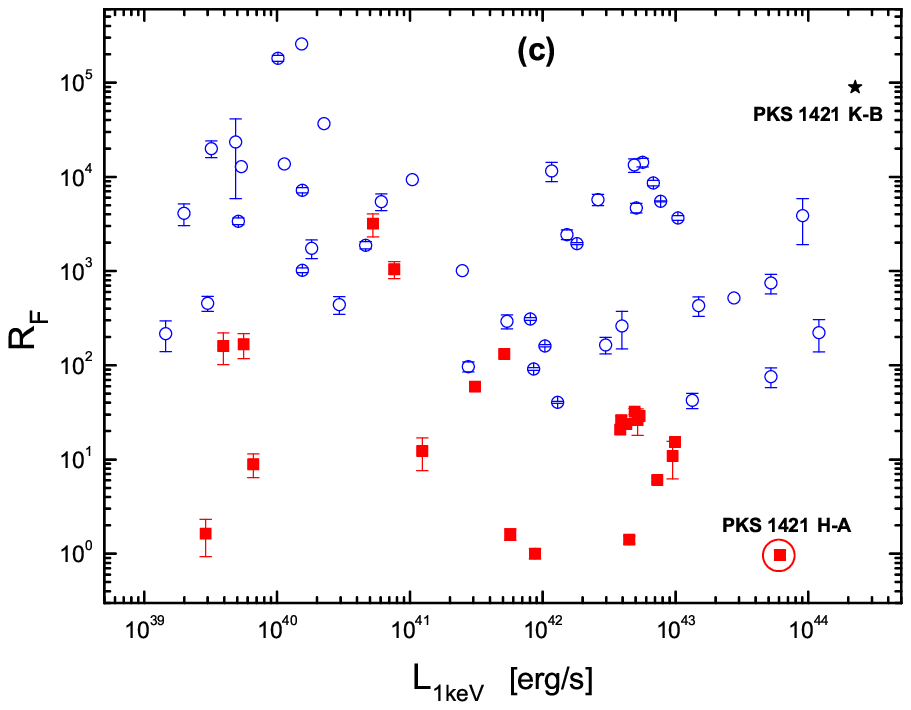}
\hfill
\includegraphics[angle=0,scale=0.80]{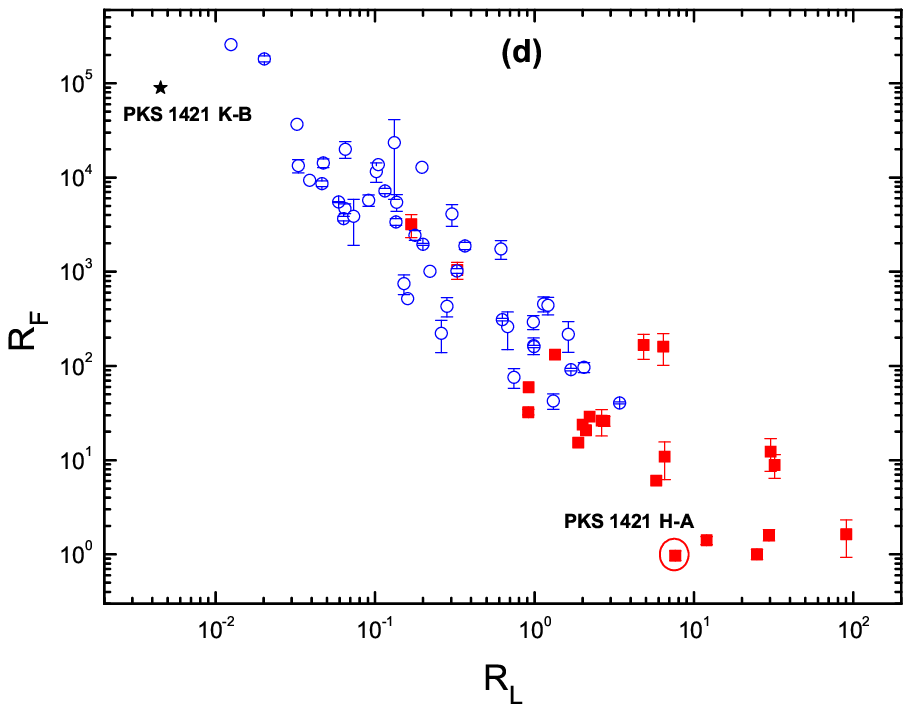}
\caption{{\em Panel a}---Distributions of the ratio ($R_{F}$) of
the observed flux density at 1 keV to that of expected
from the SSC model with $B=B^{\delta=1}_{\rm eq}$. {\em Panels
b,c,d}---$R_F$ as a function of the luminosities at 5 GHz and 1
keV bands and the radio ($R_{L}$) of the luminosities in the two
energy bands for the knots and hot spots. The symbol styles are the same as in Fig. 2. \label{Fig:5}}
\end{figure*}

\clearpage
\begin{figure*}
\includegraphics[angle=0,scale=0.80]{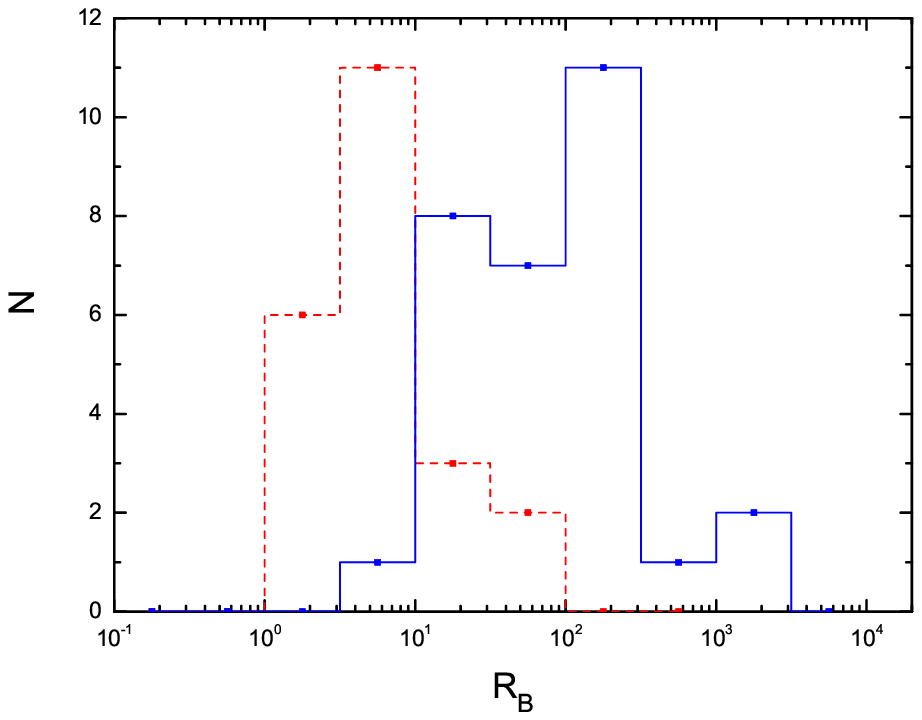}
\includegraphics[angle=0,scale=0.80]{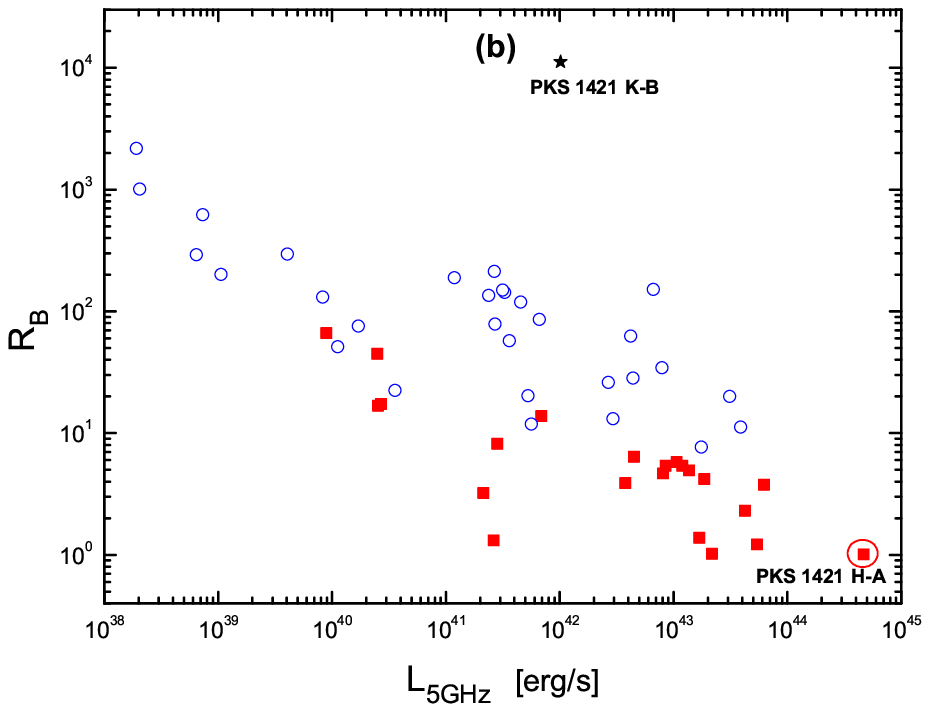}
\includegraphics[angle=0,scale=0.80]{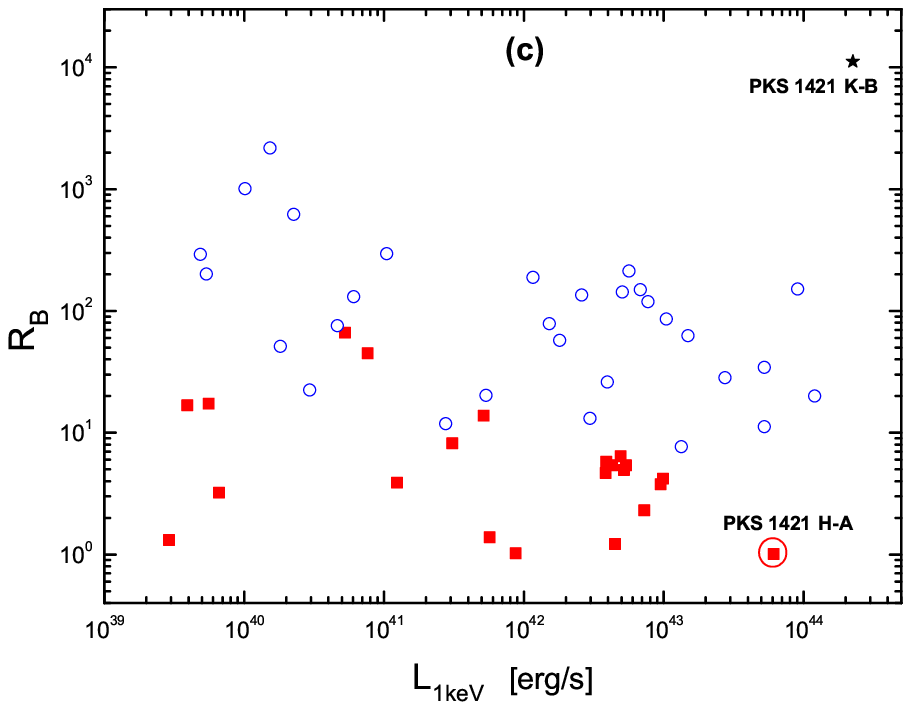}
\hfill
\includegraphics[angle=0,scale=0.80]{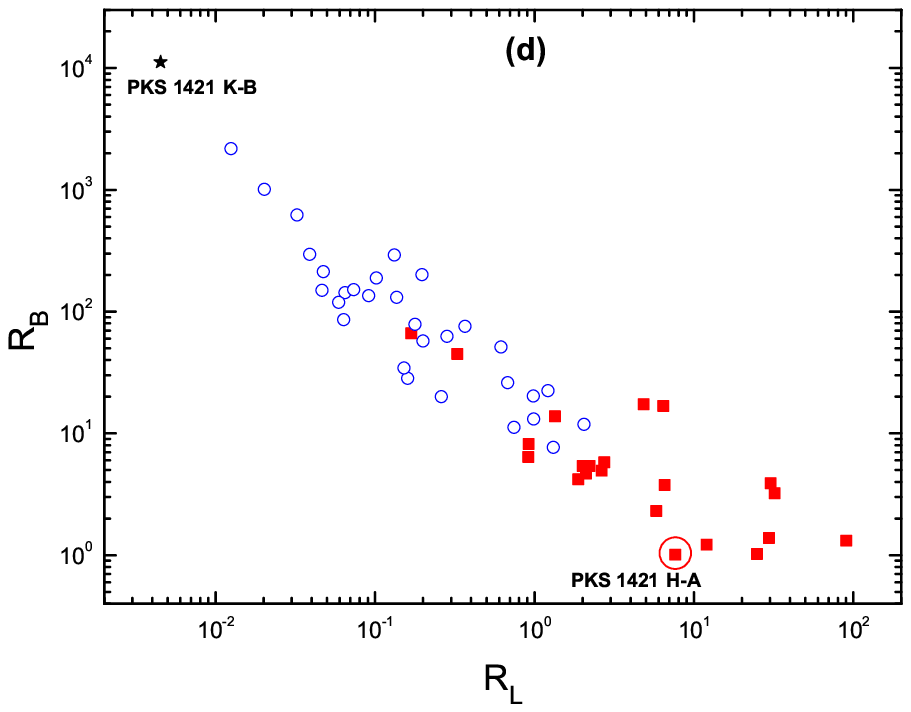}
\caption{{\em Panel a}---Distributions of the ratio ($R_{B}$) of
the equipartition magnetic field strength to $B_{\rm
ssc}^{\delta=1}$, the magnetic field strength derived from the SSC
model by assuming $\delta=1$. {\em Panels b,c,d}---$R_B$ as a
function of the luminosities at 5 GHz and 1 keV bands and the
radio ($R_{L}$) of the luminosities in the two energy bands for
the knots and hot spots. The symbol styles are the same as in Fig. 2. \label{Fig:6}}
\end{figure*}

\clearpage
\begin{figure}
\plotone{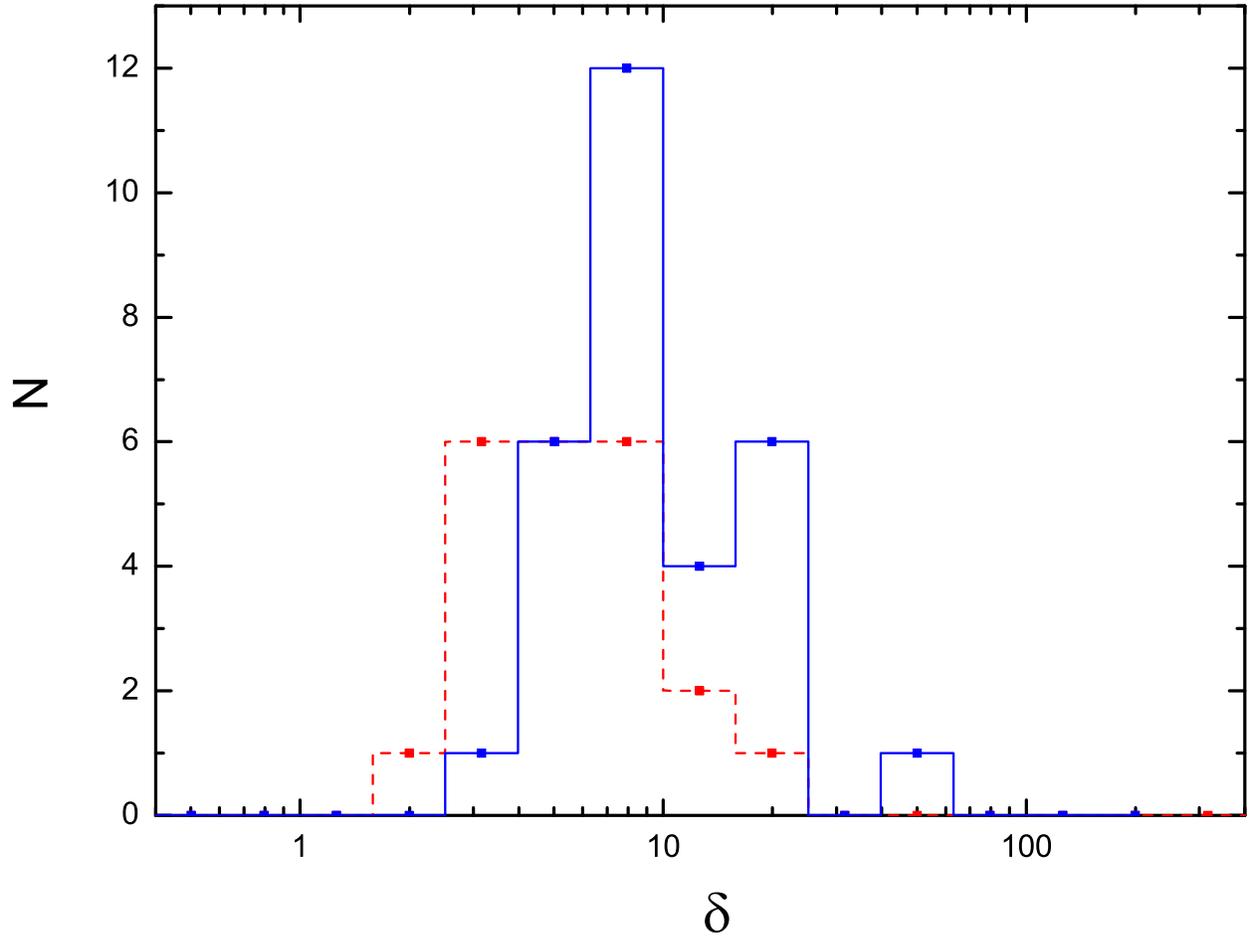} \caption{Distributions of the beaming factors for the knots and hot spots. The symbols are the same as in Fig. 2. \label{Fig:7}}
\end{figure}

\clearpage
\begin{figure}
\plotone{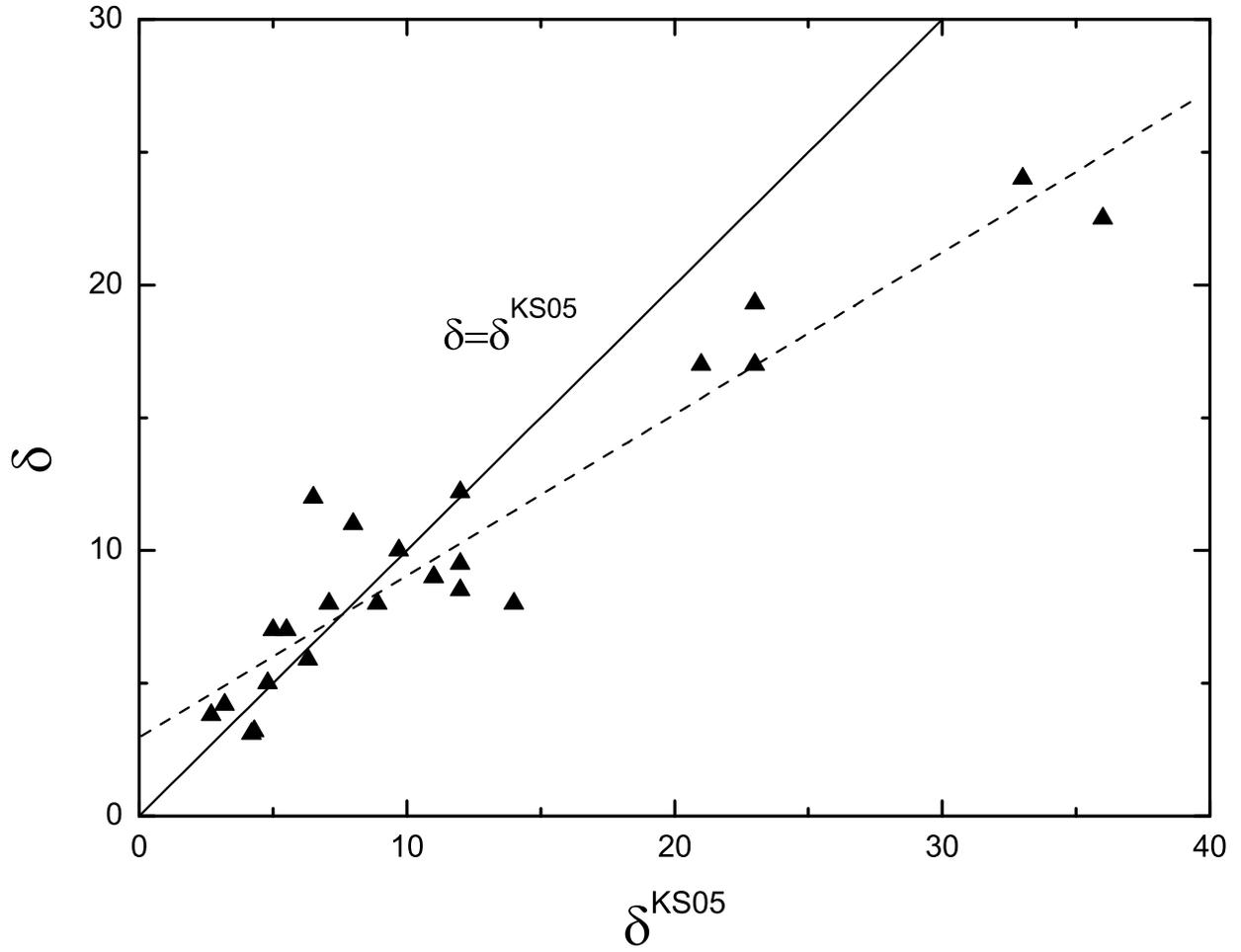} \caption{Comparison of $\delta$ between our results and that of Kataoka \& Stawarz (2005)($\delta^{\rm KS05}$). The {\em solid} line is for $\delta=\delta^{\rm KS05}$. The {\em dashed} line is the linear fit to the two quantities, with a  correlation coefficient $r=0.94$. \label{Fig:8}}
\end{figure}

\clearpage
\begin{figure*}
\includegraphics[angle=0,scale=0.8]{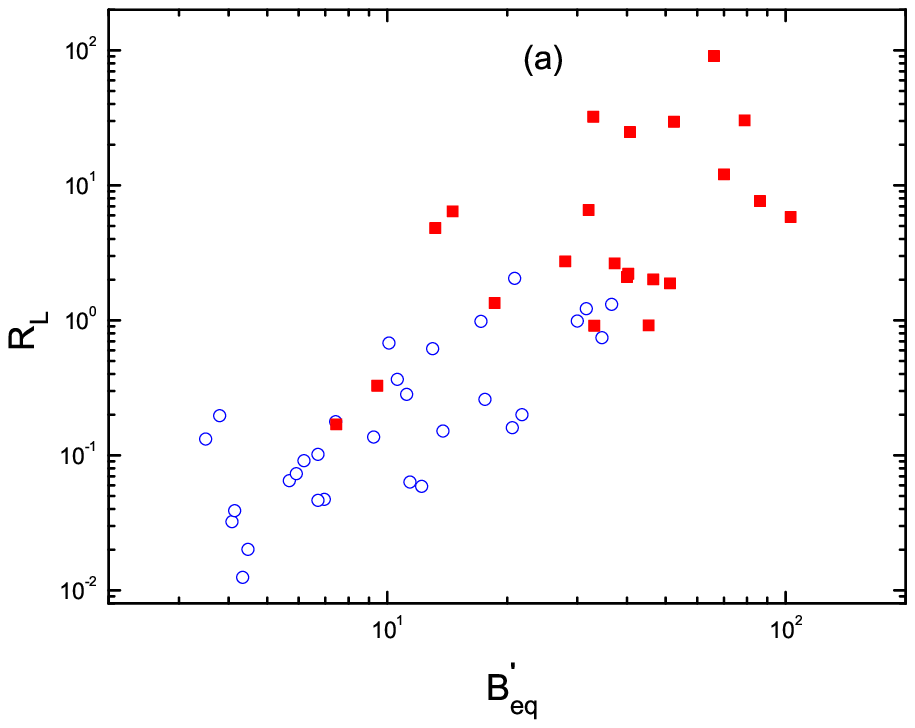}
\includegraphics[angle=0,scale=0.8]{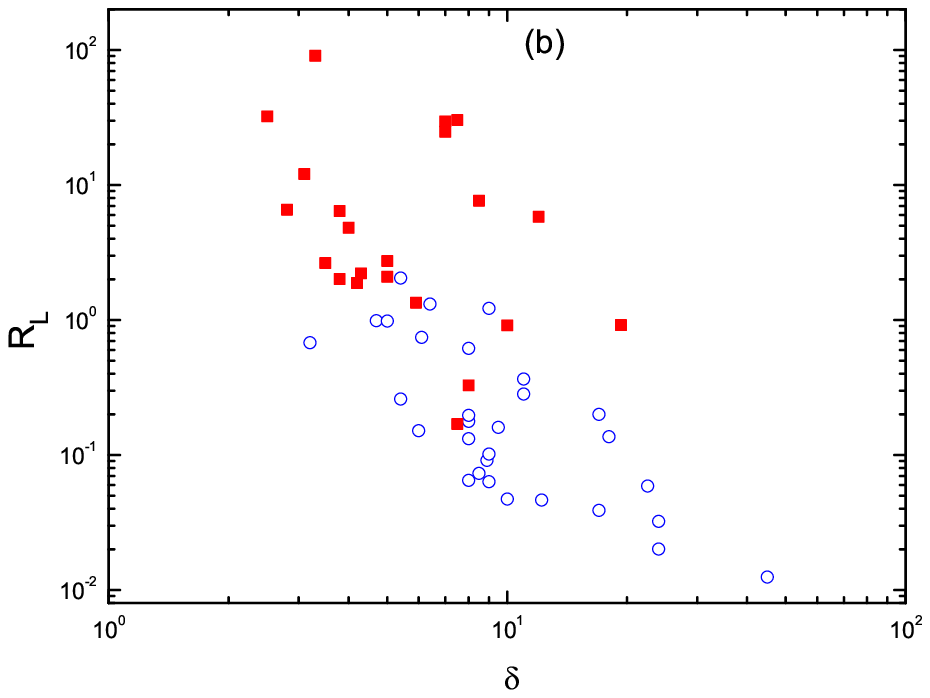}
\caption{{\em Panel b}---Correlations of $R_{L}$ with $B_{\rm eq}^{'}$ ({\em panel a}) and $\delta$ ({\em panel b})
for the knots and hot spots. The symbol styles are the same as in Fig. 2.
\label{Fig:9}}
\end{figure*}

\clearpage

\appendix

\section{The Equipartition Magnetic Field}
Under the equipartition condition, we have
\begin{equation}
\frac{B_{\rm
eq}^{2}}{8\pi}=U_{e}=m_{e}c^{2}\int^{\gamma_{\max}}_{\gamma_{\min}}N(\gamma)(\gamma-1)d\gamma.
\end{equation}
The peak frequency of the synchrotron radiation is given by
\begin{equation}
\nu_{s}=\frac{4}{3}\nu_B\gamma_{b}^{2}\frac{\delta}{1+z},
\end{equation}
where $\nu_{B}=2.8\times10^6 B$ Hz is the Larmor frequency in magnetic field
$B$. The luminosity of the synchrotron radiation is derived from
\begin{equation}
L_{\rm syn}=\int^{\gamma_{\max}}_{\gamma_{\min}}N(\gamma)P(\gamma)Vd\gamma,
\end{equation}
where $P$ is the radiation power of single electron,
$P=1.1\times10^{-15}\gamma^{2}B^{2}$ erg s$^{-1}$, and $V=\frac{4}{3}\pi R^{3}$
is the volume of radiation region. Without considering the beaming effect
($\delta=1$) and assuming $p_{2}>3>p_{1}>2$, $B_{\rm eq}$ is expressed as
\begin{equation}
B_{\rm eq}=(\frac{A_{1}L_{\rm
syn}}{A_{3}A_{2}^{(3-p_{1})/2}})^{\frac{2}{5+p_{1}}},
\end{equation}
where $A_{1}$, $A_{2}$, and $A_{3}$ are given by
\begin{equation}
A_{1}=8\pi
m_{e}c^{2}(\frac{\gamma_{\min}^{2-p_{1}}}{p_{1}-2}-\frac{\gamma_{\min}^{1-p_{1}}}{p_{1}-1}),
\end{equation}
\begin{equation}
A_{2}=\nu_{s}/3.7\times10^{6},
\end{equation}
and
\begin{equation}
A_{3}=1.1\times10^{-15}V\frac{p_{1}-p_{2}}{(3-p_{1})(3-p_{2})}.
\end{equation}
For the case of $p_{2}>3$ and $p_{1}< 2$, $B_{\rm eq}$ can be calculated by
\begin{equation}\label{Beq}
B_{\rm eq}^{7/2}=\frac{L_{\rm syn}A_{1}A_{2}^{(p_{1}-3)/2}}{A_{3}}B_{\rm
eq}^{\frac{2-p_{1}}{2}}+\frac{A_{0}L_{\rm syn}}{A_{3}A_{2}^{1/2}},
\end{equation}
where
\begin{equation}
A_{0}=8\pi m_{e}c^{2}\frac{p_{1}-p_{2}}{(2-p_{1})(2-p_{2})}.
\end{equation}
For the case of $p_{1}=2$, $A_{1}$ in Eq. (A4) is
\begin{equation}
A_{1}=8\pi m_{e}c^{2}(\frac{1}{p_2-2}-\ln\gamma_{min}-\frac{1}{\gamma_{min}}).
\end{equation}
If the beaming effect is considered, we have $L_{syn}^{'}=L_{syn}/\delta^{4}$,
$V^{'}=V/\delta$. The equipartition magnetic field hance is obtained with
\begin{equation}
B_{eq}^{'}=B_{eq}/\delta^{(p_{1}+3)/(p_{1}+5)}.
\end{equation}
This is consistent with the result presented by Stawarz et al. (2003),
$B_{eq}^{'}=B_{eq}/\delta^{5/7}$ for $p_{1}=2$. It is generally believed the
radio emission is produced by synchrotron radiation. We obtain the values of
$\alpha_{1,2}$, $\nu_s$, and $L_{\rm syn}$ by fitting the observed radio (and
optical) data using synchrotron radiation and calculate $B_{\rm eq}$ with Eqs.
(A4), (A8), and (A11).

\label{lastpage}

\end{document}